\documentclass[%
reprint,
showkeys,
superscriptaddress,
nobibnotes,
amsmath,amssymb,
pra,
]{revtex4-2}

\usepackage[normalem]{ulem}
\usepackage{graphicx}% Include figure files
\usepackage{dcolumn}% Align table columns on decimal point
\usepackage{bm}% bold math
\usepackage[dvipsnames]{xcolor}
\usepackage{hyperref}% add hypertext capabilities
\usepackage{braket}

\begin{document}
\title{Quantum synchronization of spin-1 system: Enhanced synchronization due to additive Lindblad operators}
\author{Tran Duong Anh-Tai}
\altaffiliation{These authors contributed equally to this work.}
\affiliation{Homer L. Dodge Department of Physics and Astronomy, The University of Oklahoma, 440 W. Brooks Street, Norman, Oklahoma 73019, USA}
\affiliation{Center for Quantum Research and Technology, The University of Oklahoma, 440 W. Brooks Street, Norman, Oklahoma 73019, USA}
\author{S. Zhong}
\altaffiliation{These authors contributed equally to this work.}
\affiliation{Homer L. Dodge Department of Physics and Astronomy, The University of Oklahoma, 440 W. Brooks Street, Norman, Oklahoma 73019, USA}
\affiliation{Center for Quantum Research and Technology, The University of Oklahoma, 440 W. Brooks Street, Norman, Oklahoma 73019, USA}
\author{X. Molenda}
\affiliation{Homer L. Dodge Department of Physics and Astronomy, The University of Oklahoma, 440 W. Brooks Street, Norman, Oklahoma 73019, USA}
\affiliation{Center for Quantum Research and Technology, The University of Oklahoma, 440 W. Brooks Street, Norman, Oklahoma 73019, USA}
\author{A. M. Marino}
\affiliation{Homer L. Dodge Department of Physics and Astronomy, The University of Oklahoma, 440 W. Brooks Street, Norman, Oklahoma 73019, USA}
\affiliation{Center for Quantum Research and Technology, The University of Oklahoma, 440 W. Brooks Street, Norman, Oklahoma 73019, USA}
\affiliation{Quantum Information Science Section, Computational Sciences and Engineering Division, Oak Ridge National Laboratory, Oak Ridge, TN 37831, USA.}
\thanks{This manuscript has been authored in part by UT-Battelle, LLC, under
contract DE-AC05-00OR22725 with the US Department of Energy
(DOE). The publisher acknowledges the US government license
to provide public access under the DOE Public Access Plan
(http://energy.gov/downloads/doe-public-access-plan).}
\author{D. Blume}
\affiliation{Homer L. Dodge Department of Physics and Astronomy, The University of Oklahoma, 440 W. Brooks Street, Norman, Oklahoma 73019, USA}
\affiliation{Center for Quantum Research and Technology, The University of Oklahoma, 440 W. Brooks Street, Norman, Oklahoma 73019, USA}

\date{August 2025}

\begin{abstract}
We experimentally investigate 
quantum  synchronization for one of the simplest possible quantum systems, namely an externally driven few-level system with equally spaced energy levels effectively acting as a spin-1 system.
 Coupling to excited auxiliary states, we realize additive effective Lindblad operators  that are associated with non-conventional dissipative pathways,
which are shown  to enhance, in certain parameter regimes, quantum synchronization. The experimental set-up, which utilizes cold $^{87}$Rb atoms in a MOT, and associated synchronization extraction protocol 
 are benchmarked carefully through dedicated  
 simulations. Convincing agreement is found between experiment and simulations. The dissipation engineering approach established in our work can be readily extended to
 systems with more energy levels, such as effective spin-$3/2$ or spin-$2$ systems, and has implications for quantum synchronization studies in higher-spin systems as well as for a wide range of quantum science
 studies and technology applications. 
\end{abstract}

\maketitle

\section{Introduction}
Modeling the emergence of collective behaviors in complex systems is a highly challenging task in both the classical and quantum mechanical worlds. In classical physics, synchronization---the establishment of a coordinated rhythm of self-sustained periodic oscillators due to their weak interactions---provides a route to understanding collective behaviors~\cite{pikovsky2001synchronization}. The paradigmatic Kuramoto model, e.g., features a transition from a non-phase locked (non-synchronized) to a phase locked (synchronized) regime as the normalized coupling strength is increased, either by increasing the coupling strength coefficient (``spring stiffness'') or by decreasing the damping or dissipative coefficient~\cite{Kuramoto1975,RevModPhys.77.137}. In the synchronized regime, a limit cycle stabilizes the amplitudes of the oscillations while the phases of the oscillators lock to one another or to an external driving force. Classical synchronization plays a critical role in stabilizing power grids~\cite{Motter2013,Nishikawa_2015}, the pulsing of the heart~\cite{vanderPol1927,doi:10.1137/0150098}, and neuronal networks in the brain~\cite{haken2002brain,Ma2017}.

While phase locking in classical systems occurs at well-defined  phases (geometric angles), the Heisenberg uncertainty principle limits how well  the conjugate variables phase and amplitude can be known in quantum mechanics. Phase localization in quantum mechanics is thus instead identified by analyzing probability functions such as the Wigner or Husimi-Q functions~\cite{laskar2020observation,PhysRevA.101.062104,PhysRevLett.121.053601,doi:10.1126/sciadv.ady5649,PhysRevA.105.062206,zhang2023,waechtler2026}. By comparison, the research area of quantum synchronization, which may be broadly defined  as investigating how quantum mechanics modifies classical collective behavior, is less developed (see Ref.~\cite{schmolke2026synchronizationquantumregime} for a review). Recent works suggest fundamental connections between classical synchronization in non-linear dynamical systems and quantum phenomena such as the emergence of quantum coherence and entanglement in correlated many-body systems, formulated through the lens of quantum information science concepts~\cite{PhysRevLett.121.063601,SciPostPhys.12.3.097,PhysRevLett.111.103605,Witthaut2017,PhysRevA.91.012301}. 

Adopting a bottom-up perspective, our aim in the present work is to realize quantum synchronization in one of the simplest quantum systems, namely in an isolated externally driven spin-1 system with equal energy spacings~\cite{PhysRevLett.121.053601,PhysRevA.99.043804,Tan2022halfintegervs,PhysRevResearch.2.023026}. By virtue of the finite number of energy levels (three in our case), spin systems have no classical analog. Quantum synchronization of an isolated spin-1/2 system has been investigated experimentally using  a single trapped ion~\cite{zhang2023}. In this work, the external drive and the dissipative processes were controlled independently. Specifically, the drive was implemented via a microwave horn while the dissipation was engineered by optically coupling to excited, finite-lived states. In other systems, coherent and incoherent processes were controlled independently via a trotterization or stroboscopic reset-type approach~\cite{doi:10.1126/sciadv.ady5649,waechtler2026,PhysRevResearch.2.023026} or by utilizing noise~\cite{Tao2025}. 

Our work, which utilizes---as Ref.~\cite{laskar2020observation}---the $F=1$ hyperfine ground state manifold of a $^{87}$Rb atom, employs a hybrid approach in which one coupling beam does not fully independently control the coherent or the incoherent processes but in which instead the combination of a control-synchronize and a probe beam is used to simultaneously realize coherent and incoherent processes and in which a distinct third decay beam is used to independently realize additional incoherent processes.   Our experimental results, with support from theory, demonstrate that the implemented scheme has distinct advantages, allowing for finite quantum synchronization to be observed experimentally for parameter combinations where it would be, based on a naive analysis, zero. A key realization is that the scheme in which the same lasers generate both coherent and incoherent processes can be described by an effective spin-1 master equation~\cite{PhysRevA.85.032111,molenda}. In this effective spin-1 master equation, the laser couplings that generate the external coherent drive and thus set the phase reference for the spin-1 system to synchronize to, also generate dissipators that contain interference-like terms~\cite{molenda}. The fact that interference-like terms, which are usually associated with coherent behavior, appear in the effective dissipators highlights the uniqueness of the dissipators at play.  We experimentally investigate the effects of the interference-like terms and showcase their utility for boosting quantum synchronization. 

The results presented  have implications for quantum synchronization studies of spin chains as well as systems with spin and spatial degrees of freedom. More broadly, the effective dissipators realized in our work have implications for the steadily expanding field of dissipation engineering~\cite{Verstraete2009,Harrington2022}. On the one hand, studies like the one presented in our work provide stringent tests for how to effectively, yet reliably, describe open quantum systems. On the other hand, the findings pave the way for dissipation engineering in spin ensembles, such as ensembles of cold neutral ground state or Rydberg atoms~\cite{w3x9-ll79}, which can then be used to experimentally realize dissipative quantum phase transitions~\cite{ROSSINI20211,PhysRevE.89.022102,PhysRevX.5.031028} and harness their (synchronization) properties for, e.g., enhanced quantum sensing applications~\cite{RevModPhys.89.035002,PhysRevA.111.012410,Lai2025,PhysRevLett.123.250401}. 

The remainder of this paper is organized as follows. Section~\ref{sec_overview} introduces the system, experimental set-up, and key concepts. Section~\ref{sec_dynamics} presents the dynamics of the system, presenting side-by-side comparisons of experimental data and simulation results. Section~\ref{sec_justification} provides theory and simulation support for the synchronization extraction protocol employed. Section~\ref{sec_synchro} presents quantum synchronization data  for the effective spin-1 model with interference-like terms due the presence of additive Lindbladian. Finally, Sec.~\ref{sec_conclusions} summarizes and provides an outlook.

\section{Overview of system, experimental set-up, and key concepts}
\label{sec_overview}

\begin{figure*}
    \centering
    \vspace*{-0.4in}
  \includegraphics[width=\linewidth]{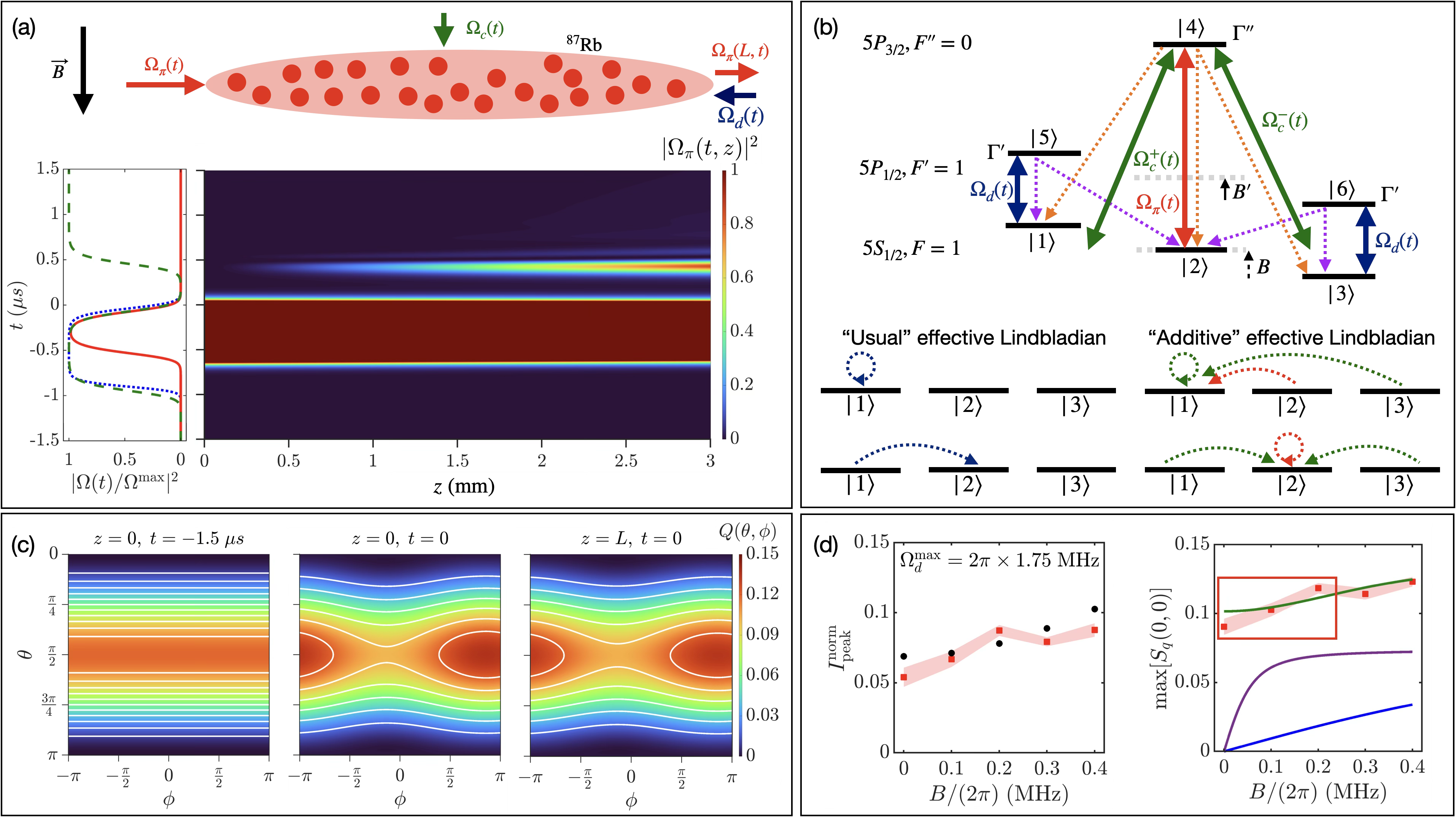}
    \caption{The upper part depicts the direction of the laser beams relative to the $^{87}$Rb atoms in a MOT. The control beam impinges on the MOT at a 90~$^{\circ}$ degree angle relative to the long MOT axis while the probe and decay beams impinge  parallel to the long MOT axis (anti-parallel to each other). The normalized incoming beam temporal profiles are shown on the left. The probe beam (solid red line) and control beam (dashed green line) act, when on simultaneously,  as the external drive as well as generators of effective dissipators. The decay beam profile (dotted blue line) closely resembles that of the control-synchronize beam. The synchronization stage ends when the probe, control-synchronize, and decay beam are turned off. When the control beam is turned on again after a variable hold time $\tau$ (here, $\tau= 0.4$~$\mu$s), we refer to it as a control-read beam. The color plot shows $|\Omega_{\pi}(t,z)|^2$ in MHz$^2$  for $B=2\pi\times 0.2$~MHz, $\Omega_d^\mathrm{max} = 2\pi\times 1.75$~MHz,  $\chi^S = 0$, and $\tau= 0.4$~$\mu$s, demonstrating that the atomic medium modifies the probe beam intensity as the light travels through the MOT.
    The retrieved probe signal is collected at the end of the MOT (in our simulations, the MOT is treated as a 1D medium, which extends from $z=0$ to $z=3$~mm). (b) The energy level diagram of the $^{87}$Rb atom is shown together with the couplings introduced by the laser beams (double arrows) and the dissipative decay channels due to the finite lifetimes of the excited $F{{''}}=0$ and $F{'}=1$ states (dotted arrows). The $F=1$ states constitute the effective spin-1 model while the $F'=1$ and $F''=0$ states serve as auxiliary states. Coupling to the auxiliary states generates effective coherent and incoherent processes within the $F=1$ ground state manifold. The structure of the effective Lindbladian due to the decay beams (bottom left) and the combination of the control-synchronize and probe beams (bottom right) is illustrated. There are three pathways that contribute to the dissipative decay into state $\ket{1}$, including one pathway that represents dissipative decay out of the limit cycle state. (c) Husimi-Q functions $Q(\theta,\phi)$ of the spin-1 system at the beginning of the MOT, prior to ($t=-1.5$~$\mu$s, left) and after ($t=0$, middle) the application of the external drive. The  Husimi-Q function  of the spin-1 system at $t=0$ at the end of the MOT (right) is very similar to that at the beginning of the MOT, indicating that the MOT set-up can be used to determine synchronization of an isolated spin-1 system. (d) Left panel: The red symbols  show the maximum $I_{\text{peak}}^{\text{norm}}$ of the experimentally measured normalized retrieved probe beam intensity at time $t=\tau$ as a function of $B$ for fixed maximum magnitude $\Omega_d^{\text{max}}$ of the decay beam. For fixed $B$, $I_{\text{peak}}^{\text{norm}}$ is determined through fits to experimental data that vary $\chi^S$.  The shaded region depicts the error band. For comparison, the black symbols show $I_{\text{peak}}^{\text{norm}}$ extracted from  simulation data. Right panel: The red symbols and error bars show the maximum of the synchronization ($\text{max}[S_q(0,0)]$), extracted from the experimental data; the red box marks the parameter combinations for which the system reached its steady state (see text for details). For comparison, the green line shows $\text{max}[S_q(0,0)]$  for the effective spin-1 model with additive Lindbladian.  The purple and blue lines show results for the ``usual'' spin-1 model [see Eq.~(\ref{eq:modified_spin1_mastereq})],  which  accounts for equal dissipative decay from $\ket{1}$ to $\ket{2}$ and from $\ket{3}$ to $\ket{2}$. The decay rates for the purple and blue lines are 
    $\Gamma_{\text{decay}}$ and $\Gamma_{\text{decay}}+\Gamma_{\text{control}}$,
    respectively, where 
    $\Gamma_{\text{decay}}$ and $\Gamma_{\text{control}}$ depend on  the decay and control-synchronize beams, respectively, but not the probe beam. The additive structure of the Lindbladian in the effective spin-1 model (green solid line), which is in good agreement with the data extracted from experiment, leads to a clear enhancement of $\text{max}[S_q(0,0)]$, including a finite synchronization for $B=0$, compared to the predictions of the ``usual'' spin-1 model. } 
      \label{fig1}
\end{figure*}

The spin-1 system is realized using the $F=1$ hyperfine ground state manifold of $^{87}$Rb, labeled by $\ket{1}=\ket{F=1,m_F=-1}$, $\ket{2}=\ket{F=1,m_F=0}$, and $\ket{3}=\ket{F=1,m_F=+1}$ [see Fig.~\ref{fig1}(b)]. 
To experimentally implement effective coherent and incoherent couplings within this manifold, lasers couple states $\ket{1}$, $\ket{2}$, and $\ket{3}$ to the auxiliary states 
$\ket{4}=\ket{F{''}=0,m_F=0}$,  
$\ket{5}=\ket{F{’}=1,m_F=-1}$, 
and $\ket{6}=\ket{F{’}=1,m_F=+1}$. 
To obtain an effective spin-1 master equation for the Hilbert space spanned by $\ket{1}$, $\ket{2}$, and $\ket{3}$, we integrate out the auxiliary states at the level of second-order time-dependent perturbation theory~\cite{PhysRevA.85.032111,molenda}, with $\Omega_{\pi}^{\text{max}}/\Omega_{c}^{\text{max}}$ serving as the small parameter; here, $\Omega_{\pi}^{\text{max}}$ and $\Omega_{c}^{\text{max}}$ denote the maximum magnitude  of the probe and control beam Rabi coupling strengths, respectively [see Fig.~\ref{fig1}(b) for the coupling scheme].
Since the drive strengths of the resulting effective spin-1 Hamiltonian depend on the product $\Omega_{c}^{\text{max}} \Omega_{\pi}^{\text{max}}$, a small $\Omega_{\pi}^{\text{max}}$ is consistent with the notion that synchronization requires a sufficiently small drive strength for the limit-cycle state not to be destroyed and not to be deformed too much by the external drive~\cite{PhysRevA.99.043804,PhysRevA.101.062104,zhang2023}. 
The classical analog of this is that the amplitude of the oscillations needs to be stable for phase synchronization to occur.  

The master equation for the full 6-state Hamiltonian is given by~\cite{molenda} 
\begin{eqnarray} 
\label{eq_masterequation_6level}
\frac{d}{dt} \rho^{\text{full}}(t)=-i [H_{\text{system}}^{\text{full}},\rho^{\text{full}}(t)] +\nonumber \\ \sum_{j=1}^7 D[L_j^{\text{full}}](\rho^{\text{full}}(t)) -
\gamma_c\sum_{k,l=1;k\ne l}^6 \rho_{kl}^{\text{full}}(t)\ket{k}\bra{l},
 \end{eqnarray}
 where $D$ denotes the dissipator,
 \begin{eqnarray}
 D[L](\rho)=L \rho L^{\dagger}-\frac{1}{2} L^{\dagger} L \rho - \frac{1}{2} \rho L^{\dagger}L.
 \end{eqnarray}
The full 6-state Hamiltonian $H_{\text{system}}^{\text{full}}$, expressed in a rotating frame, depends on the Rabi coupling strengths $\Omega_{\pi}(t)$, $\Omega^{\pm}_c(t)$, and $\Omega_d(t)$ as well as the strength of an external magnetic field, which shifts the states $\ket{1}$ and $\ket{5}$ up and the states $\ket{3}$ and $\ket{6}$ down [see Fig.~\ref{fig1}(b)].
The energy shifts of the $F=1$ and $F'=1$ states are denoted by $B$ and $B'$, respectively. Throughout, we report the strength of the external magnetic field in terms of the energy shift $B$~\cite{zhong2026light}.
The  phases and shape of $\Omega_{\pi}(t)$, $\Omega^{\pm}_c(t)$, $\Omega_d(t)$ are discussed below and in Appendix~\ref{appendix_inmedium_simulations}. 
The Lindbladian $L_j^{\text{full}}$, $j=1-7$, account for the finite lifetimes of state $\ket{4}$ (decay rate $\Gamma''$, decay of equal probability to all three $F=1$ ground states) and of states $\ket{5}$ and $\ket{6}$ (decay rate $\Gamma'$, each with decay of equal probability to two of the $F=1$ ground states). 
We also account for dephasing using the experimentally measured dephasing rate $\gamma_c$~\cite{zhong2026light}; $\gamma_c$ is more than 50 times smaller than the decay rates $\Gamma'$ and $\Gamma''$ ($\approx 2 \pi \times 0.1$~MHz versus $\approx 2 \pi \times 6$~MHz~\cite{zhong2026light}) and is found to have a perturbative effect on the extracted quantum synchronization. 

The parameters of the full Hamiltonian $H_{\text{system}}^{\text{full}}$ are fine-tuned such that the system Hamiltonian $H_{\text{system}}$ of the effective spin-1 system master equation,
\begin{eqnarray} 
\label{eq_masterequation_3level}
\frac{d}{dt} \rho(t)=-i [H_{\text{system}},\rho(t)] +\sum_{j=1}^7 D[L_j](\rho(t)) - \nonumber \\
\gamma_c\sum_{k,l=1;k\ne l}^3 \rho_{kl}(t)\ket{k}\bra{l},
\end{eqnarray}
contains equal effective coherent $\ket{1} \leftrightarrow\ket{2}$ and $\ket{2} \leftrightarrow\ket{3}$ couplings and no effective coherent $\ket{1} \leftrightarrow \ket{3}$ coupling~\cite{molenda}; this is arguably the, or one of the, simplest possible coupling schemes that allows for the perturbative build-up of finite $\rho_{12}$ and $\rho_{23}$ coherences, which are essential for phase localization~\cite{galve2017quantum,PhysRevLett.111.103605,PhysRevE.109.054207}. In addition,  the parameters of  $H_{\text{system}}^{\text{full}}$ are fine-tuned such that the Stark shifts of the effective Hamiltonian correspond to equal $\ket{1} \leftrightarrow \ket{2}$ and $\ket{2} \leftrightarrow \ket{3}$ energy spacings. As a result, the effective spin-1 Hamiltonian $H_{\text{system}}$ can be cleanly written as
\begin{eqnarray}
\label{eq_ham_spin1}
    H_{\text{system}}=
    -\Delta_{\text{eff}} \left( \ket{1}\bra{1}-\ket{3}\bra{3} \right)+  \nonumber \\
    i \frac{\Omega_{\text{eff}}\exp(-i \phi_{\text{eff}})}{\sqrt{2}} 
    \left[\ket{1}\bra{2}+\exp[-i (\alpha+\pi)]\ket{2}\bra{3}
    \right] + \nonumber \\ \mbox{h.c.}  
    ,
\end{eqnarray}
where $\Delta_{\text{eff}}$, $\Omega_{\text{eff}}$, and $\phi_{\text{eff}}$ are real quantities that depend on the parameters of the six-state model (see Refs.~\cite{molenda,footnote_notation} and Appendix~\ref{appendix_effective_spin1_models}). The phase $\alpha$ is set by the phase $\chi_0$ of the electric field associated with the probe beam and the phases $\chi_+$ and $\chi_-$ associated with the electric fields of the plus- and minus-components of the control-synchronize beam~\cite{molenda},
\begin{eqnarray}
\label{eq_alpha}
\alpha=\chi_++\chi_--2\chi_0.
\end{eqnarray}

The incoherent part of the effective spin-1 master equation is governed by seven distinct effective Lindbladian $L_j$, one for each of the Lindbladian $L_j^{\text{full}}$. The lower part of Fig.~\ref{fig1}(b) illustrates the structure of the Lindblad operators $L_j$. The Lindbladian $L_j^{\text{full}}=\sqrt{\Gamma’/3}|j\rangle\langle4|$ with $j=1-3$, e.g., map to $c_{j,1}^{(4)}| j\rangle\langle1|+c_{j,2}^{(4)}| j\rangle\langle2|+c_{j,4}^{(j)}| j\rangle\langle3|$, where the coefficients $c_{j,k}^{(l)}$ depend on the system parameters (see Appendix~\ref{appendix_effective_spin1_models}). The unusual additive structure of the effective Lindbladian reflects that, within the reduced Hilbert space, dissipative decay into state $\ket{1}$ may occur from state $\ket{1}$, $\ket{2}$, or $\ket{3}$; it is important to note that these processes are not independent from each other. Decay from state $\ket{1}$ can be interpreted  as self-dissipation, which arises by going from $\ket{1}$ to $\ket{4}$ via $\Omega_c^+(t)$ and then returning to $\ket{1}$ due to the finite lifetime of state $\ket{4}$. Decay from state $\ket{2}$ proceeds by going from $\ket{2}$ to $\ket{4}$ via $\Omega_{\pi}(t)$ and then decaying to $\ket{1}$ due to the finite lifetime of state $\ket{4}$. Decay from state $\ket{3}$ proceeds by going from $\ket{3}$ to $\ket{4}$ via $\Omega_{c}^-(t)$ and then decaying to $\ket{1}$ due to the finite lifetime of state $\ket{4}$. Plugging the additive Lindbladian $L_{j}$, $j=1$, $2$, or $3$, into the dissipator and expanding terms, it can be seen that the effective dissipator contains ``cross’’ or ``interference-like’’ terms~\cite{molenda}. 
These cross terms play an important role in the phase localization.

Since the dynamics obtained by solving the full 6-level master equation and  the effective spin-1 master equation agree well  (see Appendix~\ref{appendix_effective_spin1_models}) and since good agreement is also found between experiment and simulations for all parameter combinations considered in this work (see Sec.~\ref{sec_dynamics}, Appendix~\ref{appendix_fulldata}, and the supplemental material~\cite{SM}), we conclude that our experimental set-up realizes the effective spin-1 model introduced above. Throughout, we quantify phase localization by the synchronization measure $S_q$~\cite{PhysRevResearch.2.033422}, 
\begin{eqnarray}
\label{eq_synchro_def}
S_q(t,z)=|\rho_{12}(t,z)+\rho_{23}(t,z)|.\end{eqnarray}
Note that the $z$-dependence is included since we determine $S_q$ below at $z=0$ and $z=L$. 
Equation~(\ref{eq_synchro_def}) is obtained by evaluating the magnitude of the expectation value of a phase operator~\cite{pegg,PhysRevA.43.3795} and, as such, can be interpreted as constituting a direct generalization of the classical phase localization measure. 
In the absence of the external drive (i.e., for $\Omega_{\pi}=0$), the steady-state density matrix $\rho^{\text{ss}}$ of the effective spin-1 model is
 given by the limit cycle state $\ket{2}\bra{2}$~\cite{PhysRevLett.121.053601,Tan2022halfintegervs}. 
 This limit cycle state corresponds to a Husimi-Q function [see Fig.~\ref{fig1}(c), left panel] that has no phase preference (i.e., no $\phi$ dependence) and maximal amplitude at the equator (state $\ket{2}\bra{2}$  corresponds to $\theta=\pi/2$ while states $\ket{1}\bra{1}$ and $\ket{3}\bra{3}$ correspond to $\theta=\pi$ and $0$, respectively). 
 The effective spin-1 system is synchronized to the external drive if the steady-state density matrix is stabilized near the limit cycle of the undriven system and if $S_q$ is finite~\cite{molenda}.   
The Husimi-Q functions in the middle and on the right in Fig.~\ref{fig1}(d) are for a synchronized state at the beginning and the end of the MOT, respectively. Their similarity indicates that the presence of the atomic medium influences the synchronization of the isolated atom only perturbatively; thus, the set-up can be used to obtain the synchronization of an isolated driven spin system.

Our experiment utilizes $^{87}$Rb atoms in an elongated magneto-optical trap (MOT), as schematically shown at the top of Fig.~\ref{fig1}(a). Due to the low density of about $0.72\times10^{16}\mathrm{m}^{-3}$, the atoms within the MOT do not interact. Correspondingly, we realize many copies of the same quantum system simultaneously. The experimental set-up is very similar to that described in a previous paper written by some of us~\cite{zhong2026light}. An  improvement of the experimental set-up compared to Ref.~\cite{zhong2026light} is the optimization of polarization gradient cooling during the preparation stage (optimization of the cooling-beam detuning and the cooling duration),
which results in a lower temperature of the atoms compared to our previous work. The lower temperature leads to a somewhat smaller dephasing rate $\gamma_c$ than in our previous work [$\gamma_c=2 \pi \times 0.101(5)$], thereby further enlarging the time scale separation between the engineered dissipation rates and the ``unwanted''  dephasing rate $\gamma_c$. In addition to ensuring that $S_q$ is, to a very good approximation, independent of $\gamma_c$, this improvement allows us to obtain reliable retrieved probe signals for somewhat larger hold times than in Ref.~\cite{zhong2026light}. 
As discussed in more detail below, obtaining a retrieved probe signal for hold times as large as $1/(2 \pi \times B)$ aids with the extraction of the quantum synchronization. Recall that the magnetic field strength $B$ is reported as the equal energy splittings of the $\ket{1}\leftrightarrow \ket{2}$ and $\ket{2}\leftrightarrow \ket{3}$ states (this work considers $B=0$ to $B=2\pi \times 0.4$~MHz). 

To determine $S_q$, we employ  state preparation protocol~I (see Appendix~\ref{appendix_experiment} for details), which aims to prepare the atoms in the limit cycle state $\ket{2}\bra{2}$, and set $B$ to the desired value.  After a wait time, the control-synchronize, probe, and decay beams are turned on adiabatically, using the pulse sequence shown on the left in Fig.~\ref{fig1}(a). The probe beam is $\pi$-polarized and impinges on the MOT at a 90$^{\circ}$ angle relative to the control beam. The decay beam is also $\pi$-polarized and impinges on the MOT anti-parallel to the probe beam. The probe beam travels along the weakly-confined direction of the MOT and its intensity profile changes as it travels through the MOT [Fig.~\ref{fig1}(a) shows example simulation results]. The intensity profiles of the control-synchronize, control-read, and decay beams, in contrast, are not modified by the atomic medium. We arrive at this conclusion by analyzing the overlap for each of the experimentally measured intensity profiles with and without the MOT.

The laser frequencies of the three beams are kept fixed for all the experimental runs. Specifically, they are set such that the detuning is zero when $B=0$: the probe beam is on resonance with the $\ket{2}\leftrightarrow \ket{4}$ transition, 
the ``plus- and minus-components'' of the control
beam are on resonance with the $\ket{1}\leftrightarrow \ket{4}$ and $\ket{3}\leftrightarrow \ket{4}$ transitions, respectively, and
the decay beam is on resonance with the $\ket{1}\leftrightarrow \ket{5}$ and $\ket{3}\leftrightarrow \ket{6}$ transitions for $B=0$. The implementation of the decay beam, which was not used in our previous atomic memory work~\cite{zhong2026light}, is discussed in Appendix~\ref{appendix_experiment}. Dedicated calibration measurements and simulation-experiment comparisons  show that the maximum magnitudes $\Omega_{\pi}^{\text{max}}$ and $\Omega_{c}^{\text{max}}$ of the Rabi coupling strengths  as well as the atom number and atomic density are the same as in Ref.~\cite{zhong2026light} within the estimated uncertainties, indicating long-term stability of the apparatus. For our analysis, we use $\Omega_{\pi}^{\text{max}}= 2\pi \times 1$~MHz, $\Omega_{c}^{\text{max}}=2\pi \times 9.5$~MHz, and  $c\mu_a=2.118 \times 10^7$~$\mu$s$^{-2}$, where $c$ denotes the speed of light in vacuum and $\mu_a$ is related to the optical depth and can be thought of as the effective coupling strength of the $\ket{2} \leftrightarrow \ket{4}$ transition (see Appendix~\ref{appendix_inmedium_simulations}). The maximum magnitude $\Omega_d^{\text{max}}$ of the decay beam Rabi coupling strength is calibrated through comparison with simulations for one laser power and then scaled accordingly for the other laser powers. This paper shows results for $\Omega_d^{\text{max}}$ between $0$ and $2 \pi \times 2.75$~MHz.

When the control-synchronize and probe beams are both on, the spin system is being synchronized to the external drive.
The phase reference is set by the external drive. Since our aim is to measure the steady-state synchronization, the beams must be on for a sufficiently long time. For the  initial state considered for synchronization measurements (state preparation approach~I; see Appendix~\ref{appendix_experiment}), the steady state is reached within less than a microsecond for most parameter combinations considered  (see below for details). To experimentally characterize the steady state, we utilize a spin wave-based   interferometric measurement protocol that produces interference patterns in the retrieved probe signal from which we can deduce information about the steady-state atomic density matrix elements $\rho_{12}$ and $\rho_{23}$~\cite{laskar2020observation}. 

Specifically, we turn all three  beams off at $t \approx 0$. This turn-off stores $\rho_{12}(0)$ and $\rho_{23}(0)$ in two spin waves or matter-like dark state polaritons, which evolve  essentially freely during the hold time $\tau$~\cite{laskar2020observation,zhong2026light,PhysRevA.71.041801,PhysRevA.65.022314,PhysRevA.95.013818}. Note that although the spatial dependence is not explicitly indicated, it is understood that the $\rho_{jk}(0)$ refer, for the remainder of this section, to the $t=0$ density matrix elements at the end of the MOT. The two dark state polaritons can be thought of as being associated with two distinct three-level $\Lambda$ systems, one made up of states $\ket{1}$, $\ket{2}$, and $\ket{4}$ and the other made up of states $\ket{2}$, $\ket{3}$, and $\ket{4}$. During the free evolution time, the dark state polaritons acquire an additional phase that is, just as one might expect naively from the free evolution of the Schr\"odinger equation,  set by the $\ket{1} \leftrightarrow \ket{2}$ and $\ket{2} \leftrightarrow \ket{3}$ bare energy splittings of $2 \pi \times B$~\cite{laskar2020observation,zhong2026light}. After a variable hold time $\tau$, the control beam is turned back on (control-read beam) to convert the atomic information stored in the polaritons to the light degree of freedom~\cite{laskar2020observation,zhong2026light}. As a result, the retrieved probe signal at the end of the MOT (i.e., at $z=L$)  encodes information about the atomic coherences $\rho_{12}(0)$ and $\rho_{23}(0)$ that characterize the effective spin-1 system in the steady state (i.e., prior to turning off the beams at $t \approx 0$).

Assuming that the polaritons evolve freely and that the processes of ``storing'' at $t=0$ and ``retrieving'' at $t=\tau$ occur instantaneously, the retrieved intensity $I_{\pi}(t,L)$, normalized to the maximum intensity of the input probe pulse $I_{\pi}^{\text{max}}$, 
\begin{eqnarray}
    I_{\pi}^{\text{norm}}(t)=\frac{I_{\pi}(t,L)}{I_{\pi}^{\text{max}}},
    \end{eqnarray}
    at the end of the MOT as a function of time $t=\tau$ can be expressed as~\cite{laskar2020observation}
\begin{eqnarray}
\label{eq_retrieved_intensity}
    I_{\pi}^{\text{norm}}(\tau)
    = A_I^2 
    \exp \left( - 2 \gamma_c \tau \right) 
    \bigg[
    |\rho_{12}(0)|^2 + |\rho_{23}(0)|^2
    + \nonumber \\
    2 |\rho_{12}(0) \rho_{23}(0)| \cos \left( \varphi_{\text{R}}
   \right)
    \bigg], \nonumber \\
\end{eqnarray}
where the $t=0$ coherences
$\rho_{12}(0)$ and $\rho_{23}(0)$  are written as
\begin{eqnarray}
    \rho_{12}(0)&=&|\rho_{12}(0)| \exp \left[i\left( \varphi_{12}(0)+\chi_+-\chi_0\right)\right], \\
        \rho_{23}(0)&=&|\rho_{23}(0)| \exp \left[i \left(\varphi_{23}(0)-\chi_-+\chi_0\right)\right] ,
\end{eqnarray}
where the phase difference $\varphi_{\text{R}}$ is defined through 
\begin{eqnarray}
   \varphi_{\text{R}}=  \varphi_{12}(0)+\varphi_{23}(0) +\chi^S - 2 \pi \times 2B \tau
\end{eqnarray}
with
\begin{eqnarray}
    \chi^S=\chi_+-\chi_-,
\end{eqnarray}
and where the off-diagonal coherences are assumed to evolve freely for $0 \le t \le \tau$,
\begin{eqnarray}
\label{eq_coherences_free}
\rho_{12}(t)&=&\rho_{12}(0) \exp(-i 2\pi \times B t - \gamma_c t),\\
\rho_{23}(t)&=&\rho_{23}(0) \exp(-i 2\pi \times B t - \gamma_c t).
\end{eqnarray}

The quantum synchronization $S_q(t,L)$ at the end of the MOT at time $t=0$ can then, within first-order perturbation theory, be expressed as~\cite{laskar2020observation} 
\begin{eqnarray}
\label{eq_synchronization}
    S_q(0,L)
    = A_S \times  \nonumber \\
    \bigg[
    |\rho_{12}(0)|^2 + |\rho_{23}(0)|^2
    + 
    2 |\rho_{12}(0) \rho_{23}(0)| \cos \left( \varphi_{\text{sync}}
   \right)
    \bigg]^{1/2}, \nonumber \\
\end{eqnarray}
where
\begin{eqnarray}
    \varphi_{\text{sync}}=\varphi_{12}(0)-\varphi_{23}(0) +\alpha . 
\end{eqnarray}
The proportionality factors $A_I^2$ and $A_S$ in Eqs.~(\ref{eq_retrieved_intensity}) and Eq.~(\ref{eq_synchronization}) depend on the optical depth.

While $I_{\pi}^{\text{norm}}(\tau)$ depends on the phase difference $\chi^S$
of the plus- and minus-components of the control-synchronize beam,
$S_q(0,L)$ depends on the sum $\chi_+ + \chi_-$.
Experimentally, $\chi^S$ can  be controlled by tuning the linear polarization angle of the control beam to within a few degrees accuracy~\cite{zhong2026light}. Since our current experimental set-up does not have an absolute phase reference, we do not know $\chi_++\chi_-$ and $\chi_++\chi_--2\chi_0$. However, recording $I_{\pi}^{\text{norm}}(\tau)$ as functions of $\chi^S$ 
and $\tau$, we can determine the phases $\varphi_{\text{R}}$ for which $\cos(\varphi_{\text{R}})$ is equal to $-1$ and $+1$ and, correspondingly, $I_{\text{peak}}^{\text{norm}}$ [the largest value of  $I_{\pi}^{\text{norm}}(\tau)$] and $I^{\text{norm}}_{\text{bg}}$ [the smallest value  of  $I_{\pi}^{\text{norm}}(\tau)$] for a given $B$ and $\Omega_d^{\text{max}}$
(recall, $\Omega_{c}^{\text{max}}$ and $\Omega_{\pi}^{\text{max}}$ are kept fixed throughout).   
We can then 
subsequently obtain, except for an overall scaling factor that needs to be supplied by theory, the maximum and minimum of $S_q$ from the experimentally measured retrieved signal (see Sec.~\ref{sec_synchro} for details). As alluded to above, since we do not know $\chi_++\chi_-$, we can ``only'' measure the maximal possible value and the minimal possible value of $S_q$ but not for which phases the maximum and minimum are realized.  Note that if $|\rho_{12}(0)|$ and $|\rho_{23}(0)|$ are equal, 
then $I^{\text{norm}}_{\text{bg}}$ is equal to $0$. 

The red squares in the left panel of
Fig.~\ref{fig1}(d) show examples of  $I_{\text{peak}}^{\text{norm}}$ extracted from experimental data 
while the right panel shows the corresponding maximum of the quantum synchronization, extracted using the approach discussed in Sec.~\ref{sec_synchro}. 
The agreement with the theory predictions for the effective spin-1 model, which features additive Lindbladian, is quite good (green line). In contrast, two other spin-1 models (purple and blue lines), which do not properly ``renormalize'' the dissipative decay pathways,  disagree with the experimental data. Figure~\ref{fig1}(d) demonstrates that our experimental set-up realizes the effective spin-1 model and that the additive structure of the Lindbladian enhances the quantum synchronization
for the parameter combinations considered.

\section{Side-by-side comparison of
experimental and simulation data}
\label{sec_dynamics}

\begin{figure}[!htbp]
    \centering
    \includegraphics[scale=.14]
{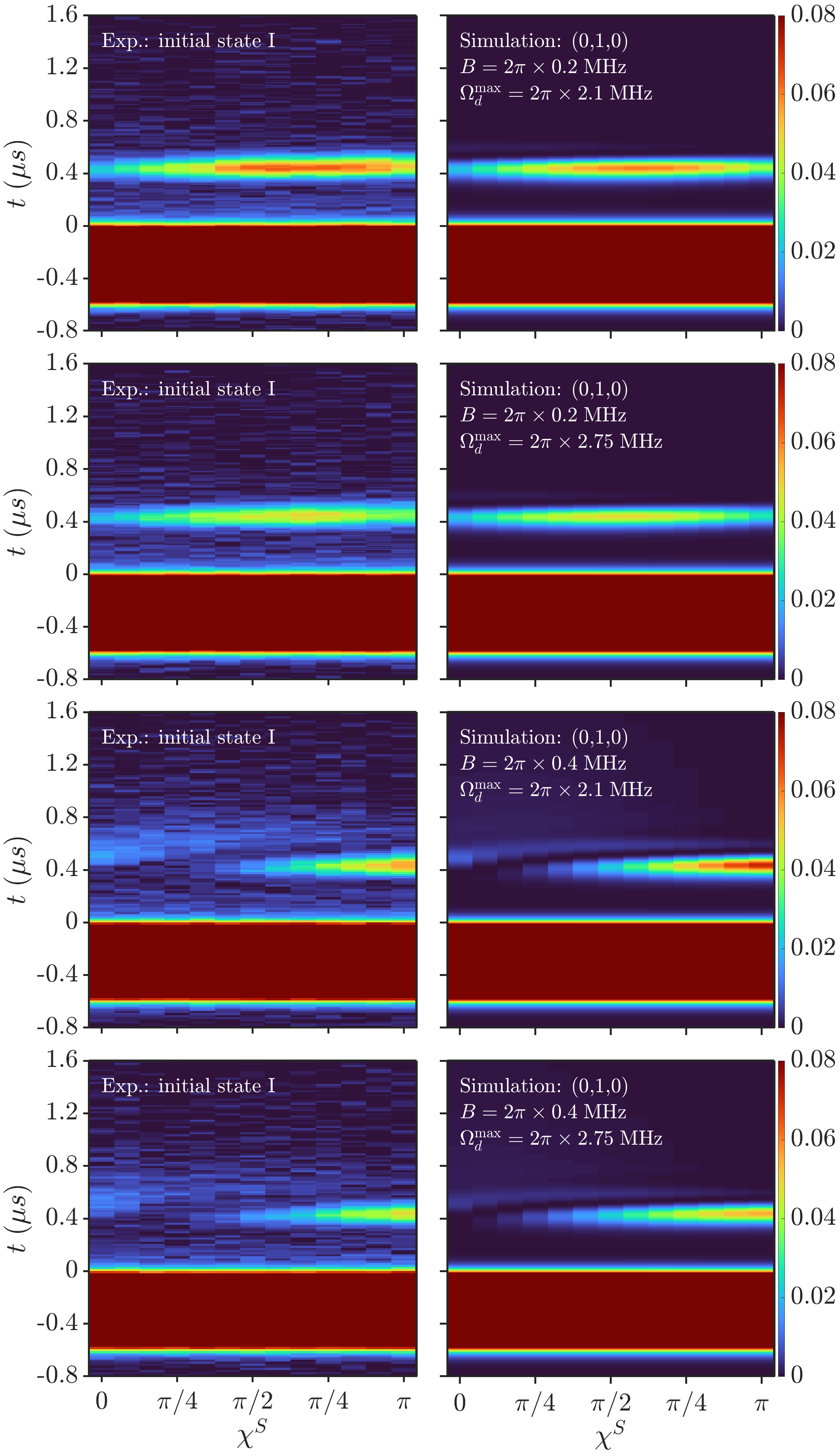}
    \caption{Normalized probe beam intensity $I_{\pi}^{\text{norm}}(t)$ at the end of the MOT (color coded) for the hold time $\tau=0.4$~$\mu$s as functions of the time $t$ and the relative phase $\chi^S$ of the plus- and minus-components of the control-synchronize beam. The left column shows experimental data obtained using the initial state preparation~I while the right column shows simulation results for the initial state $\ket{2}\bra{2}$ [denoted by $(0,1,0)$ in the figure labels]. 
    Two magnetic field strengths are used, namely $B=2 \pi \times 0.2$~MHz (top two rows) and $B=2 \pi \times 0.4$~MHz (bottom two rows), and two different maximal magnitudes $\Omega_d^{\text{max}}$ of the Rabi coupling strength of the  decay beam. The labels in the figure report the values used in the simulations. 
        The decay Rabi coupling strength is extracted through comparisons with simulation data. The agreement between the experimental data and simulation results is very good.  }
    \label{fig2}
\end{figure}

In our first experiment, we aim to pump the atoms into the   limit cycle state $\ket{2}\bra{2}$ prior to the application of the synchronization and read-out beams using our initial state preparation approach~I (see Appendix~\ref{appendix_experiment}). We measure the probe beam intensity as a function of  time $t$ for various relative phases $\chi^S$ between the plus- and minus-components of the control beam after the beam has passed through the MOT. 
Figure~\ref{fig2} compares experimental data (left column) with simulation results (right column) for two different magnetic field strengths, namely for $B=2 \pi \times 0.2$~MHz (the top two rows) and for $B=2 \pi \times 0.4$~MHz (the bottom two rows), and two different maximum magitudes $\Omega_d^{\text{max}}$ of the Rabi coupling strength of the decay beam. The normalized probe beam intensity $I_{\pi}^{\text{norm}}(t)$ at the end of the MOT is color-coded in the figures. For the examples shown in Fig.~\ref{fig2}, the beams that govern the synchronization process are turned off at $t = 0$ and the control-read pulse  is turned on at $t = 0.4$~$\mu$s, i.e., the hold time is set to $\tau=0.4$~$\mu$s.
The figure demonstrates overall good agreement between the experimental data and the simulation results, which are obtained by self-consistently solving for the modification by the atomic medium  of the probe beam Rabi coupling strength and the atomic density matrix elements (states $\ket{1}$ through $\ket{6}$ are included in the simulations; see Ref.~\cite{zhong2026light} and Appendix~\ref{appendix_inmedium_simulations} for details).

Inspection of Fig.~\ref{fig2} reveals two key features. The first key feature is that the angle $\chi^S$ at which the retrieved signal is maximal depends primarily on $B$ and only secondarily on $\Omega_d^{\text{max}}$. The near-absence of the retrieved signal at $\chi^S \approx 3 \pi/8$ for $B=2 \pi \times 0.4$~MHz, e.g., is the result of a destructive interference between the 
two spin waves that are created at $t \approx 0$. As discussed in the previous section, these two spin waves  evolve essentially freely and independently during the hold time. Information carried by them, including the phase acquired during the hold time (free evolution time), is mapped to the light degree of freedom through the application of the control-read beam. In contrast to our earlier work, which used the same apparatus as the one in the present work to demonstrate a storage-retrieval protocol in a tripod system~\cite{zhong2026light}, the plus- and minus-components of the control-read beam have the same phase (the phase is, instead, added during the synchronization stage). Our experimental observations agree with Eq.~(\ref{eq_retrieved_intensity}), which states that the interference fringes are  governed by the $\chi^S$- and $B$-dependence of the  cosine. This statement is corroborated below through a quantitative fit analysis. Together, this suggests that the spin-wave picture, which plays a critical role in our synchronization extraction protocol, is applicable.  The second key feature revealed in Fig.~\ref{fig2}  is that,
for fixed magnetic field strength, 
the retrieved signal decreases with increasing $\Omega_d^{\text{max}}$, with the interference pattern
being weakly or possibly not at all impacted by the Rabi coupling strength of the decay beam. 
Intuitively this can be understood by viewing the decay beams as perturbing the spin wave formation, thereby leading to a reduction of the retrieved probe beam intensity.  

\begin{figure}[!thbp]
    \centering
    \includegraphics[width=\linewidth]{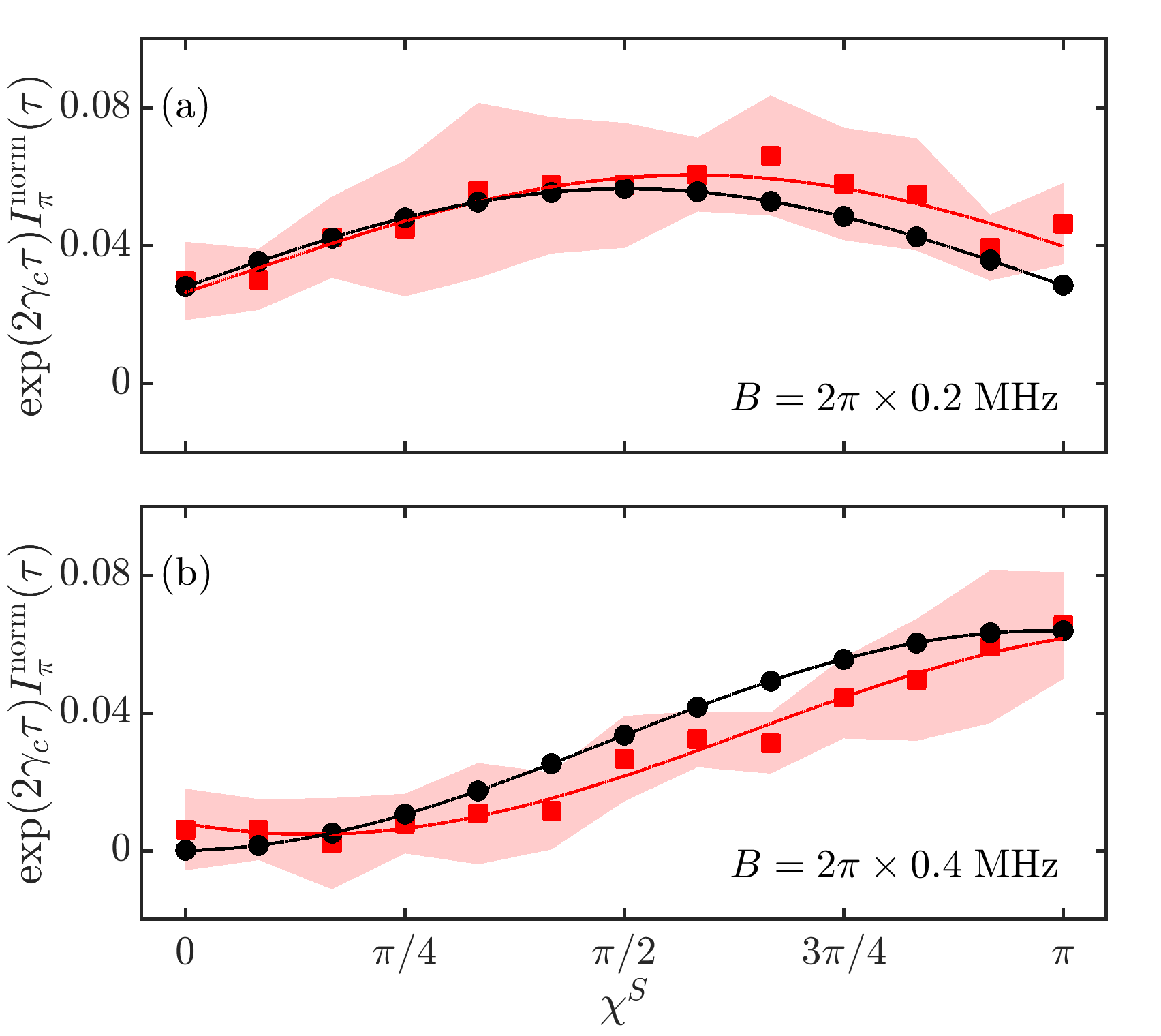}
    \caption{Red squares and black circles show the ``undamped" normalized probe beam intensity $\exp(2\gamma_c\tau)I_{\pi}^{\text{norm}}(\tau)$  as a function of the relative phase $\chi^S$ of the control-synchronize beam extracted from the experimental and simulation data, respectively, for $\tau=0.4$~$\mu$s and $\Omega_d^{\text{max}}= 2 \pi \times 2.75$~MHz. Panels (a) and (b) are for different magnetic field strengths (see labels in the figure).
    The red shaded error bands depict the error of the experimental data, combining the statistical uncertainty from 10 independent measurements and the uncertainty of $\gamma_c$. The solid red and solid black lines show fits of the weighted experimental data and the simulation data, respectively, to Eq.~(\ref{eq_fit}). 
    }
    \label{fig3}
\end{figure}

To extract quantum synchronization information from the retrieved probe signal, we isolate the oscillatory nature of the retrieved probe signal by ``undoing'' the exponential decay due to the dephasing, i.e., we multiply $I_{\pi}^{\text{norm}}(\tau)$  by $\exp(2 \gamma_c \tau)$. Figure~\ref{fig3} shows two examples, one for $B=2 \pi \times 0.2$~MHz and the other for $B=2 \pi \times 0.4$~MHz. Both examples are   for the largest decay beam Rabi coupling strength considered in our work, namely for $\Omega_d^{\text{max}}=2 \pi \times 2.75$~MHz.   
The red symbols show our experimental data, with the red shaded region indicating the error bars that account for both  the statistical uncertainty of ten independent measurements and  the uncertainty of the dephasing rate $\gamma_c$ ($\gamma_c$ is, as in Ref.~\cite{zhong2026light}, extracted from $B=0$ measurements).  The red solid line shows a least-square fit to the weighted experimental data using an oscillatory fit function of the form
\begin{eqnarray}
\label{eq_fit}
    \exp(2 \gamma_c \tau) I_{\pi}^{\text{norm}}(\tau)=\nonumber \\
    I_{\text{peak}}^{\text{norm}}\cos^2\left( 2\pi \times B \tau + \chi^S/2+\phi_{\text{shift}} \right) + I_{\text{bg}}^{\text{norm}},
\end{eqnarray}
where $I_{\text{peak}}^{\text{norm}}$, $\phi_{\text{shift}}$, and $I_{\text{bg}}^{\text{norm}}$ are treated as fitting parameters.
Our fits yield values for the background $I_{\text{bg}}^{\text{norm}}$ that are consistent with zero. Since we expect  that $I_{\text{bg}}^{\text{norm}}$ is zero (see also below), we interpret the fact that the fit results for  $I_{\text{bg}}^{\text{norm}}$  are consistent with zero as a nice consistency check. 

While the experimental data do not necessarily reach the maximum of the undamped retrieved signal [they do for the example shown in Fig.~\ref{fig3}(a) but not for that shown in Fig.~\ref{fig3}(b)], the use of an oscillatory fit function with amplitude $I_{\text{peak}}^{\text{norm}}$ allows us to extrapolate beyond the regime where experimental data are available. 
It is worthwhile noting that the full oscillatory behavior can be measured, in principle, even for relatively weak magnetic field strengths if the hold time is increased. When scanning $\tau$, one needs to keep in mind though that the impact of the finite dephasing rate $\gamma_c$ increases during the free evolution time, thereby reducing the signal-to-noise ratio of the retrieved probe signal.

\begin{figure}[!thbp]
    \centering
    \includegraphics[width=\linewidth]{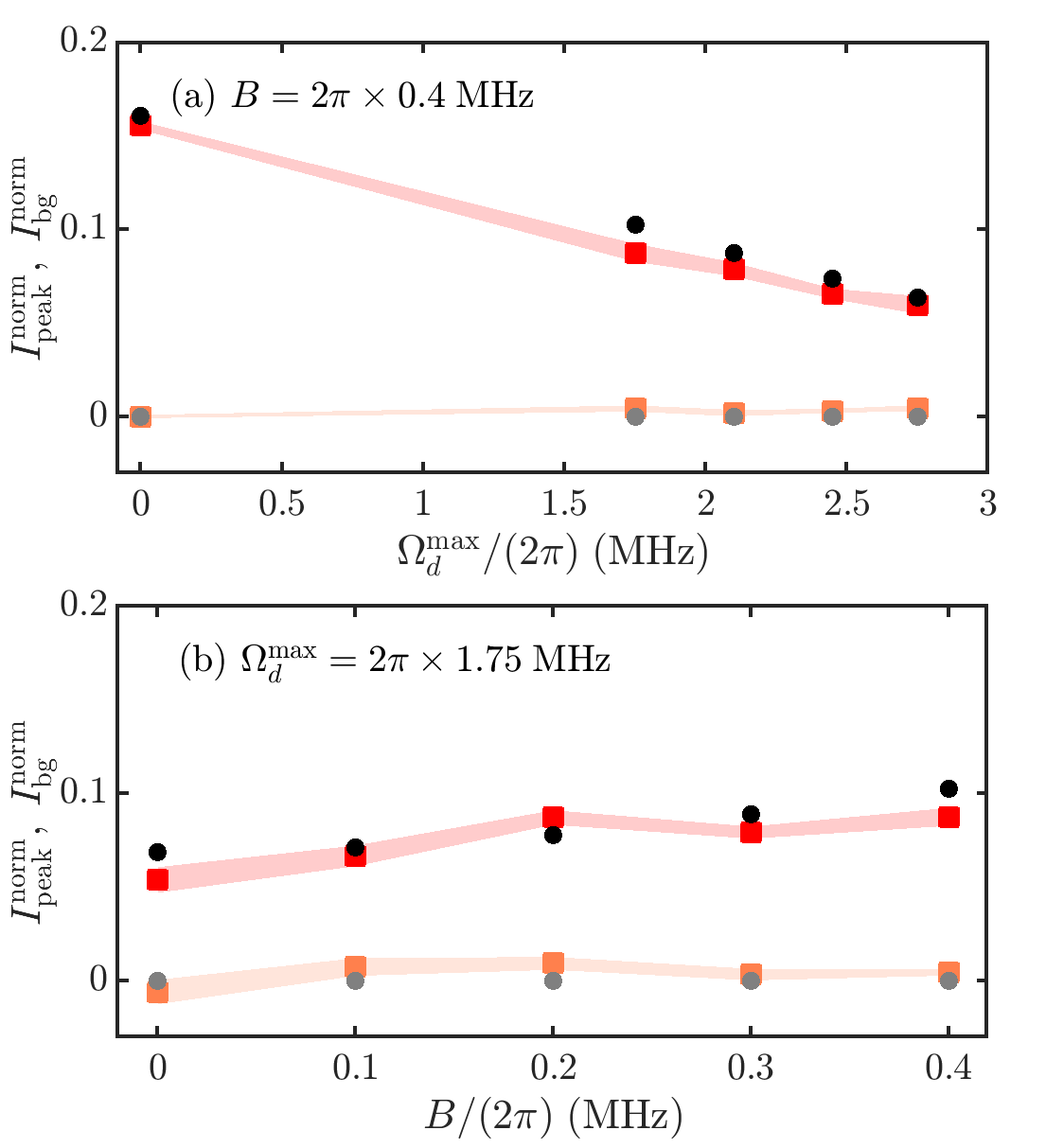}
    \caption{Fit parameters $I_{\text{peak}}^{\text{norm}}$ and $I_{\text{bg}}^{\text{norm}}$ as a function of (a) the maximum magnitude $\Omega_d^{\text{max}}$ of the decay beam Rabi coupling strength (fixed $B$) and (b) the magnetic field strength $B$ (fixed $\Omega_d^{\text{max}}$). Squares with error bands show the results for the fits to the experimental data (red and orange squares are for $I_{\text{peak}}^{\text{norm}}$ and $I_{\text{bg}}^{\text{norm}}$, respectively)  while circles show the results for the fits to the simulation data (black and grey circles are for $I_{\text{peak}}^{\text{norm}}$ and $I_{\text{bg}}^{\text{norm}}$, respectively). In both panels, the upper and lower set of curves correspond to $I_{\text{peak}}^{\text{norm}}$ and $I_{\text{bg}}^{\text{norm}}$, respectively; $I_{\text{bg}}^{\text{norm}}$ is, within error bars,  consistent with zero. The results extracted from the experimental and simulation data agree quite well. }
    \label{fig4}
\end{figure}

The full experimental data sets for the synchronization sequence are shown in Appendix~\ref{appendix_fulldata} and the supplemental material~\cite{SM} for both $\chi^S$ and $\tau$ scans.
The red symbols in Fig.~\ref{fig4} show examples of the fit parameters $I_{\text{peak}}^{\text{norm}}$ and $I_{\text{bg}}^{\text{norm}}$  extracted from our experimental data. Figure~\ref{fig4}(a) is for varying $\Omega_d^{\text{max}}$ and $B=2 \pi \times 0.4$~MHz  while Fig.~\ref{fig4}(b) is for varying  
$B$ and  $\Omega_d^{\text{max}}=2 \pi \times 1.75$~MHz [note that examples of $I_{\text{peak}}^{\text{norm}}$, extracted from experiment, are also shown by the red symbols in the left panel of Fig.~\ref{fig1}(d)].

For comparison, the circles in  Figs.~\ref{fig3} and \ref{fig4}  
show results from our simulations, which  account for  the modification of the probe-beam Rabi coupling strength by the atomic medium. Figure~\ref{fig3} shows 
the quantity $\exp(2 \gamma_c \tau)  I_{\pi}^{\text{norm}}(\tau)$  while Fig.~\ref{fig4} shows the fit parameters $I_{\text{peak}}^{\text{norm}}$ and $I_{\text{bg}}^{\text{norm}}$, extracted by performing, just as in the analysis of the experimental data,  fits to Eq.~(\ref{eq_fit}). The agreement between the experimentally measured data and the simulations is very good throughout.
 The fit parameters $I_{\text{peak}}^{\text{norm}}$ and $I_{\text{bg}}^{\text{norm}}$ are used in Sec.~\ref{sec_synchro} to determine the minimum and the maximum of the quantum synchronization $S_q$.

In the experiments discussed so far, the atoms are prepared in the limit cycle state prior to the application of the synchronization sequence. We now discuss results from comparative measurements, where we prepare two different initial states for fixed hold time $\tau$, fixed magnetic field $B$, and various $\chi^S$. State preparation protocol~I (same as employed above) aims to prepare the limit cycle state $\ket{2}\bra{2}$ while state preparation protocol~II aims to prepare a superposition of all three states with equal populations, namely 
$(\ket{1}\bra{1}+\ket{2}\bra{2}+\ket{3}\bra{3})/3$. For notational simplicity, we refer to the states as $(0,1,0)$ and $(1/3,1/3,1/3)$. While we  do not currently have the capability to measure populations very precisely, detailed comparisons with simulation data suggest that our state preparation protocols work as intended.

\begin{figure}[!thbp]
    \centering
    \includegraphics[scale=.14]
{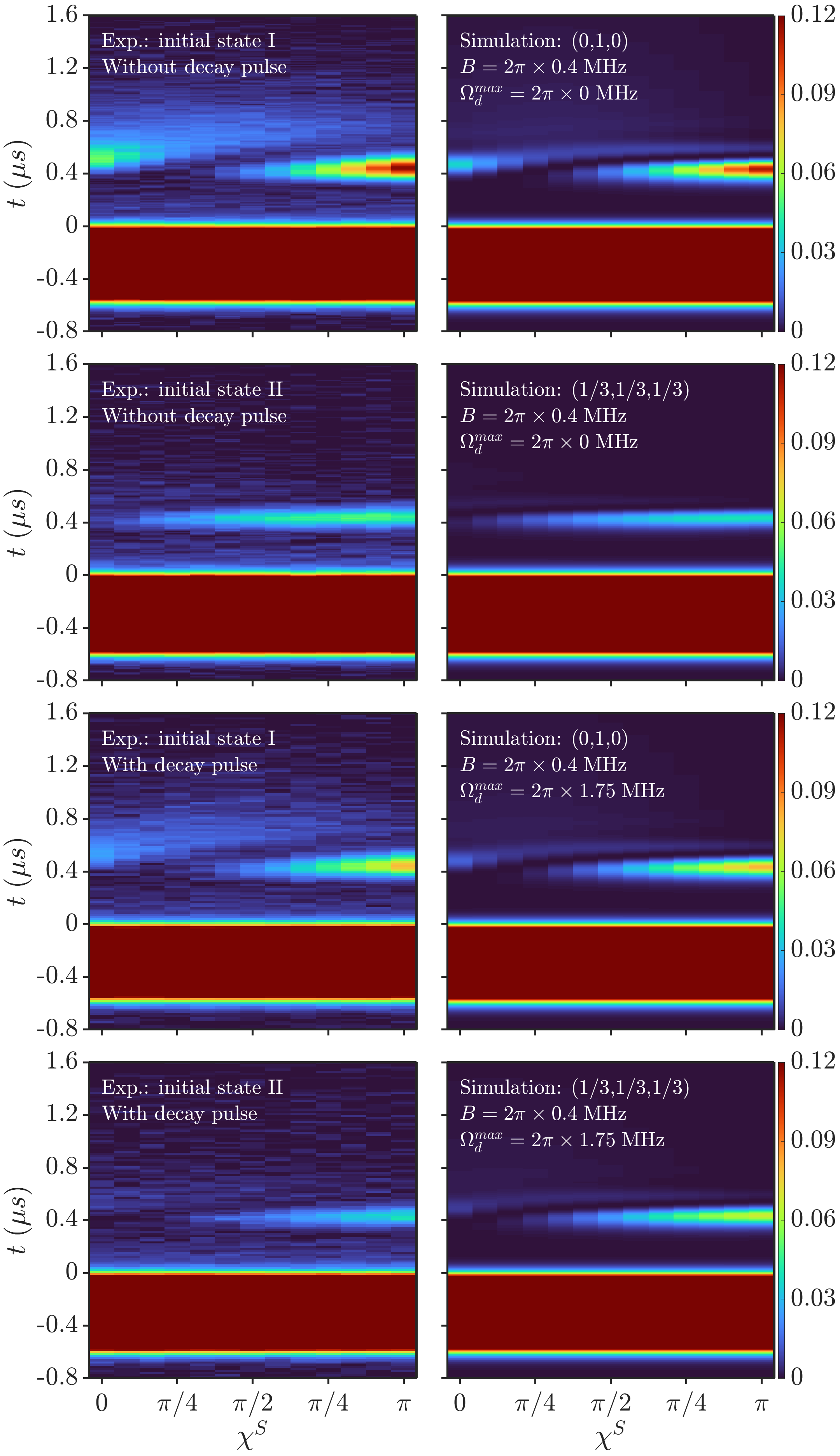}
    \caption{Dependence of the normalized probe beam intensity $I_{\pi}^{\text{peak}}(t)$ on the initial state preparation approach; specifically, the initial state preparation~I and initial state preparation~II approaches are compared for $\tau=0.4$~$\mu$s. The layout of the figure is similar to Fig.~\protect\ref{fig2}, with  the left column showing experimental data and the right column simulation results. The magnetic field strength is fixed at $B=2 \pi \times 0.4$~MHz and the maximum magnitude $\Omega_d^{\text{max}}$ of the Rabi coupling strength of the decay pulse is indicated in the sub-figures that show simulation results.}
    \label{fig5}
\end{figure}

Figure~\ref{fig5} compares experimental measurements without the decay beam and with the decay beam (left column), alongside simulation results (right column). The aim of this study is to show that the atomic state at the time at which the probe and control-synchronize beams (and the decay beam for the case where $\Omega_d^{\text{max}} \ne 0$) are being turned off depends  
on the initial state. The initial-state dependence arises because the time required to reach the equilibrium state, starting in the equally populated state (state~II), is longer than the time interval over which the probe and control-synchronize beams are on simultaneously. The separation of time scales is large for $\Omega_d^{\text{max}}=0$ and significantly less extreme
for $\Omega_d^{\text{max}}=2 \pi \times 1.75$~MHz.
 These arguments are supported by our experimental and simulation data. The retrieved signals for the $\Omega_d^{\text{max}}=0$ case, shown in the first and second rows of Fig.~\ref{fig5}, display marked differences. Specifically, we see that the phase $\chi^S$ at which the destructive interference occurs depends appreciably on the initial state. If the system is started too far from equilibrium, the beams need to be on for a longer time period  for the spin-1 system to reach a stationary state. 
 For $\Omega_d^{\text{max}}=2 \pi \times 1.75$~MHz
 (two bottom-most rows in Fig.~\ref{fig5}), the dependence on the initial state is significantly weaker. In this case, the main difference for the state~I and state~II data is not at which $\chi^S$ the destructive interference occurs but the change of the  intensity of the retrieved probe signal. The dependence of the time scales for reaching the steady-state density matrix on the initial state are discussed in more detail in the next section.
 
\section{Validation of assumptions}
\label{sec_justification}

The  extraction protocol of the maximum of $S_q$  of the effective spin-1 model from the measured retrieved probe beam intensity relies on the following assumptions and observations:
\begin{itemize}
    \item It is assumed that the off-diagonal density matrix elements of the atoms in the MOT  evolve with time such that their behavior can be related straightforwardly to the time evolution of an isolated effective spin-1 system.  
    \item It is assumed that the probe beam is, for the initial state $\ket{2}\bra{2}$, on for a sufficiently long time period for the system  to reach a steady state.
    \item It is assumed that the magnitude of the atomic density matrix elements $\rho_{12}$ and $\rho_{23}$ is the same. 
    \item It is assumed that the time evolution of the off-diagonal atomic density matrix elements (atomic coherences) during the hold time follows free-particle evolution.
    \item It is assumed that the change from steady state to free-particle evolution and subsequently from free-particle evolution to read out by the control-read pulse occurs instantaneously. 
\end{itemize}
This section justifies these assumptions through dedicated simulations.

\begin{figure}[!thbp]
    \centering
    \includegraphics[width=\linewidth]{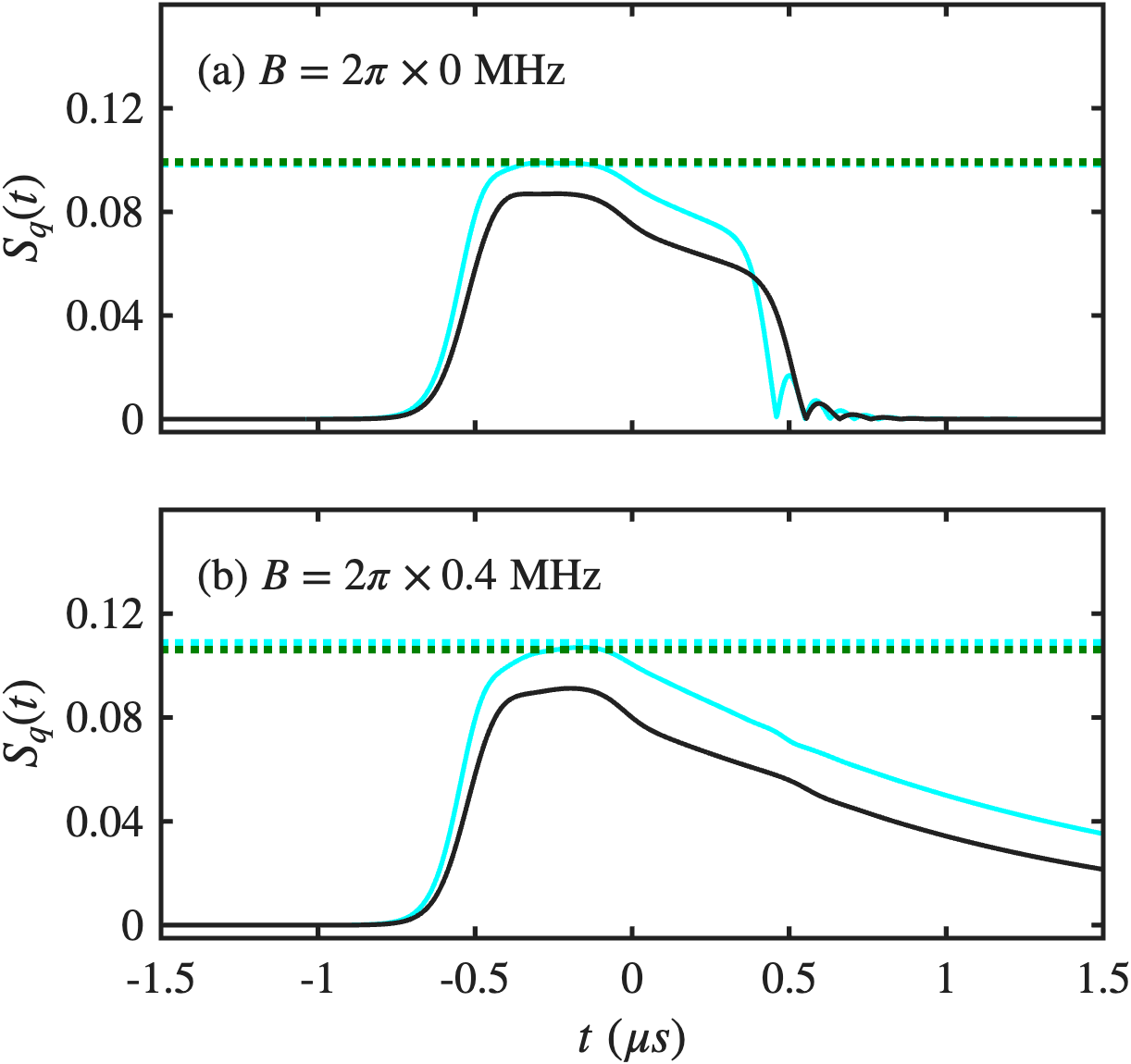}
    \caption{Simulation results for the initial state $\ket{2}\bra{2}$. $S_q(t,z)$ as a function of time  for $\tau=0.4$~$\mu$s, $\chi_+=\chi_0=\chi_-=0$, $\Omega_d^{\text{max}}=2 \pi \times 2.75$~MHz, and two different $B$ (see the legend in the figure) using the beam sequence employed in the experiment. The black and turquoise lines show $S_q(t,L)$ and $S_q(t,0)$, respectively; these lines are  obtained by solving a set of Heisenberg-Langevin equations that account for the medium (see Appendix~B of Ref.~\cite{zhong2026light}) and the master equation for the isolated 6-level system [Eq.~(\ref{eq_masterequation_6level})]. 
    The results for the effective spin-1 master equation (not shown) agree extremely well  with the latter. 
    It can be seen that the ratio between the solid turquoise and solid black lines is roughly constant for $-0.5~\mu\mbox{s} \lesssim t \lesssim \tau$ and that both $S_q(t,L)$ and $S_q(t,0)$ are approximately constant for $-0.5~\mu\mbox{s} \lesssim t \lesssim 0$. 
    For comparison, the  horizontal dashed turquoise and dotted green lines show the steady-state synchronization for the isolated six-level system and the effective spin-1 system, respectively [steady state solutions of Eqs.~(\ref{eq_masterequation_6level}) and (\ref{eq_masterequation_3level}), respectively]. For both $B$ values considered, it can be seen that the solid turquoise line is essentially flat for $t$ just a bit smaller than $0$ and equal to the dashed turquoise line, indicating that the synchronization beam stage is long enough for the system to reach its steady state. 
    }
    \label{fig6}
\end{figure}

Figure~\ref{fig6} shows  $S_q(t,z)$ as a function of time for  $\Omega_d^{\text{max}}= 2 \pi \times 2.75$~MHz for the initial state $\ket{2}\bra{2}$. The top panel is for $B=0$ while the bottom panel is for $B=2\pi \times 0.4$~MHz. Even though the concept of synchronization requires that a steady state is established and that the steady state is robust against perturbations, we analyze the  time-dependence of $S_q(t,z)$; if the system is in a stable  steady state, the quantity $S_q(t,z)$ is a  quantum synchronization measure.

Figure~\ref{fig6} shows results for two  time-dependent models, corresponding to $z=0$ and $z=L$. Both use the same beam sequence as used in the experiment.
(i) The turquoise solid line shows the results for the full 6-level system without the medium, i.e., $S_q(t,0)$. For reference, the horizontal dashed turquoise line shows $S_q^{\text{ss}}$ for the steady-state solution to the master equation for the isolated 6-level system. The solid turquoise line approaches the horizontal dashed turquoise line at the end of the synchronization stage, indicating that a steady state is reached---for the parameters considered and the beam sequence employed. 
As an additional reference, the horizontal dotted green line shows $S_q^{\text{ss}}$ for the stationary solution to the master equation of the isolated effective spin-1  system. The agreement between this line and the stationary solution for the isolated 6-state model is excellent.
In fact, the $S_q(t,0)$ obtained from the solution to the effective spin-1 master equation for the time-dependent beam sequence (not shown) agrees excellently with the $S_q(t,0)$ for the isolated 6-level system (turquoise line). The excellent agreement indicates that the effective spin-1 model provides a faithful description of our system.
(ii) The solid black line shows $S_q(t,L)$ for the full 6-level system at the end of the MOT. The results are obtained using the same theoretical model as employed in  the previous section.  It can be seen that
$S_q(t,L)$ has a very similar shape as $S_q(t,0)$ for $-0.5~\mu\mbox{s} \lesssim t \lesssim \tau$, indicating that $S_q(t,0)$ and $S_q(t,L)$ are related by a simple scaling factor.  We denote the ratio $S_q({t},0)/S_q({t},L)$ for $t=0$ by $R_{\text{m-f}}$ (``m-f'' stands for ``medium-to-free''), $R_{\text{m-f}}=S_q(0,0)/S_q(0,L)$.

\begin{figure*}[!thbp]
    \centering
    \includegraphics[width=\linewidth]{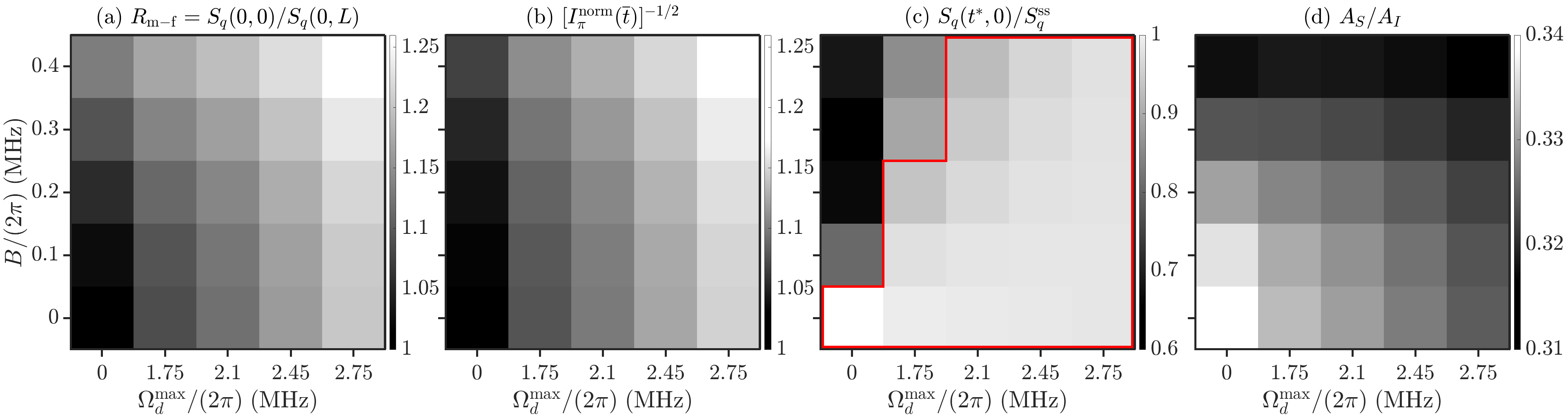}
    \caption{Simulation results for the initial state  (0,1,0) as functions of $B$ and $\Omega_d^{\text{max}}$ for $\chi_+=\chi_0=\chi_-=0$. Note that  $B$ and $\Omega_d^{\text{max}}$ each take five discrete values; the grid in $B$ is equidistantly spaced while that in $\Omega_d^{\text{max}}$ is not equidistantly spaced.  (a) The grey scale shows the ratio $R_{\text{m-f}}$. (b)  The grey scale shows the ratio $[I_{\pi}^{\text{norm}}(\bar{t})]^{-1/2}$.
    It can be seen that the ratio $R_{\text{m-f}}$ is approximated reasonably well by  $[I^\text{norm}_{\pi}(\bar{t})]^{-1/2}$. (c)  The grey scale shows the ratio $S_q(t^*,0)/S_q^{\text{ss}}$ for $t^*=-0.1$~$\mu$s,
     where $S_q^{\text{ss}}$ is the quantum synchronization of the steady state of the isolated 6-state model. The cases where the ratio is $\ge 90$~$\%$ are encircled by a red box. For finite $B$, a finite $\Omega_d^{\text{max}}$ shortens the time required for the system to reach its steady state. (d) The grey scale shows the ratio $A_S/{A_I}$, which enters into Eq.~(\ref{eq_max_sync_scale}). The ratio varies by less than $10$~\% for the parameter combinations considered.}
    \label{fig7}
\end{figure*}

Figure~\ref{fig7}(a) shows the ratio $R_{\text{m-f}}$ for several $(B,\Omega_d^{\text{max}})$ combinations; specifically five equidistantly spaced $B$ values and five non-equidistantly spaced $\Omega_d^{\text{max}}$ values. The ratio varies between $\approx 1$ and $\approx 1.25$   
for the parameter combinations considered. As can be seen by comparing Figs.~\ref{fig7}(a) and \ref{fig7}(b), the ratio $R_{\text{m-f}}$ is, to a fairly good approximation, captured by the ratio $|\Omega_{\pi}^{\text{max}}/\Omega_{\pi}(\bar{t},L)|=[I_{\pi}^{\text{norm}}(\bar{t})]^{-1/2}$, where $\bar{t}$ denotes the time at which the transmitted probe beam intensity takes its maximum. We find that the maximal deviation between $R_{\text{m-f}}$ and $[I_{\pi}^{\text{norm}}(\bar{t})]^{-1/2}$
is, for the parameter combinations considered, $\approx 6.7$~\% for
$(B,\Omega_d^{\text{max}})=(2 \pi \times 0.4~\mbox{MHz},2 \pi \times 2.75~\mbox{MHz})$ while the smallest deviation is $\approx 0.8$~\% for $(B,\Omega_d^{\text{max}})=(2 \pi \times 0.1~\mbox{MHz},2 \pi \times 0~\mbox{MHz})$.
We conclude that we can, to a good approximation, obtain $S_q(0,0)$ by multiplying $S_q(0,L)$ by $[I_{\pi}^{\text{norm}}(\bar{t})]^{-1/2}$.

Figure~\ref{fig7}(c) quantifies how close the state that is prepared by the synchronization beam sequence is to the true stationary state. To quantify this, we plot the ratio $S_q(t^*,0)/S_q^{\text{ss}}$, with both $S_q(t^*,0)$ and $S_q^{\text{ss}}$ being calculated for the full 6-state model without the medium. The time $t^*$ is chosen to be slightly smaller than $0$ to account for the fact that $S_q(t,0)$ starts to decrease, as can be seen in Fig.~\ref{fig6}, for values slightly smaller than $t=0$ due to the smooth turn-off of the beams. Figure~\ref{fig7}(c) uses $t^*=-0.1$~$\mu$s. It can be seen that the smallest ratio is  $0.653$ for the parameter combination  $(B,\Omega_d^{\text{max}})=(2\pi\times 0.3\,\text{MHz},0)$. The red lines encircle the parameter combinations for which the synchronization beam sequence prepares a state for which the quantum synchronization deviates by 10~\% or less from that of the stationary state. We use this (arbitrary) 10~\% threshold to delineate the regime where the beam sequence prepares, for the initial state $(0,1,0)$, a fully synchronized system from that where the beam sequence does not prepare a truly synchronized system.

\begin{figure}[!thbp]
    \centering
    \includegraphics[width=\linewidth]{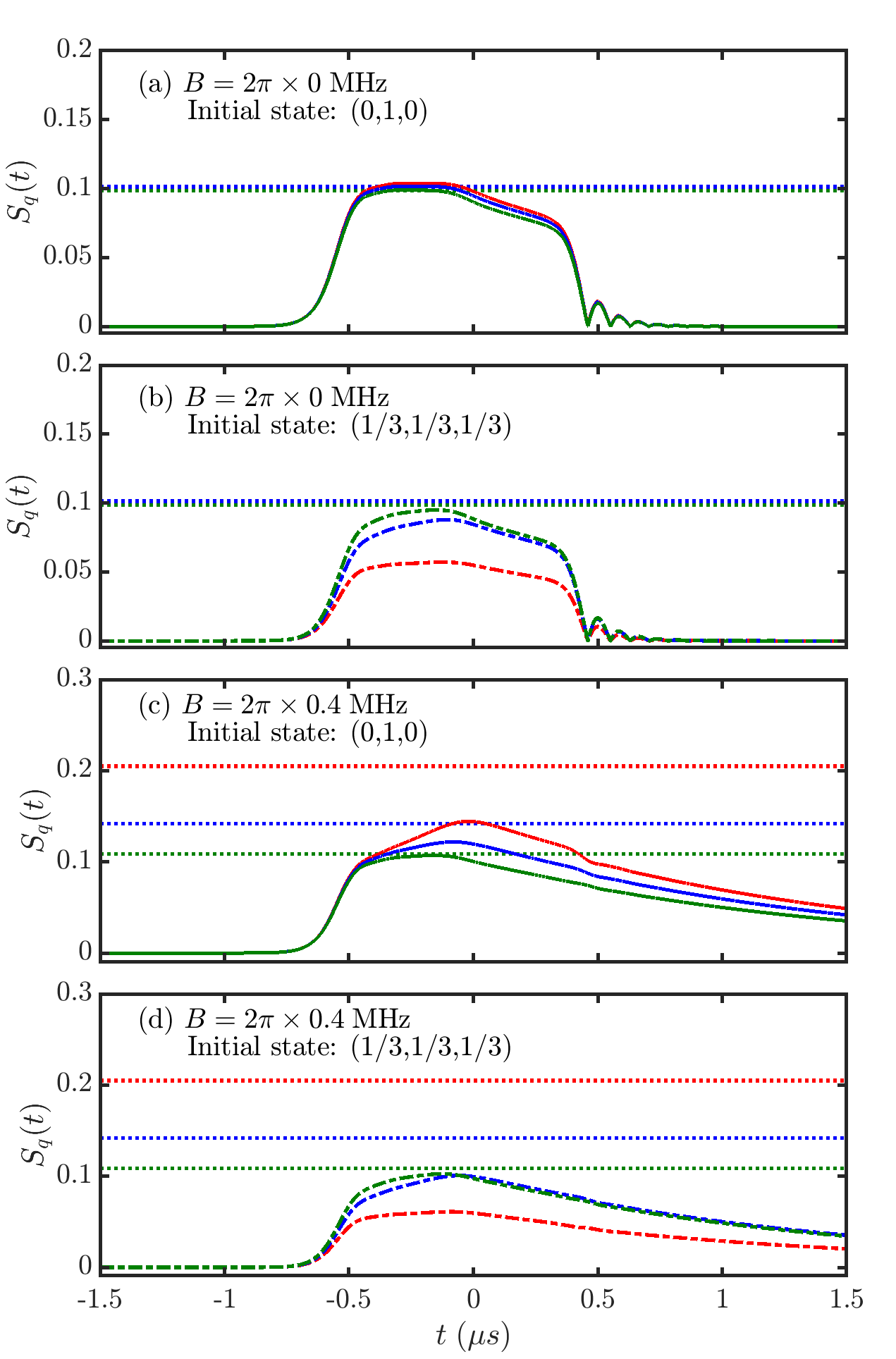}
    \caption{Simulation results for the full 6-state system without the medium, comparing $S_q(t,0)$ for the initial states  $(0,1,0)$ and $(1/3,1/3,1/3)$ for $\tau=0.4$~$\mu$s, two values of $B$ (see the legend),  $\chi_+=\chi_0=\chi_-=0$, and varying $\Omega_d^{\text{max}}$. The red, blue, and green lines are for $\Omega_d^\mathrm{max} = 2\pi\times 0$~MHz, $2\pi\times 1.75$~MHz, and $2\pi\times 2.75$~MHz, respectively. The solid lines are for the beam sequence used in the experiment while the horizontal dotted lines show steady-state values.}
    \label{fig8}
\end{figure}

\begin{figure}
    \centering
    \includegraphics[width=\linewidth]{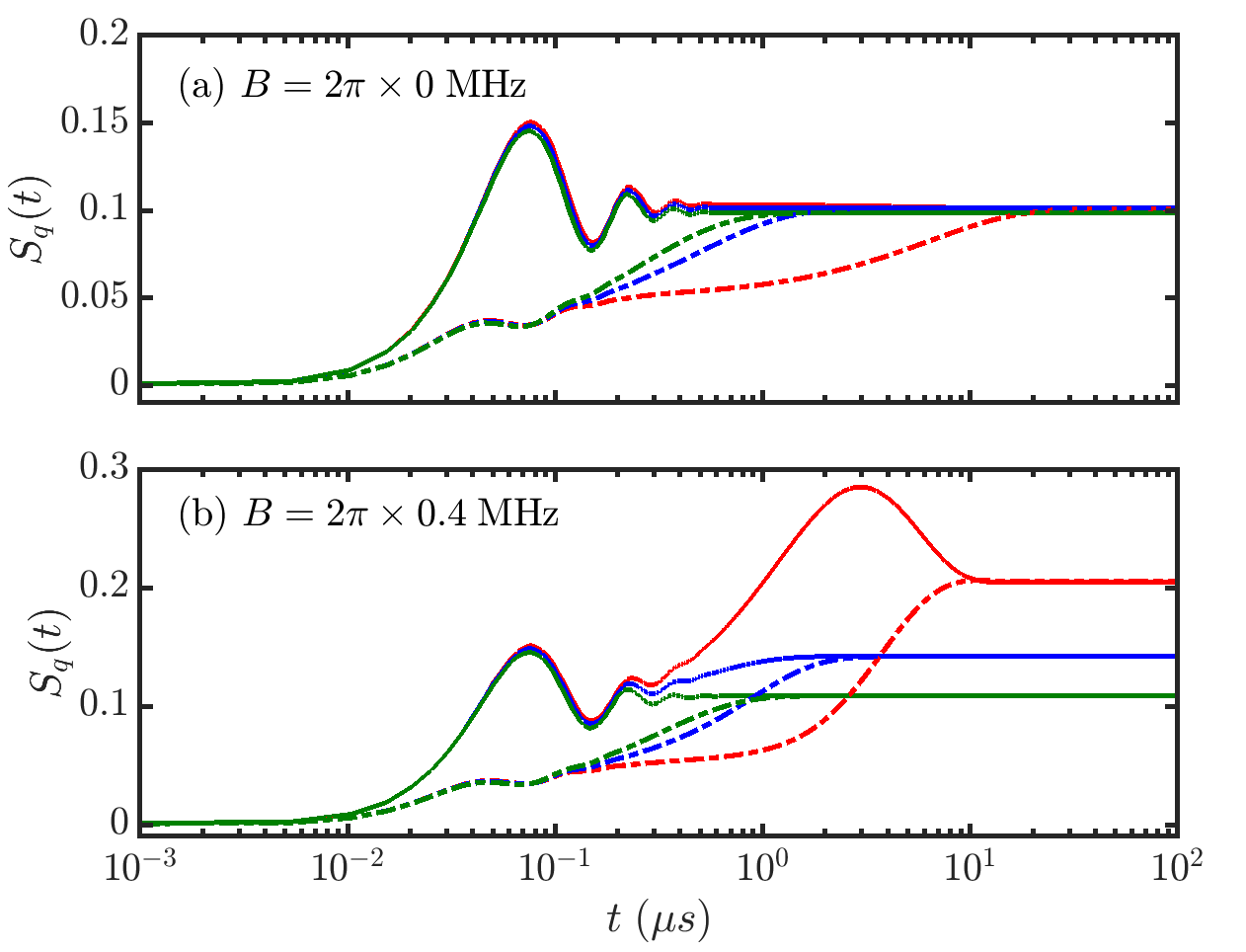}
    \caption{Simulation results for the full 6-state system without the medium for constant Rabi coupling strengths (i.e., the beams are turned on instantaneously at $t= 0$ but are otherwise independent of time), comparing $S_q(t,0)$ for the initial states  $(0,1,0)$ (solid lines) and $(1/3,1/3,1/3)$ (dash-dotted lines) for  two values of $B$ (see the legend),  $\chi_+=\chi_0=\chi_-=0$, and varying $\Omega_d^{\text{max}}$. As in Fig.~\ref{fig8}, the red, blue, and green lines are for $\Omega_d^\mathrm{max} = 2\pi\times 0$~MHz, $2\pi\times 1.75$~MHz, and $2\pi\times 2.75$~MHz, respectively. }
    \label{fig9}
\end{figure}

To further elucidate the importance of the time scales, Fig.~\ref{fig8} shows simulation results for two different initial states, namely $(0,1,0)$ and $(1/3,1/3,1/3)$, obtained by solving the master equation for the full 6-state model without the medium. The top two rows are for $B=0$ while the bottom two rows are for $B=2 \pi \times 0.4$~MHz. In each panel, three different $\Omega_d^{\text{max}}$ are considered, namely, $\Omega_d^{\text{max}}=0$ (red lines), 
$\Omega_d^{\text{max}}=2 \pi \times 1.75$~MHz (blue lines), and
$\Omega_d^{\text{max}}=2 \pi \times 2.75$~MHz (green lines).
 For the initial state $(0,1,0)$ and $B=0$ [Fig.~\ref{fig8}(a)], the  $S_q(t,0)$ reach the steady state values (color-matched horizontal straight lines) at the end of the synchronization stage for all three values of $\Omega_d^{\text{max}}$. For the initial state $(1/3,1/3,1/3)$ [Fig.~\ref{fig8}(b)], in contrast, the steady state value is only reached for the largest decay beam Rabi coupling strength considered. For $\Omega_d^{\text{max}}=0$, the values of $S_q(t,0)$ are roughly constant during the synchronization stage. This does, in this case, not correspond to reaching the steady state but instead indicates that the time scale  is set by $\gamma_c$, i.e.,  the steady state would only be reached if the synchronization stage lasted for several tens of microseconds. This is further highlighted in Fig.~\ref{fig9}, which shows the time dynamics for constant beam intensities over a much longer time scale (note the logarithmic time scale).
 For $B=2 \pi \times 0.4$~MHz [Figs.~\ref{fig8}(c) and \ref{fig8}(d)], a  dependence on the initial state is also clearly visible. In this case, the steady state is only reached for the largest decay beam strength considered. For $\Omega_d^{\text{max}}=0$, the time scale is, as in the $B=0$ case, set by $\gamma_c$. For finite $B$, the time scale for the same finite $\Omega_d^{\text{max}}$ is slightly larger than for $B=0$.

\begin{figure}[!thbp]
    \centering
    \includegraphics[width=\linewidth]{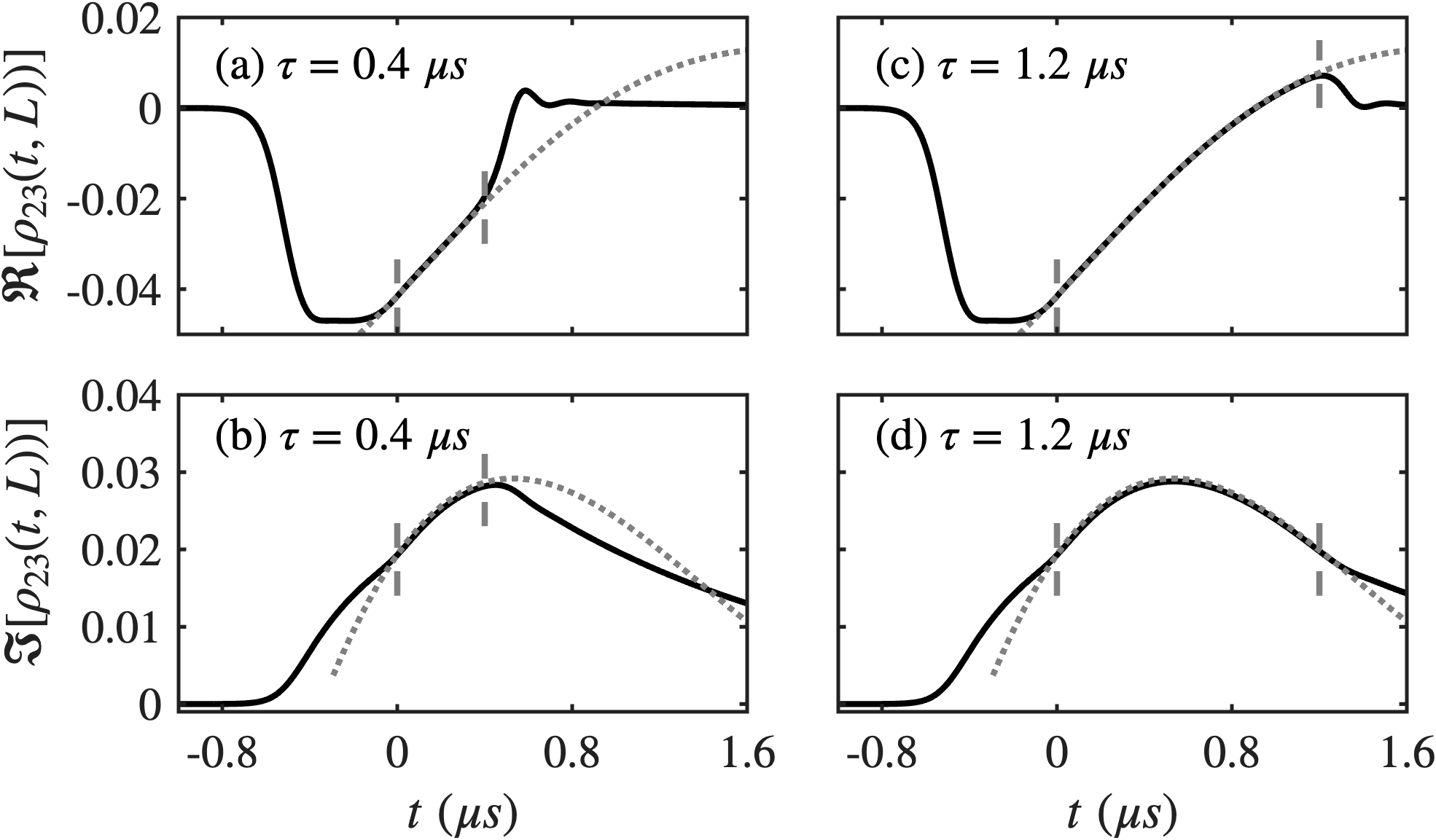}
    \caption{Simulation results for the 6-state model in the presence of the medium for the initial state $(0,1,0)$, $B=2\pi\times 0.2$~MHz, $\chi_+=\chi_0=\chi_- = 0$, and $\Omega_d^{\text{max}} = 2\pi\times 1.75$~MHz for two hold times $\tau$ (see the legend). Solid lines in the top and bottom rows show the  real and imaginary parts, respectively,  of the density matrix element  $\rho_{23}(t,L)$ for the beam sequence used in the experiment. For comparison, the dotted lines show the analytical free-evolution expressions [Eq.~(\ref{eq_coherences_free})]. It can be seen that the analytical expressions provide an excellent description of the numerical results for $0 \le t \le \tau$. As a guide to the eye, the dashed vertical lines mark the times $t=0$ and $t=\tau$.}
    \label{fig10}
\end{figure}

To validate our read-out approach, we analyze the coherences at the end of the MOT. Since we are working with an energy level scheme for which the hyperfine states with positive and negative projection quantum numbers are shifted by the same amount but in opposite directions, since the detunings exhibit an analogous ``symmetry,'' and since the ``positive and negative projection quantum number contributions'' are the same for our initial states, it follows straightforwardly that $|\rho_{12}(t,z)|$ and $|\rho_{23}(t,z)|$ are equal. Using Eq.~(\ref{eq_synchronization}), it follows 
\begin{eqnarray}
\label{eq_synchronization_min}
\min [S_q(0,L)]=0
\end{eqnarray}
and 
\begin{eqnarray}
\label{eq_synchronization_max}
\max [S_q(0,L)]=2 A_S |\rho_{12}(0,L)| ,
\end{eqnarray}
where the min and max functions are evaluated by scanning over $\varphi_{\text{sync}}$ (in practice, this corresponds to scanning $\chi^S$ and/or $\tau$). 
Equation~(\ref{eq_synchronization_min}) is confirmed by the fact that the values of $I_{\text{bg}}^{\text{norm}}$, extracted from our experimental data, are consistent with zero within error bars (see Sec.~\ref{sec_dynamics}).  Equation~(\ref{eq_synchronization_min}) thus shows that our measurement of $I_{\text{bg}}^{\text{norm}}=0$ corresponds to $\text{min}[S_q(0,L)]=0$. As discussed in the next section, this is a clean signature of the synchronization blockade~\cite{PhysRevLett.118.243602,Tan2022halfintegervs,PhysRevA.99.043804,PhysRevA.108.022216} of the effective spin-1 model. 
To confirm that the off-diagonal coherences follow the free evolution equation,
solid lines in Fig.~\ref{fig10} show the real and imaginary parts of the density matrix element  $\rho_{23}(t,L)$  as a function of time for two different $\tau$, namely, $\tau=0.4$~$\mu$s and $\tau=1.2$~$\mu$s. For comparison, dotted lines show the analytical free-evolution expressions [Eq.~(\ref{eq_coherences_free})], which assume the formation of two polaritons. To plot the dotted lines, $\Re[\rho_{23}(0,L)]$ and $\Im[\rho_{23}(0,L)]$  are taken from the simulations. 
The agreement during the hold time is very good, thereby validating the read-out step of the synchronization extraction protocol.

\section{Synchronization for spin-1 system governed by additive Lindbladian}
\label{sec_synchro}

Having validated various assumptions in the previous section, this section determines $\text{min}[S_q(t,z)]$ and $\text{max}[S_q(t,z)]$ for $t=0$ and $z=0$, i.e., at the
beginning of the MOT, which corresponds to the synchronization of an isolated spin system. 
Our focus is on the steady-state synchronization, i.e., on the parameter combinations
that are encircled by the red lines in Fig.~\ref{fig7}(c).

\begin{figure*}
    \centering
    \includegraphics[width=\linewidth]{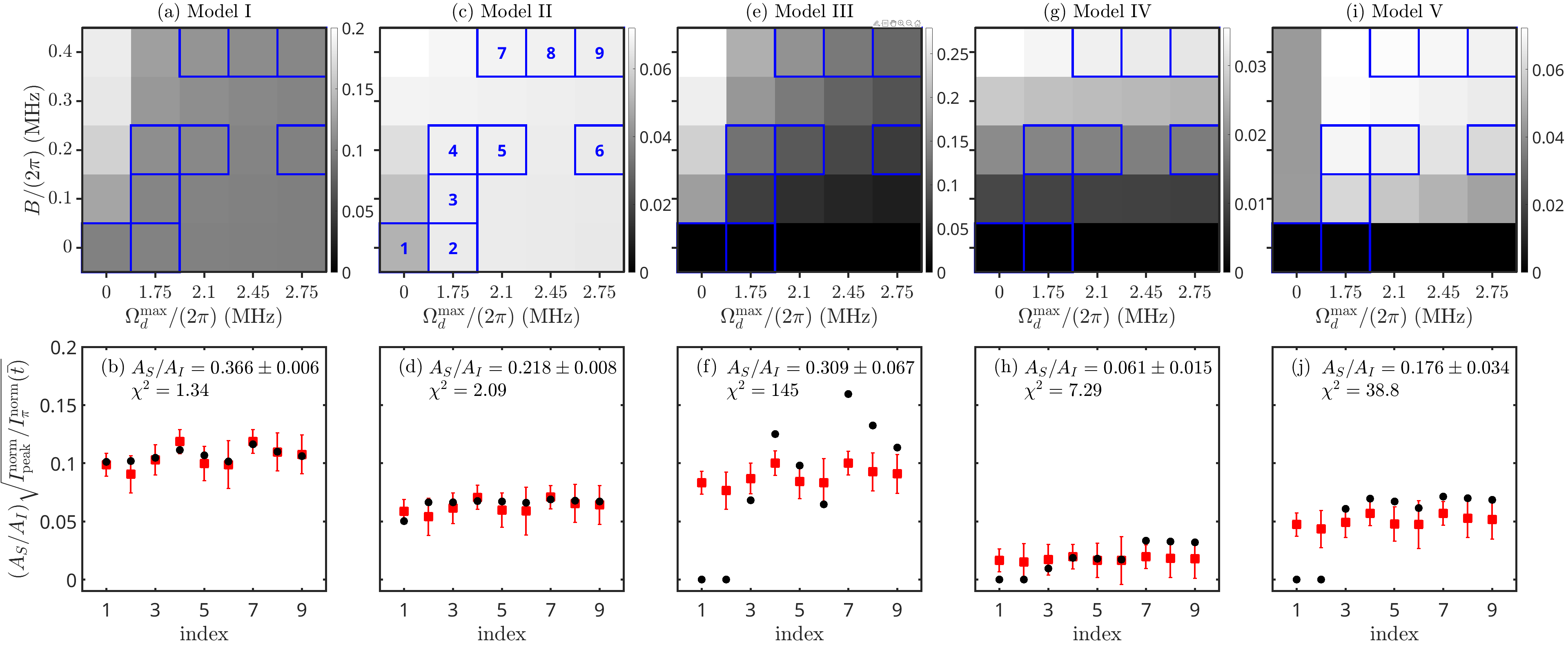}
    \caption{Experimentally determined $\text{max}[S_q(0,0)]$. 
    Top row: The maximum steady-state synchronization $\max(S_q^{\text{ss}})$, obtained from simulations, is shown as  functions of the magnetic field $B$ and the maximum $\Omega_d^{\text{max}}$ of the Rabi coupling strength  of the decay pulse  for five different models   of the isolated spin-1 system; the models are introduced in the text. The blue boxes mark the nine parameter combinations for which the bottom row shows  experimental data. 
    Bottom row: 
    The black circles show the values of $\max(S_q^{\text{ss}})$ from the simulations that correspond to the parameter combinations encircled by the blue boxes; each index on the $x$-axis corresponds to a specific   $(B,\Omega_d^\text{max})$ combination [see panel~(c) for the index convention].  The red squares show the quantity $(A_S/A_I)[I_\text{peak}^\text{norm}/I_{\pi}^\text{norm}(\bar{t})]^{1/2}$, where $[I_\text{peak}^\text{norm}/I_{\pi}^\text{norm}(\bar{t})]^{1/2}$ is extracted from experiment and $A_S/A_I$ is obtained by fitting $(A_S/A_I)[I_\text{peak}^\text{norm}/I_{\pi}^\text{norm}(\bar{t})]^{1/2}$, treating $A_S/A_I$ as a fit parameter, to $S_q^{\text{ss}}$ for each of the five models.  The value of $A_S/A_I$ and the corresponding  $\chi^2$ are reported in the figures. The fit to Model~I, which is the effective spin-1 model with additive Lindblad operators, is characterized by the lowest $\chi^2$, indicating that our experimental data are best described by Model~I. Correspondingly, the red squares shown in (b) should be interpreted as the experimentally measured maximum of $S_q^{\text{ss}}$ of an isolated effective spin-1 system with additive Lindblad operators.}
     \label{fig11}
\end{figure*}

Approximating $R_{\text{m-f}}$ by $[I_{\pi}^{\text{norm}}(\bar{t})]^{-1/2}$ and using Eqs.~(\ref{eq_retrieved_intensity}) and (\ref{eq_synchronization}),
we find
\begin{eqnarray}
\label{eq_min_sync_scale}\text{min} [S_q(0,0)]=\frac{A_S}{{A_I}}  \sqrt{\dfrac{I_{\text{bg}}^{\text{norm}}}{I_\pi^\mathrm{norm}(\bar{t})}}
\end{eqnarray}
and
\begin{eqnarray}
\label{eq_max_sync_scale}\text{max} [S_q(0,0)]=\frac{A_S}{{A_I}}  \sqrt{\dfrac{I_{\text{peak}}^{\text{norm}}}{I_\pi^\mathrm{norm}(\bar{t})}}.
\end{eqnarray}
As discussed in the previous sections, we can extract  $I_{\pi}^{\text{norm}}(\bar{t})$
and ${I_{\text{peak}}^{\text{norm}}}$ from experiment but not the ratio ${A_S}/{A_I}$. 
Figure~\ref{fig7}(d) shows the ratio ${A_S}/{A_I}$, extracted from our simulation data. It can be seen that $A_S/A_I$ varies by only about $10$~\% for the parameter combinations considered in Fig.~\ref{fig7}(d). For the analysis that follows, we assume that $A_S/A_I$ is constant. Specifically, 
motivated by the small spread of $A_S / A_I$, we fit our experimental data to the $\text{max}[S_q(0,0)]$  simulation results for five different spin-1 models,
allowing for an overall scaling factor for each model; as shown  by the blue boxes in Fig.~\ref{fig11}, we took measurements for nine
$(B,\Omega_{\pi}^{\text{max}})$ combinations that fall inside the red encircled steady-state region. These nine parameter combinations are indexed in Fig.~\ref{fig11}(c). The top row of Fig.~\ref{fig11} shows simulation results for the five different models, referred to as Model~I to Model~V,  while the bottom row of Fig.~\ref{fig11} shows the steady-state synchronization of an isolated spin-1 system extracted from the experimental data, using the indexing introduced in Fig.~\ref{fig11}(c), separately assuming applicability of each of the five models, and treating $A_S/A_I$ as a fit parameter.

To appreciate the implications of Fig.~\ref{fig11}, we need to introduce the key features of Model~I to Model~V; details can be found in Appendices~\ref{appendix_inmedium_simulations}-\ref{appendix_ideal}. Models~I-III all refer to variants of the effective three-level system with additive Lindblad operators. 
    Model~I refers to the effective three-level system with additive Lindblad operators.  Model~II  artificially sets the self-dissipation of state $|2\rangle$ to zero while Model~III  artificially sets all dissipative coupling out of state $|2\rangle$ as well as the self-dissipation of state  $\ket{2}$ to zero. 
    Figure~\ref{fig11} shows that neglecting the self-dissipation term lowers the synchronization by roughly the same amount for all parameter combination considered. In contrast, additionally neglecting the dissipative processes out of state $\ket{2}$ either lowers or increases the synchronization, depending on the parameter combination. Notably, Model~III leads to a vanishing synchronization for $B=0$, in contrast to what is observed experimentally.
    Model~IV and V both refer to the ideal spin-1 system but with different dissipation coefficients, namely $\gamma_d= \gamma_g = \Gamma_\text{control} + \Gamma_\text{decay}$ for Model~IV and $\gamma_d= \gamma_g = \Gamma_\text{decay}$ for Model~V, respectively. It can be seen that the synchronization for these models is lower than for the effective spin-1 model (Model~I). This indicates that the effective spin-1 model is characterized by a boosted synchronization, compared to the ideal spin-1 model. Figure~\ref{fig11} shows that Model~I features the smallest $\chi^2$, indicating that our experimental data are best described by Model~I, i.e.,  the effective spin-1 model with additive Lindblad operators.

As discussed in Sec.~\ref{sec_dynamics}, our fits to the experimental data yield values for $I_{\text{bg}}^{\text{norm}}$ that are consistent with $0$. 
Since $A_S/A_I$ is not infinitely large and $I_{\pi}^{\text{norm}}(\bar{t})$ is finite, it follows 
    from Eq.~(\ref{eq_synchronization_min}) that our measurements yield $\text{min}[S_q(0,0)]=0$. Since our coupling scheme is well described by the effective spin-1 model with additive dissipators (see above), this provides strong experimental evidence that the effective spin-1 model possesses, just as the ideal spin-1 model, a synchronization blockade~\cite{PhysRevLett.118.243602,Tan2022halfintegervs,PhysRevA.99.043804,PhysRevA.108.022216}. This finding agrees with the theory analysis of the effective spin-1 model with equal energy spacings and additive dissipators, which shows that the model exhibits a synchronization blockade for $\varphi_{\text{sync}}=\pi$~\cite{molenda}.

While the duration  for which the probe and control-synchronize beams are on simultaneously for our experimental data taken for  $\Omega_d^{\text{max}}=0$ and finite $B$ is too short for the system to reach its steady state during the synchronization stage, we can determine lower limits of the steady-state synchronization based on our measurements if we assume that $S_q(L,t)$ increases monotonically between the time that the probe pulse is turned on and $t^*=-0.1$~$\mu$s. Applying the same analysis approach as before,
we find
$\text{max}[S_q(0,0)] \ge 0.123(2)$,
$0.134(3)$, $0.137(1)$, and $0.157(2)$ for $B=2 \pi \times 0.1$~MHz,
$2 \pi \times 0.2$~MHz,
$2 \pi \times 0.3$~MHz, and
$2 \pi \times 0.4$~MHz,
respectively. For comparison, the corresponding theory values for the steady-state synchronization  for the effective spin-1 model are 
$S_q^{\text{ss}}= 0.129$,
$0.164$, $0.181$, and $0.184$ for $B=2 \pi \times 0.1$~MHz,
$2 \pi \times 0.2$~MHz,
$2 \pi \times 0.3$~MHz, and
$2 \pi \times 0.4$~MHz,
respectively. 
The values extracted from experiment are lower than the theory predictions, consistent with our argument that the values extracted from experiment should be interpreted as lower bounds. 
These results are indicative of the tunibility of the maximum synchronization as well as the role of the decay beam: for the parameter combinations considered in this work, the addition of the decay beam leads to a reduction of the synchronization but ensures that the limit-cycle state is reached faster.

It is instructive to compare our work to the seminal work by Laskar {\em{et al.}}~\cite{laskar2020observation}, where synchronization of a spin-1 system was, for the first time,  investigated using the $F=1$ hyperfine ground state manifold. Even though we use, as that work, the $F=1$ ground state manifold, there exist notable differences between our and their set-ups. Our decay beams couple the $F=1$ ground state manifold to the $F'=1$  excited state hyperfine manifold while Ref.~\cite{laskar2020observation} realizes dissipative coupling by coupling the $F''=0$ manifold, i.e., the same manifold that is coupled to in order to realize---in their and our case---the coherent drive. Moreover, the turn-on and turn-off of our decay beam is nearly the same as that of the control-synchronize beam while the decay beam in Ref.~\cite{laskar2020observation} is on for the entire experimental sequence. Laskar~{\em{et al.}} interpret their experimental results within the framework of a variant of the ideal spin-1 model, which allows for non-equal energy splittings; while the coherent part is derived by adiabatic elimination, the incoherent part is not and instead added assuming cross terms can be neglected. For our parameter combinations, these cross terms have been shown to play an important role. Neglecting the cross terms may be justified in the larger $B$-field regime considered by Laskar {\em{et al.}} or in the case where the decay beam Rabi coupling strength dominates, but likely not in general. Another notable distinction is our focus on steady-state synchronization. The length of our probe beam is $t_{\pi,2}-t_{\pi,1}$, which corresponds to $0.53$~$\mu$s. Taking into account that our probe beam intensity falls off faster than that of the beam profile used by Laskar {\em{et al.}}, we estimate that our synchronization stage is 2-3 times longer than that used by     Laskar {\em{et al.}}, allowing for more time to reach the steady state.

\section{Conclusions and outlook}
\label{sec_conclusions}

This paper presents a joint experiment-theory study of quantum synchronization of an effective spin-1 system. Our experiment utilizes atoms in a MOT, where the atomic density is so low that interactions between different spins can be neglected.
While the atomic medium plays a critical role for how our measurements are conducted, namely, we record the medium-dependent retrieved probe beam intensity~\cite{laskar2020observation,zhong2026light,PhysRevA.71.041801,PhysRevA.65.022314,PhysRevA.95.013818}, the synchronization extraction protocol that we applied converts the retrieved probe beam intensity to the synchronization of an isolated effective spin-1 system. Our approach contains one free parameter, which we determine through a fit to simulation data.

The key accomplishments of our work are as follows. 
(i) We experimentally implemented a driven effective spin-1 system using cold $^{87}$Rb atoms in a MOT that features two distinct dissipative couplings. The dissipative couplings due to the decay beams are fully decoupled from the coherent drive while the dissipative couplings due to the probe and control-synchronize beams are not independent of the coherent drive. 
(ii) Our simulations, which account for the atomic medium and propagate both the probe beam Rabi coupling strength and the atomic density matrix elements, provide an excellent description of the retrieved probe beam intensity. With $\Omega_{\pi}^{\text{max}}$, $\Omega_{c}^{\text{max}}$,
$\Omega_{d}^{\text{max}}$,
and $c \mu_a$ fixed through dedicated simulation-theory comparisons~\cite{zhong2026light}, our simulations have no free parameter. 
(iii) To analyze and interpret our measurements, we employed a driven  effective spin-1 model, which is derived within second-order perturbation theory from first principles~\cite{molenda,PhysRevA.85.032111} and does not have any free parameters. This effective spin-1 model features unconventional additive Lindblad operators, distinct from standard spin-1 systems considered in the literature~\cite{PhysRevLett.121.063601,PhysRevLett.121.053601,Tan2022halfintegervs}. We found that our experimental data are described very well by this effective spin-1 model but not by the ideal spin-1 model, which only features dissipative decay into the limit cycle state.
(iv) We demonstrated a boost of the maximum quantum synchronization of the effective spin-1 model with additive Lindblad operators relative to the ideal spin-1 model.
(v) We demonstrated that the effective spin-1 model  with additive Lindblad operators features, just as the ideal spin-1 model, a synchronization blockade, i.e., a regime where synchronization vanishes due to destructive interference.

The coupling schemes employed in our work, which lead to effective Lindblad operators with non-conventional dissipators, arise due to the fact that the probe and control-synchronize beams couple to the same excited state. Analogous coupling schemes can be implemented for higher-spin systems as well as other energy level structures. Dissipative couplings play a central role in dissipative phase transitions as well as in sensing schemes that utilize exceptional points. Due to the broad general interest in dissipation engineering, it is expected that our work will be of relevance broadly to the atomic, quantum optics, quantum information science, quantum sensing, quantum simulation, and quantum many-body communities.

\section*{Acknowledgments}
We thank A.~J. Sudler for discussions. This work is supported by the W. M. Keck Foundation. T.D.A.T gratefully acknowledges support through a Dodge Postdoctoral Fellowship. The computing for this project was performed at the OU Supercomputing Center for Education \& 
Research (OSCER) at the University of Oklahoma
(OU).  

\section*{Data Availability}
The data that supports the findings of this article cannot be made publicly available because the data is too large and no suitable public repository exists for hosting in this field of study. The data are available upon reasonable request from the authors.

\appendix

\section{6-level Hamiltonian and associated master equation}
\label{appendix_inmedium_simulations}

Employing the rotating-wave approximation to eliminate the rapidly oscillating terms, the full 6-level Hamiltonian $H_{\text{system}}^{\text{full}}(t)$ in the rotating frame reads~\cite{zhong2026light,molenda} 
\begin{widetext}
\begin{equation}
    H_{\text{system}}^{\text{full}}(t) = \frac{1}{2}\begin{pmatrix}
2B & 0 & 0 & -\Omega_c^+(t) & - \Omega_d(t) & 0 \\
0 & 0 & 0 & - \Omega_\pi(t) & 0 & 0 \\
0 & 0 & -2B & - \Omega_c^-(t) & 0 & - \Omega_d(t) \\
-[\Omega_c^+(t)]^* & - [\Omega_\pi(t)]^* & - [\Omega_c^-(t)]^* & 0 & 0 & 0 \\
-[\Omega_d(t)]^* & 0 & 0 & 0 & 2B' & 0 \\
0 & 0 & -[\Omega_d(t)]^* & 0 & 0 & -2B'
\end{pmatrix},
\end{equation}
\end{widetext}
where $\Omega_c^\pm(t)$, $\Omega_d(t)$, and $\Omega_\pi(t)$, respectively, denote the Rabi coupling strengths of the control, decay, and probe pulses and $B$ and $B'$ are the angular frequencies associated with the linear Zeeman splittings of the $F=1$ ground-state and $F'=1$ excited-state manifolds, respectively. Both energy shifts are experimentally controlled by a homogeneous external magnetic field, which is parallel to the polarization direction of the probe beam and defines the quantization axis [see Fig.~\ref{fig1}(a)].

The parameters that describe $\Omega_c^\pm(t)$, $\Omega_d(t)$, and $\Omega_\pi(t)$ are determined by fitting the experimentally recorded intensities $I_{c}(t)$, $I_{d}(t)$, and $I_{\pi}(t)$ to the magnitude square of the expressions 
\begin{align}
\label{eq_beam_pr}
    \Omega_{c}^{\pm}(t) = \Omega_{c}^{\max} \left[\sum_{\ell=1}^{2}(-1)^{\ell -1}\tanh \left(\frac{t-t_{c,\ell}}{\tau_{c,\ell}} \right)/2 \right] \times \nonumber \\ 
    \exp(i\chi_{\pm}) + \nonumber \\
    \Omega_{c}^{\max} \left[\sum_{\ell=3}^{4}(-1)^{\ell -1}\tanh \left(\frac{t-t_{c,\ell}}{\tau_{c,\ell}} \right)/2 \right],
\end{align}
\begin{equation}
\label{eq_beam_decay}
    \Omega_d(t) = \Omega_{d}^{\max} \left[\sum_{\ell=1}^2(-1)^{\ell -1} \tanh \left(\frac{t-t_{d,\ell}}{\tau_{d,\ell}} \right) /2 \right],
\end{equation}
and
\begin{eqnarray}
\label{eq_beam_pi}
    \Omega_\pi(t) = \Omega_{\pi}^{\max} \left[\sum_{\ell=1}^2(-1)^{\ell -1} \tanh \left(\frac{t-t_{\pi,\ell}}{\tau_{\pi,\ell}} \right) /2\right] 
    \times \nonumber \\ 
    \exp(i \chi_0).
\end{eqnarray}
The control-synchronize beam uses
\begin{align}
\label{eq_beam_control_synchronize}
    \Omega_{c}^{\pm}(t) = \Omega_{c}^{\max} \left[\sum_{\ell=1}^{2}(-1)^{\ell -1}\tanh \left(\frac{t-t_{c,\ell}}{\tau_{c,\ell}} \right)/2 \right] \times \nonumber \\ \exp(i\chi_{\pm})
\end{align}
while the control-read beam uses 
\begin{align}
\label{eq_beam_control_read}
    \Omega_{c}^{\pm}(t) = \Omega_{c}^{\max} \left[\sum_{\ell=3}^{4}(-1)^{\ell -1}\tanh \left(\frac{t-t_{c,\ell}}{\tau_{c,\ell}} \right) /2\right].
\end{align}
Strictly speaking, the electric field associated with the decay beam also has a phase. However, since none of the observables of interest in this work depend on this phase, it is not included in Eq.~(\ref{eq_beam_decay}).  
The fits of the experimental data to the magnitude square of Eqs.~(\ref{eq_beam_pr})-(\ref{eq_beam_pi}) determine the time points $t_{c,\ell}$, 
$t_{d,\ell}$, and $t_{\pi,\ell}$ as well as the time intervals $\tau_{c,\ell}$, 
$\tau_{d,\ell}$, and $\tau_{\pi,\ell}$ (see Table~\ref{tab1}) but not the maximal Rabi coupling strengths 
$\Omega_c^{\text{max}}$, $\Omega_d^{\text{max}}$, and $\Omega_{\pi}^{\text{max}}$. The chosen functional forms provide an excellent description of the experimental measured intensity profiles; on the scale of Fig.~\ref{fig1}(a), the experimental data and fit would be indistinguishable. 
The beam sequence is designed so that the rise and fall of the decay beam Rabi coupling strength closely follows that of the control-synchronize beam Rabi coupling strength. Moreover, the fall of the probe beam Rabi coupling strength closely follows that of the decay and control-synchronize Rabi coupling strengths. The premise is that the simultaneous fall-off of all three beams provides the best approximation to an instantaneous transfer of information from the light to the matter degrees of freedom.

\begin{table}[ht]
    \centering
    \begin{tabular}[t]{cc}
        Parameter & Value  \\ \hline
        $\tau_{c,1}$ & $0.1035(1)$~$\mu\text{s}$ \\
        $\tau_{c,2}$ & $0.1019(1)$~$\mu\text{s}$ \\
        $\tau_{c,3}$ & $0.1030(1)$~$\mu\text{s}$ \\ 
        $\tau_{c,4}$ & $0.1016(1)$~$\mu\text{s}$ \\ 
        $t_{c,1}$ & $-0.985(1)$~$\mu$\text{s} \\
        $t_{c,2}$ & $-0.031(1)$~$\mu$\text{s} \\
        $t_{c,3}$ & $0.016(1)~\mu\text{s} + \tau$\\
        $t_{c,4}$ & $1.575(1)~\mu\text{s} + \tau$\\
        $\tau_{d,1}$ & $0.0678(1)$~$\mu\text{s}$ \\
        $\tau_{d,2}$ & $0.0861(1)$~$\mu\text{s}$ \\
        $t_{d,1}$ & $-0.9268(1)$~$\mu$\text{s} \\
        $t_{d,2}$ & $-0.0042(1)$~$\mu$\text{s} \\  
        $\tau_{\pi,1}$ & $0.0955(1)$~$\mu\text{s}$ \\
        $\tau_{\pi,2}$ & $0.0972(1)$~$\mu\text{s}$ \\
        $t_{\pi,1}$ & $-0.5671(1)$~$\mu$\text{s} \\
        $t_{\pi,2}$ & $-0.036(1)$~$\mu$\text{s} \\  
     \end{tabular}
    \label{tab1}
    \caption{Fit parameters and their uncertainties [see Eqs.~\eqref{eq_beam_pr}-\eqref{eq_beam_pi}]. The times $t_{c,3}$ and $t_{c,4}$ depend on the delay time $\tau$.}
\end{table}

The spontaneous decay from the   $F''=0$ and $F'=1$ excited-state manifolds (lifetimes $\Gamma''$ and $\Gamma'$, respectively) to the $F=1$ ground-state manifold is described by the Lindblad jump operators $L_{j}^{\text{full}}$ with $j=1-7$,
\begin{align}
    L_{j}^{\text{full}} &= \sqrt{\Gamma''/3}\,|j\rangle\langle4|\; \mbox{ for } \; j=1,2,3, \\
    L_4^{\text{full}}&= \sqrt{\Gamma'/2}\,|1\rangle\langle5|, \\
    L_5^{\text{full}} &= \sqrt{\Gamma'/2}\,|2\rangle\langle5|,\\
    L_6^{\text{full}} &= \sqrt{\Gamma'/2}\,|2\rangle\langle6|, \\
    L_7^{\text{full}} &= \sqrt{\Gamma'/2}\,|3\rangle\langle6|.
\end{align}
 In addition, environmental dephasing is described by Lindblad jump operators that yield the terms that are proportional to $\gamma_c$ in Eq.~(\ref{eq_masterequation_6level}). 
The master equation for the 6-level systems is given by Eq.~(\ref{eq_masterequation_6level}).
It should be noted that the hyperfine states $\ket{5}$ and $\ket{6}$ can, in principle,  decay to $F=2$ ground state levels. These decay channels are neglected in our analysis since they contribute only weakly since  the populations of states $\ket{5}$ and $\ket{6}$ are generally small and since the decay beam is only on for a relatively short time (less than a microsecond). 

Equation~(\ref{eq_masterequation_6level}) describes a single externally driven 6-level system that is coupled to the environment. In our experiment, the
atoms are held in a MOT. A key aspect is that the probe beam Rabi coupling strength is modified as it travels through the atomic  medium~\cite{laskar2020observation,zhong2026light,PhysRevA.71.041801,PhysRevA.65.022314,PhysRevA.95.013818}.
 To model the propagation of the probe pulse through the atomic medium, we solve the time-dependent Maxwell-Bloch equations by self-consistently evolving the atomic density matrix and the Rabi coupling strength of the probe beam. Specifically, this means that $\Omega_{\pi}(t)$ becomes spatially dependent, i.e., $\Omega_{\pi}(t) \rightarrow \Omega_{\pi}(t,z)$, with $z$ denoting the position along the 1D MOT, which is assumed to be characterized by a homogeneous atomic density.
Under the slowly varying envelope approximation, the propagation of the probe field is described by the Maxwell equation~\cite{zhong2026light} 
\begin{align}
\label{eq:maxwelleq}
\dfrac{\partial \Omega_\pi(t,z)}{\partial t} + c\dfrac{\partial \Omega_\pi(t,z)}{\partial z} & = - \dfrac{i}{2} c\mu_a\rho_{24}(t,z),
\end{align}
where  $c$ is the speed of light in vacuum and $\mu_a$ denotes the effective coupling between the states $|2\rangle$ and $|4\rangle $.  Equation~\eqref{eq:maxwelleq} is subject to the boundary condition $\Omega_\pi(t,z=0) = {\Omega}_\pi(t)$. The coupled Eqs.~\eqref{eq_masterequation_6level}, with each of the density matrix elements now being dependent on both $t$ and $z$, and \eqref{eq:maxwelleq} are then numerically solved self-consistently using the algorithm described in Ref.~\cite{zhong2026light}.

The maximal Rabi coupling strengths of the control and probe beams as well as the effective coupling between states $\ket{2}$ and $\ket{4}$ were determined in our previous paper~\cite{zhong2026light}. We  repeated the calibration for the present work and found that the experimental system has been stable, i.e., corrections to the previously determined values were not needed.  Owing to the optimization of the polarization gradient cooling, the MOT temperature is  reduced compared to our previous work. Using the experimental $B=0$ data for fixed $\chi^S$ and various $\tau$, we find from a fit to the maximum of the retrieved probe signal~\cite{zhong2026light} that $\gamma_c$ is equal to $2 \pi \times 0.101(5)$~MHz. 

To calibrate the maximal Rabi coupling strength of the decay beam, we scan $\chi^S$ for three different $B$ values, namely $B=0$, $B=2 \pi \times 0.2$~MHz, and $B=2 \pi \times 0.4$~MHz, for $\tau=0.4$~$\mu$s and fixed power of the decay beam. To determine $\Omega_d^{\text{max}}$ for this power, we perform in-medium simulations for various $\Omega_d^{\text{max}}$, increasing $\Omega_d^{\text{max}}$ in steps of $2 \pi \times 0.25$~MHz. Focusing primarily on the normalized retrieved probe signal and secondarily on the transmitted probe signal, we determine  $\Omega_d^{\text{max}}$ by identifying the best simulation-experiment agreement. Using the fact that the Rabi coupling strength scales with the square root of the power, we determine the Rabi coupling strengths for the power settings used in this work.   

\section{Effective spin-1 model}
\label{appendix_effective_spin1_models}
The dynamics of the $5S_{1/2},F=1$ ground-state manifold are described by the effective master equation given in Eq.~(\ref{eq_masterequation_3level}), which is obtained from the full six-level master equation by eliminating the excited-state manifolds using second-order perturbation theory~\cite{molenda}. The perturbative treatment assumes $\gamma_c=0$. The explicit expressions for $\Delta_{\text{eff}}$, 
$\Omega_{\text{eff}}$ and
$\phi_{\text{eff}}$
read
\begin{align}
\label{eq_effective_delta}
    \Delta_{\text{eff}}=-B-\frac{B|\Omega_c^\pm|^2}{(\Gamma'')^2+4B^2}-\frac{(B-B')|\Omega_d|^2}{(\Gamma')^2+4(B-B')^2},
\end{align}
\begin{eqnarray}
    \Omega_{\text{eff}}=\frac{\sqrt{2}}{8}\frac{|B\Omega_\pi\Omega_c^\pm|}{(\Gamma'')^2/4+B^2}\sqrt{1+\left(\frac{2B}{\Gamma''}\right)^2},
\end{eqnarray}
and
\begin{align}
    \phi_{\text{eff}}=- \chi_++\chi_0+\arctan \left(\frac{2B}{\Gamma''}\right)+\pi-\frac{\pi}{2}\mathrm{sgn}(B),
\end{align}
respectively.
It should be noted that $\Delta_{\text{eff}}$, 
$\Omega_{\text{eff}}$ and
$\phi_{\text{eff}}$ depend on time.

The Lindbladian $L_j$ with $j=1-7$ are
given by
\begin{align}
\label{eq_lindblad_gammac}
    L_j&=\sum_{k=1}^3c^{(4)}_{j,k}|j\rangle\langle k|, j=1,2,3\\
    L_4&=c_{1,1}^{(5)}|1\rangle\langle 1|,\\
    L_5&=c_{2,1}^{(5)}|2\rangle\langle 1|,\\
    L_6&=c_{2,3}^{(6)}|2\rangle\langle 3|,\\
    L_7&=c_{3,3}^{(6)}|3\rangle\langle 3|,
\end{align}
where $L_1$, $L_2$, and $L_3$ are additive Lindbladian.
The prefactors $c_{j,k}^{(l)}$ are given by
\begin{align}
    c_{j,k}^{(l)}=\frac{\sqrt{\Gamma_{j,l}}\left(H^\text{full}_{\text{system}}\right)_{l,k}}{\left(H^\text{full}_{\text{system}}\right)_{l,l}-\left(H^\text{full}_{\text{system}}\right)_{k,k}-i\Gamma_l/2},
\end{align}
where $\Gamma_l$ is the total decay rate out of $|l\rangle$ ($\Gamma_4=\Gamma''$, $\Gamma_5=\Gamma_6=\Gamma'$) and $\Gamma_{j,l}$ is the decay rate from $l$ to $j$, which satisfies $\sum_j\Gamma_{j,l}=\Gamma_l$ ($\Gamma_{j,4}=\Gamma''/3$, $\Gamma_{1,5}=\Gamma_{2,5}=\Gamma_{2,6}=\Gamma_{3,6}=\Gamma'/2$). All $\Gamma_l$ and $\Gamma_{j,l}$ not explicitly defined here are zero. $|c^{(l)}_{j,k}|^2$ is the effective dissipation rate from the ground state $|k\rangle$ to the ground state $|j\rangle$ via the excited state $|l\rangle$.
To account for dephasing,
we add  Lindblad jump operators that yield the terms that are proportional to $\gamma_c$ in Eq.~(\ref{eq_masterequation_3level}) in an {\em{ad hoc}} manner.
This {\em{ad hoc}} treatment is consistent with the notion that terms proportional to $\gamma_c$ describe dephasing (i.e., account for dissipative processes that are not accounted for otherwise).
In the main text, the model described by Eqs.~(\ref{eq_masterequation_3level}), (\ref{eq_ham_spin1}), and (\ref{eq_effective_delta})-(\ref{eq_lindblad_gammac}) are referred to as the effective spin-1 model.

It was shown in Ref.~\cite{molenda} that the effective spin-1 model provides a very good description of the dynamics of the full 6-state system for the parameter combinations considered. To quantify the deviations, Table~\ref{tab_synch} lists the maximum of the steady-steady synchronization of the effective
spin-1 model, together with its deviation from the maximum of the steady-state synchronization of the isolated 6-state model. For the nine parameter combinations for which we took experimental steady-state data, the maximum of the synchronization is, on average, $2$~$\%$ lower for the effective spin-1 model than for the 6-state model, with the maximal and minimal percentage deviations being $7.5$~\% and $0.1$~\%, respectively.
Since the agreement between the two models deteriorates with increasing $B$ ($B\ge 2 \pi \times 0.1$~MHz), we restrict ourselves to $B \le 2\pi \times 0.4$~MHz in this work.

\begin{widetext}

\begin{table}
    \centering
    \begin{tabular}[t]{c||c|c|c|c|c}
        $B/(2 \pi)$ & $\Omega_d^{\text{max}}=0$ & $\Omega_d^{\text{max}}=2 \pi \times 1.75$~MHz & $\Omega_d^{\text{max}}=2 \pi \times 2.1$~MHz& $\Omega_d^{\text{max}}=2 \pi \times 2.45$~MHz& $\Omega_d^{\text{max}}=2 \pi \times 2.75$~MHz\\ \hline
        $0.4$~MHz & 0.184 / 90.0 & 0.125 / 87.9 & 0.116 / 92.6 & 0.110 / 95.9 & 0.106 / 97.7 \\ \hline
        $0.3$~MHz & 0.181 / 93.9 & 0.119 / 93.8 & 0.112 / 96.6 & 0.107 / 98.4 & 0.104 / 99.3 \\\hline
        $0.2$~MHz & 0.164 / 95.9 & 0.111 / 98.0 & 0.107 / 99.1 & 0.104 / 99.8 & 0.102 / 99.8 \\ \hline
        $0.1$~MHz & 0.129 / 98.2 & 0.105 / 99.9 & 0.103 / 99.8 & 0.101 / 99.6 & 0.100 / 99.5 \\ \hline
        $0$~MHz   & 0.101 / 99.8 & 0.102 / 99.7 & 0.101 / 99.6 & 0.100 / 99.5 & 0.099 / 99.4
     \end{tabular}
    \caption{Comparison of $\text{max}(S_q^{\text{ss}})$ in the steady state between the effective spin-1 model with additive Lindblad operators and the full isolated 6-state model. For each table entry, the first number denotes $\text{max}(S_q^{\text{ss}})$ in the steady state for the effective spin-1 model with additive Lindblad operators while the second number denotes the ratio of $\text{max}(S_q^{\text{ss}})$ for the effective spin-1 model and $\text{max}(S_q^{\text{ss}})$ for the isolated 6-state model, multiplied by $100$.}
    \label{tab_synch}
\end{table}

\end{widetext}

Section~\ref{sec_synchro} does not only show results for the effective spin-1 model but also for two variants of the effective spin-1 model, labeled Model II and Model III. Model II drops the self-dissipation of state $\ket{2}$, i.e., it sets the quantity $c_{2,2}^{(4)}$ to zero in an {\em{ad hoc}} manner.
Model III drops the dissipative couplings out of state $\ket{2}$ as well as the dissipative coupling of $\ket{2}$ with itself, i.e.,  it sets the quantities $c_{j,2}^{(4)}$ to zero for $j=1-3$ in an {\em{ad hoc}} manner.
We do this to show that these terms, which emerge as a consequence of the non-trivial additive structure of the effective Lindbladian, contribute appreciably.

\section{Ideal spin-1 model}
\label{appendix_ideal}

The paradigmatic spin-1 system is characterized by the system Hamiltonian $H_{\text{system}}$, Eq.~(\ref{eq_ham_spin1}), with $\alpha=0$
and the 
Lindblad operators $L_g$ and $L_d$,
where 
$L_g=\sqrt{\gamma_g/2} F_+F_z$
and
$L_d=\sqrt{\gamma_d/2} F_-F_z$~\cite{PhysRevLett.121.063601,PhysRevLett.121.053601,Tan2022halfintegervs}. Here, $F_+$ and $F_-$ denote the usual raising and lowering operators, treating the states $\ket{1}$, $\ket{2}$, and $\ket{3}$ as basis states, i.e., $F_z \ket{j}=m_F \ket{j}$.
For $\gamma_g=\gamma_d$, the system exhibits a so-called synchronization blockade in which destructive interference between the off-diagonal coherences results, despite finite driving, in zero synchronization at the level of first-order perturbation theory~\cite{PhysRevLett.118.243602,Tan2022halfintegervs,PhysRevA.99.043804,PhysRevA.108.022216}.
Even though the dissipators in the effective spin-1 model have a significantly more intricate structure than those of the paradigmatic spin-1 model, the effective spin-1 model is characterized by equal  dissipative rates for the effective $(\ket{1} \leftrightarrow \ket{2})$-couplings and the  $(\ket{3} \leftrightarrow \ket{2})$-couplings. Correspondingly, the vanishing of the synchronization for $\alpha=0$ of the effective spin-1 model discussed in the main text can be interpreted as a synchronization blockade.

Since the central observable in the main text is the maximal possible synchronization, we are interested in the situation where $\alpha=0$. 
Thus, to contrast the dynamics of the effective spin-1 model to the literature, we add a phase to the paradigmatic spin-1 model, i.e., we set $\alpha=0$ (we denote the Hamiltonian by $H_{\text{system}}^{\alpha=0}$), and we consider equal dissipative coupling strengths, i.e., we set $\gamma_g=\gamma_d=\gamma_{\text{eff}}$. We refer to the model described by   
\begin{align}
\label{eq:modified_spin1_mastereq}
    \dot{{\rho}}=-i[{H}_{\text{system}}^{\alpha=0},{\rho}]+\frac{\gamma_{\text{eff}}}{2}D[{F}_+{F}_z]({\rho})+\frac{\gamma_{\text{eff}}}{2}D[{F}_-{F}_z]({\rho})
\end{align}
as the ideal spin-1 model. 
The main text compares the synchronization of the effective spin-1 model with that of the ideal spin-1 model, using $\gamma_{\text{eff}}=\Gamma_{\text{decay}}$ and $\gamma_{\text{eff}}=\Gamma_{\text{decay}}+\Gamma_{\text{control}}$,
where $\Gamma_{\text{decay}}$ is governed by the decay beam and 
$\Gamma_{\text{control}}$ by the control-synchronize beam,
\begin{eqnarray}
    \Gamma_{\text{decay}}= |c^{(5)}_{2,1}|^2=|c^{(6)}_{2,3}|^2
\end{eqnarray}
and
\begin{eqnarray}
    \Gamma_{\text{control}}= |c^{(4)}_{2,1}|^2=|c^{(4)}_{2,3}|^2.
\end{eqnarray}

\section{Husimi-Q function, phase distribution, and synchronization}

The Husimi-Q function $Q(\theta,\phi)$ for a spin-F system is defined through~\cite{PhysRevA.12.1019}
\begin{eqnarray}
Q(\theta,\phi)=\frac{2F+1}{4 \pi} \bra{\theta,\phi} \rho \ket{\theta,\phi},    
\end{eqnarray}
where
\begin{eqnarray}
    \ket{\theta,\phi}=\exp(-i \phi F_z) \exp(-i \theta F_y) \ket{m_F=F}.
\end{eqnarray}
Defining 
\begin{eqnarray}
    P(\phi)=\int_0^{\pi} Q(\theta,\phi) \sin \theta d \theta,
\end{eqnarray}
the normalization can be written as 
\begin{eqnarray}
    \int_0^{2 \pi} P(\phi) d\phi=1.
\end{eqnarray}
The quantity $2 \pi \max_{\phi}P(\phi)-1$, which measures the localization of the Husimi-Q function, is frequently used as an alternative quantum-synchronization measure~\cite{PhysRevA.91.061401}. Correspondingly, inspection of the Husimi-Q function provides information about the quantum synchronization of the system.
The determination of $Q(\theta,\phi)$ for the 3-level system, which is characterized by $F=1$, is straightforward. For the 6-level system, we work in a regime where the populations of states $\ket{4}$, $\ket{5}$, and $\ket{6}$ are small. To calculate the Husimi-Q function, we take the sub-space spanned by the states $\ket{1}$, $\ket{2}$, and $\ket{3}$ and set the other blocks of the density matrix artificially to zero.

\section{Experimental details}
\label{appendix_experiment}
 Except for the decay beam, the experimental apparatus is unchanged and described in detail in our previous work~\cite{zhong2026light}, including the MOT preparation, magnetic-field calibration, atom number determination, and determination of the temperature of the atomic cloud. This appendix summarizes the experimental procedure used to realize  synchronization.

\begin{figure}
    \centering
    \vspace{0.05in}
    \includegraphics[width=0.9\linewidth]{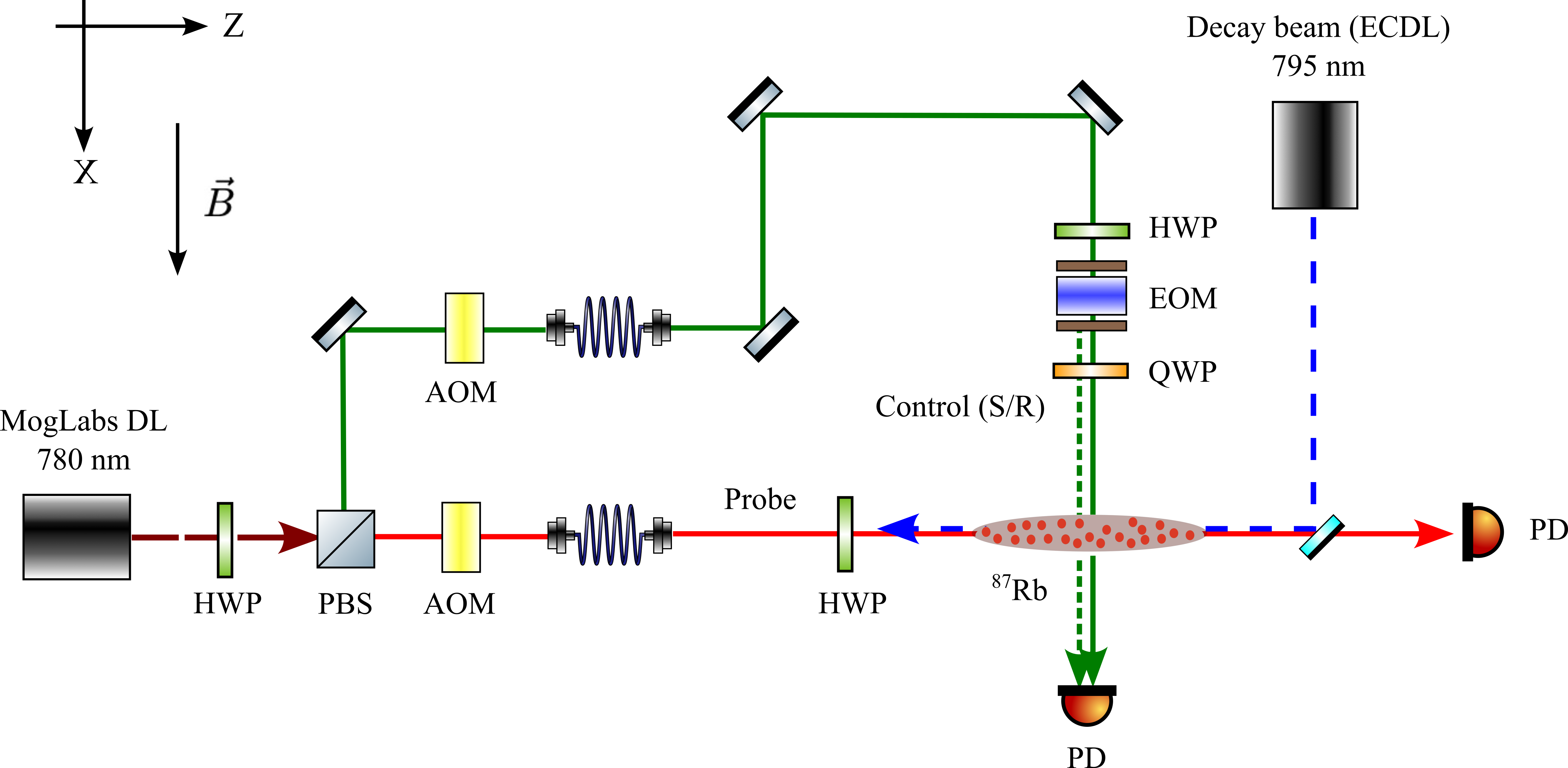}
    \caption{Experimental layout. The optical setup is identical to that reported in our previous work~\cite{zhong2026light}, except for the addition of the decay beam (blue dashed lines). The decay beam is generated by an external-cavity diode laser (ECDL), which operates at 795~nm and is resonant with the $^{87}$Rb D$_1$ transition.  A homogeneous bias magnetic field $\vec{B}$ is applied along the $x$-axis (this is the quantization axis). 
    The abbreviations stand for diode laser (DL), half-wave plate (HWP), acousto-optic modulator (AOM), quarter wave plate (QWP), electro-optic modulator (EOM), and photo diode (PD). }
    \label{fig:layout}
\end{figure}

The experimental layout is shown in Fig.~\ref{fig:layout}. The control beam (green lines) is linearly polarized and propagates along the quantization axis, with a $1/e^2$ beam diameter of $6.45~\mathrm{mm}$ and a power of $11.44~\mathrm{mW}$. The probe beam (red lines in Fig.~\ref{fig:layout}) is $\pi$-polarized, with $1/e^2$ beam diameters of $0.64~\mathrm{mm}$ and a power of $9.68~\mathrm{\mu W}$. The decay beam (blue lines in Fig.~\ref{fig:layout}) is generated with an external-cavity diode laser (ECDL) that operates at $795$~nm, resonant with the $^{87}$Rb D$_1$ line (transition from $|F=1, m_F=\pm1\rangle$ to $|F'=1, m_F=\pm1\rangle$). The decay beam 
is 
$\pi$-polarized and counter-propagates with respect to the probe beam, with a $1/e^2$ diameter of approximately 
$2.0~\mathrm{mm}$,
which is larger than the probe beam to ensure that all atoms that are addressed by the probe beam are also illuminated by the decay beam. 

\begin{figure}
    \centering
    \vspace*{0.1in}
    \includegraphics[width=0.9\linewidth]{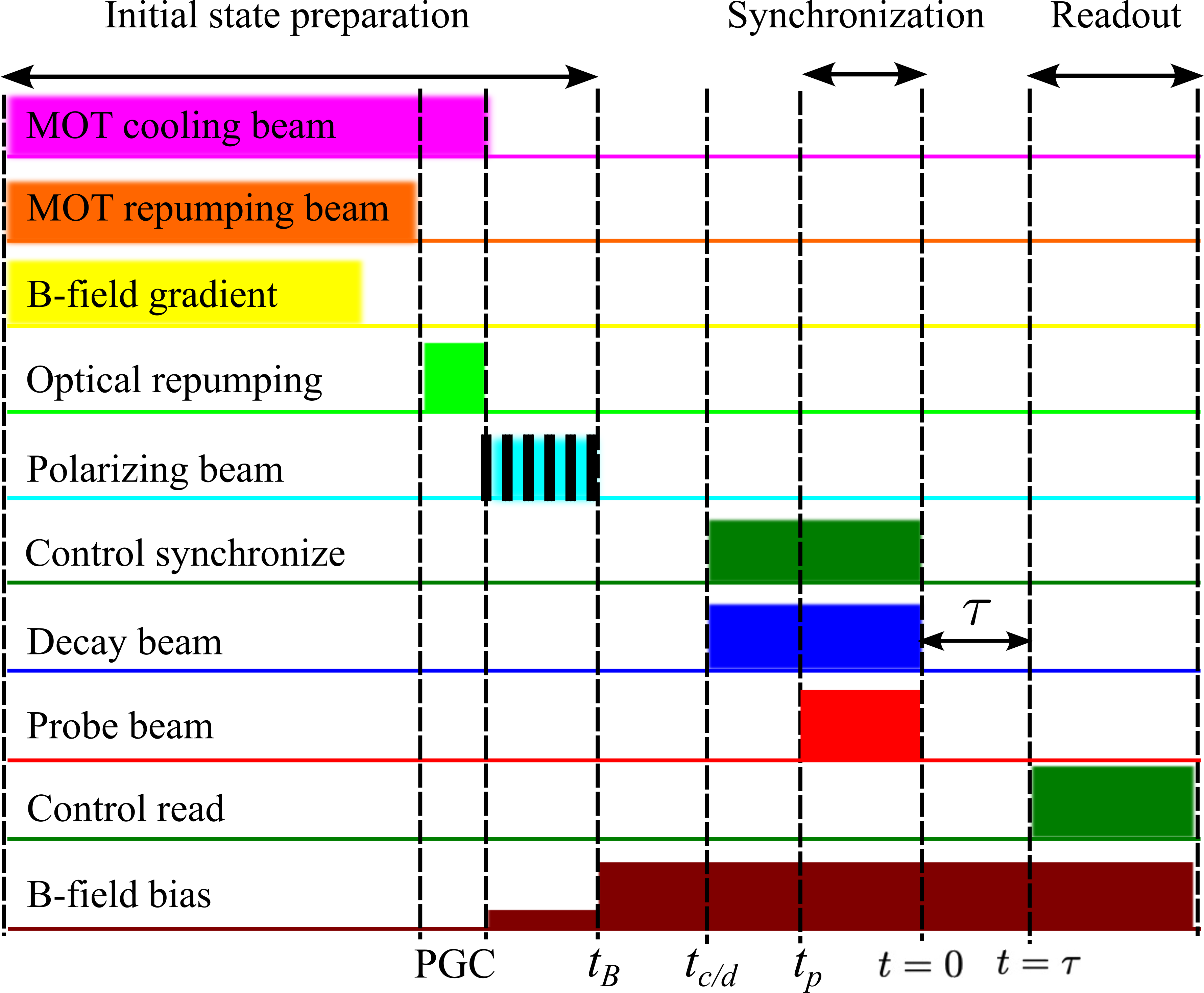}
    \caption{The experimental timing sequence (not drawn to scale) includes the initial-state preparation, synchronization, and readout stages. Two different initial states are prepared in this work. For initial state preparation protocol I, a polarizing beam (cyan-striped region) is applied after polarization gradient cooling (PGC) has concluded. For initial state preparation protocol II, the polarizing beam is off for all times. At $t_B$, i.e., after the initial state preparation stage, the magnitude of the homogeneous magnetic bias field is increased and held for $40~\mu\mathrm{s}$ to allow the magnetic field to stabilize. The control and decay beams are turned on approximately simultaneously at $t_{c/d}$ and the probe beam at $t_p$. The probe, control, and decay beams are switched off approximately simultaneously at $t=0$. Following a variable delay time $\tau$, the control-read is applied to transfer information from the spin waves to the light. The retrieved optical signal is detected with a photo diode detector.}
    \label{fig:timeseq}
\end{figure}

The experimental timing sequence 
is illustrated in Fig.~\ref{fig:timeseq}. Our state preparation protocols I and II prepare two distinct  initial atomic states. In both protocols, the MOT is loaded by applying MOT cooling beams (red-detuned by $15~\mathrm{MHz}$ from the $|F=2\rangle \rightarrow |F''=3\rangle$ transition), MOT repumping beams, and a magnetic-field gradient for $2~\mathrm{s}$. The atomic ensemble is subsequently cooled by polarization-gradient cooling (PGC) for $400~\mu\mathrm{s}$.  During the PGC stage, the MOT cooling beams are red-detuned by $40~\mathrm{MHz}$ from the $|F=2\rangle \rightarrow |F''=3\rangle$ transition. 
Moreover, the MOT repumping beam is switched off and an optical repumping beam is turned on subsequently, which is resonant with the $|F=2\rangle \rightarrow |F''=2\rangle$ transition and efficiently transfers the atomic population into the $F=1$ ground-state manifold. The PGC reduces the temperature of the atomic ensemble from several hundred microkelvin to approximately $20~\mu\mathrm{K}$.
State preparation protocol I applies a polarizing beam to transfer atoms from the $|F=1,m_F=\pm1\rangle$ states to the $|F=1,m_F=0\rangle$ state via the excited state $|F''=0,m_F=0\rangle$. The polarizing beam is designed to prepare a population distribution close to $(0,1,0)$ in the $m_F=-1,0,+1$ basis. 
While the polarizing beam is on, a bias magnetic field (directed along the $x$-axis) of small magnitude is applied  to lift the degeneracy of the $m_F=\pm1$ states, thereby suppressing unwanted Raman coupling between these two Zeeman sublevels while efficiently transferring atomic population from the $m_F=\pm1$ states to the $m_F=0$ state.
State preparation protocol II does {\em{not}} apply the  polarizing beam. Consequently, the atomic population is approximately equally distributed among the three Zeeman sublevels of the $F=1$ manifold, yielding a population distribution close to $(1/3,1/3,1/3)$. 

In preparation for the synchronization stage, the magnitude of the homogeneous magnetic bias field is increased. After a $40~\mu\mathrm{s}$ wait time during which the magnetic field stabilizes, the control and decay beams are turned on essentially simultaneously using acousto-optic modulators. Once both beams are stable (this takes less than a $\mu$s), the probe beam is turned on for approximately $0.6~\mu\mathrm{s}$. During the synchronization stage, defined by the probe and control beams being on simultaneously, the atomic density matrix reaches, for most parameter combinations, a steady state.   At the end of the synchronization stage, the probe, control, and decay (if present) beams are switched off approximately simultaneously. Information is read out from the %excited 
spin-waves by applying a control-read beam after a variable delay time $\tau$. The retrieved optical signal is detected using a photo diode, and its intensity is recorded and analyzed to extract the synchronization of the spin system at time $t=0$.

   \section{Experimental and simulation data for $\tau$ scan}
\label{appendix_fulldata}

\begin{figure*}[!htbp]
    \centering
    \includegraphics[width=\linewidth]{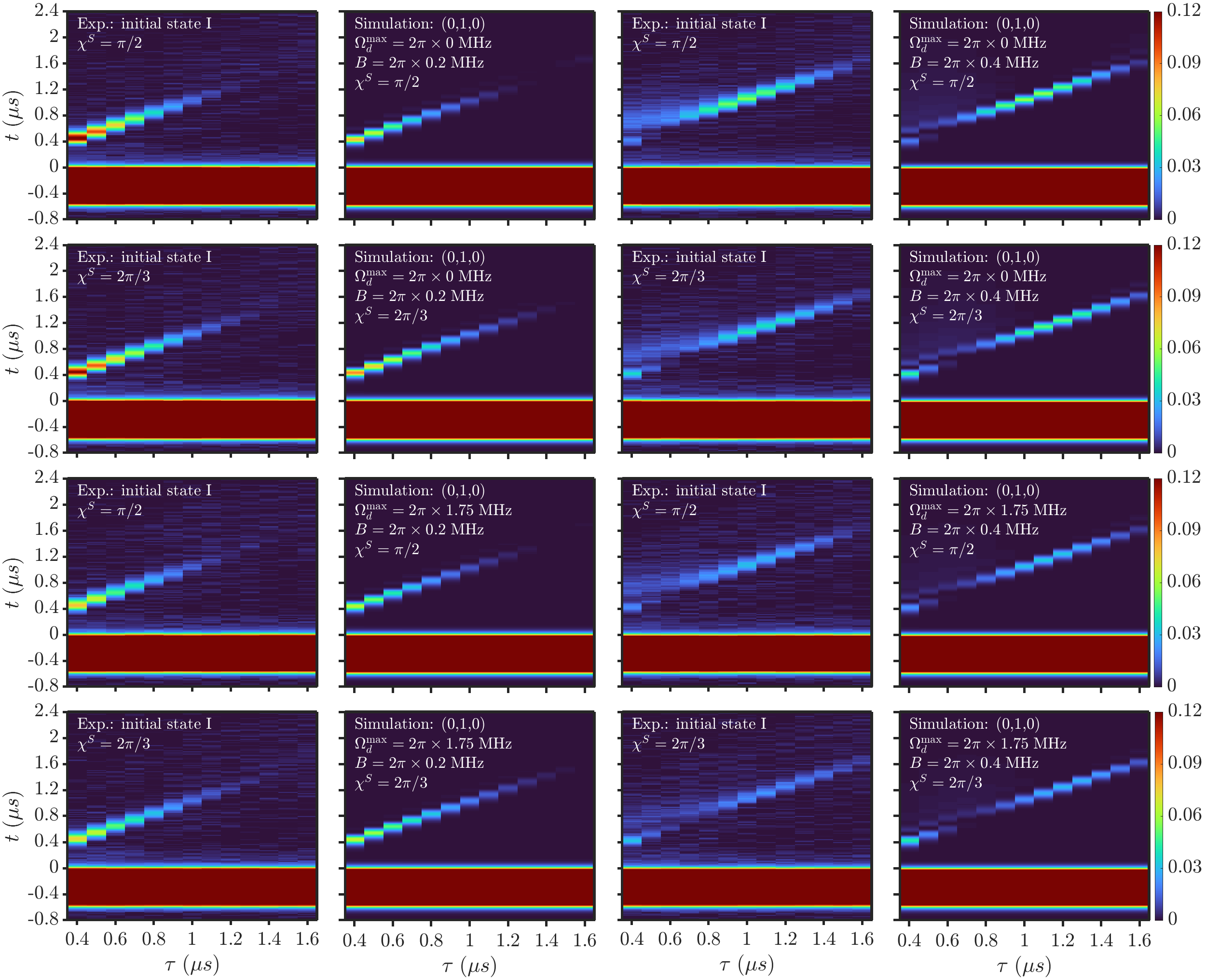}
    \caption{ Normalized probe beam intensity $I_{\pi}^{\text{norm}}(t)$ at the end of the MOT (color coded)  as functions of the time $t$ and the hold time $\tau$ for the relative phases $\chi^S = \pi/2$ (first and third rows) and $\chi^S = 2\pi/3$ (second and fourth rows). The first and third columns show experimental data obtained using the initial state preparation~I, while the second and fourth columns show the simulation results for the initial state $\ket{2}\bra{2}$ [denoted by $(0,1,0)$ in the figure labels] for two magnetic field strengths $B$ and two maximum Rabi coupling strengths $\Omega_d^\text{max}$ as shown in the labels. The agreement between the experimental data and simulation results is very good. }
    \label{figextra_figure_exp_3}
\end{figure*}

Figure~\ref{figextra_figure_exp_3} shows experimental data as functions of $\tau$ and $t$ for fixed $B$, $\Omega_d^{\text{max}}$, and $\chi^S$. The agreement between experiment (first and third columns) and the simulations (second and fourth columns) is very good.
The supplemental material~\cite{SM} contains additional experimental data as well as corresponding simulation data.

\section{Supplemental material}
\begin{figure*}[!htbp]
    \centering
    \includegraphics[width=\textwidth]{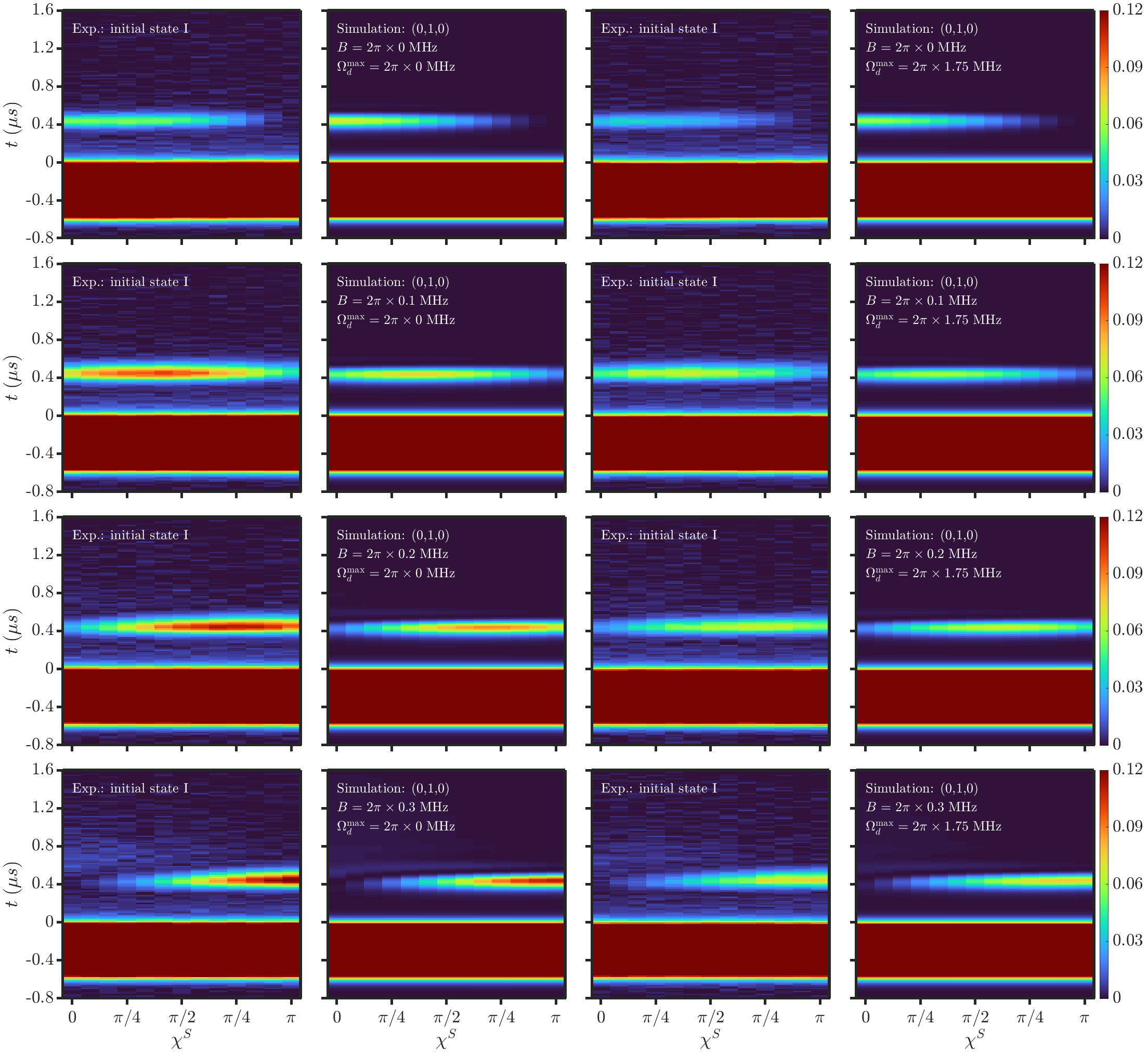}
    \caption{Normalized probe beam intensity $I_{\pi}^{\text{norm}}(t)$ at the end of the MOT (color coded) for the hold time $\tau=0.4$~$\mu$s as functions of the time $t$ and the relative phase $\chi^S$ of the plus- and minus-components of the control-synchronize beam. The first and third columns show experimental data obtained using the initial state preparation~I while the second and fourth columns show simulation results for the initial state $\ket{2}\bra{2}$ [denoted by $(0,1,0)$ in the figure labels]. Two different maximal magnitudes $\Omega_d^{\text{max}}$ of the Rabi coupling strength of the decay beam are used, namely $\Omega_d^\text{max}=2 \pi \times 0$~MHz (two leftmost columns) and $\Omega_d^\text{max}=2 \pi \times 1.75$~MHz (two rightmost columns) for different magnetic field strengths (see the labels). The agreement between the experimental data and simulation results is very good.  
    For $B=0$~MHz, $\Omega_d^{\text{max}}=2\pi\times 1.75$~MHz, and $\chi^S=5\pi/12$, 
        the experimental data  shown are computed by averaging the experimental results for $\chi^S=6\pi/12$ and $\chi^S=4\pi/12$. }
    \label{figextra_figure_exp_1}
\end{figure*}

\begin{figure}[!htbp]
    \centering
    \includegraphics[width=\linewidth]{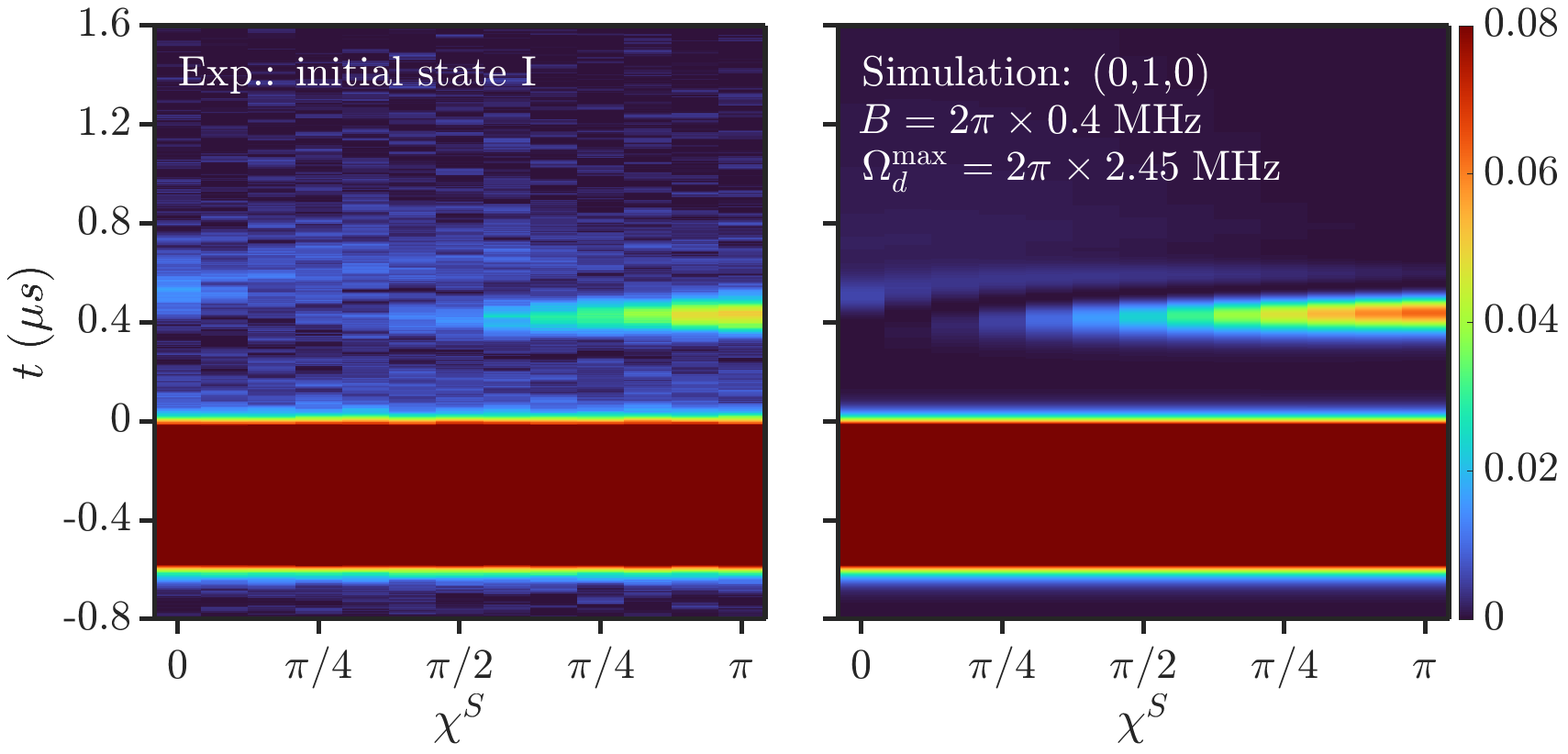}
    \caption{Normalized probe beam intensity $I_{\pi}^{\text{norm}}(t)$ at the end of the MOT (color coded) for the hold time $\tau=0.4$~$\mu$s as functions of the time $t$ and the relative phase $\chi^S$ of the plus- and minus-components of the control-synchronize beam for $B=2 \pi \times 0.4$~MHz and $\Omega_d^\text{max}=2 \pi \times 2.45$~MHz. The left panel shows experimental data obtained using the initial state preparation~I, while the right panel shows simulation results for the initial state $\ket{2}\bra{2}$ [denoted by $(0,1,0)$ in the figure labels]. The agreement between the experimental data and simulation results is very good.  }
    \label{figextra_figure_exp_2}
\end{figure}

\begin{widetext}

Table~\ref{tab2} summarizes the figure numbers that contain the experimentally measured normalized probe pulse intensity for various  $(B,\Omega_d^{\text{max}})$ combinations. This supplemental material shows Figs.~\ref{figextra_figure_exp_1} and \ref{figextra_figure_exp_2}, which adopt the same format as used in the main text and Appendix~\ref{appendix_fulldata}. 
\begin{table}
    \centering
    \begin{tabular}[t]{c||c|c|c|c|c}
        $B/(2 \pi)$ & $\Omega_d^{\text{max}}=0$ & $\Omega_d^{\text{max}}=2 \pi \times 1.75$~MHz & $\Omega_d^{\text{max}}=2 \pi \times 2.1$~MHz& $\Omega_d^{\text{max}}=2 \pi \times 2.45$~MHz& $\Omega_d^{\text{max}}=2 \pi \times 2.75$~MHz\\ \hline
        $0.4$~MHz & scan of $\chi^S$: Fig.~\ref{fig5} & scan of $\chi^S$: Fig.~\ref{fig5} & \textcolor{red}{scan of $\chi^S$: Fig.~\ref{fig2}} & \textcolor{red}{scan of $\chi^S$: Fig.~\ref{figextra_figure_exp_2}} & \textcolor{red}{scan of $\chi^S$: Fig.~\ref{fig2}} \\
        & scan of $\tau$: Fig.~\ref{figextra_figure_exp_3}  & scan of $\tau$: Fig.~\ref{figextra_figure_exp_3}  &&& \\ \hline
        $0.3$~MHz & scan of $\chi^S$: Fig.~\ref{figextra_figure_exp_1} & scan of $\chi^S$: Fig.~\ref{figextra_figure_exp_1} && \\\hline
        $0.2$~MHz & scan of $\chi^S$: Fig.~\ref{figextra_figure_exp_1} & \textcolor{red}{scan of $\chi^S$: Fig.~\ref{figextra_figure_exp_1}} & \textcolor{red}{scan of $\chi^S$: Fig.~\ref{fig2}} &  & \textcolor{red}{scan of $\chi^S$: Fig.~\ref{fig2}} \\
        & scan of $\tau$: Fig.~\ref{figextra_figure_exp_3}  & \textcolor{red}{scan of $\tau$: Fig.~\ref{figextra_figure_exp_3}}  &&& \\ \hline
        $0.1$~MHz & scan of $\chi^S$: Fig.~\ref{figextra_figure_exp_1} & \textcolor{red}{scan of $\chi^S$: Fig.~\ref{figextra_figure_exp_1}}&&& \\ \hline
        $0$~MHz & \textcolor{red}{scan of $\chi^S$: Fig.~\ref{figextra_figure_exp_1}}& \textcolor{red}{scan of $\chi^S$: Fig.~\ref{figextra_figure_exp_1}} &&& 
     \end{tabular}
    \caption{Summary of experimental data. Entries in red color refer to  experimental data for which steady-state synchronization is observed.}
    \label{tab2}
\end{table}

\end{widetext}

%

%\bibliography{references.bib}

\begin{thebibliography}{53}%
\makeatletter
\providecommand \@ifxundefined [1]{%
 \@ifx{#1\undefined}
}%
\providecommand \@ifnum [1]{%
 \ifnum #1\expandafter \@firstoftwo
 \else \expandafter \@secondoftwo
 \fi
}%
\providecommand \@ifx [1]{%
 \ifx #1\expandafter \@firstoftwo
 \else \expandafter \@secondoftwo
 \fi
}%
\providecommand \natexlab [1]{#1}%
\providecommand \enquote  [1]{``#1''}%
\providecommand \bibnamefont  [1]{#1}%
\providecommand \bibfnamefont [1]{#1}%
\providecommand \citenamefont [1]{#1}%
\providecommand \href@noop [0]{\@secondoftwo}%
\providecommand \href [0]{\begingroup \@sanitize@url \@href}%
\providecommand \@href[1]{\@@startlink{#1}\@@href}%
\providecommand \@@href[1]{\endgroup#1\@@endlink}%
\providecommand \@sanitize@url [0]{\catcode `\\12\catcode `\$12\catcode
  `\&12\catcode `\#12\catcode `\^12\catcode `\_12\catcode `\%12\relax}%
\providecommand \@@startlink[1]{}%
\providecommand \@@endlink[0]{}%
\providecommand \url  [0]{\begingroup\@sanitize@url \@url }%
\providecommand \@url [1]{\endgroup\@href {#1}{\urlprefix }}%
\providecommand \urlprefix  [0]{URL }%
\providecommand \Eprint [0]{\href }%
\providecommand \doibase [0]{https://doi.org/}%
\providecommand \selectlanguage [0]{\@gobble}%
\providecommand \bibinfo  [0]{\@secondoftwo}%
\providecommand \bibfield  [0]{\@secondoftwo}%
\providecommand \translation [1]{[#1]}%
\providecommand \BibitemOpen [0]{}%
\providecommand \bibitemStop [0]{}%
\providecommand \bibitemNoStop [0]{.\EOS\space}%
\providecommand \EOS [0]{\spacefactor3000\relax}%
\providecommand \BibitemShut  [1]{\csname bibitem#1\endcsname}%
\let\auto@bib@innerbib\@empty
%</preamble>
\bibitem [{\citenamefont {Pikovsky}\ \emph {et~al.}(2001)\citenamefont
  {Pikovsky}, \citenamefont {Rosenblum},\ and\ \citenamefont
  {Kurths}}]{pikovsky2001synchronization}%
  \BibitemOpen
  \bibfield  {author} {\bibinfo {author} {\bibfnamefont {A.}~\bibnamefont
  {Pikovsky}}, \bibinfo {author} {\bibfnamefont {M.}~\bibnamefont
  {Rosenblum}},\ and\ \bibinfo {author} {\bibfnamefont {J.}~\bibnamefont
  {Kurths}},\ }\href@noop {} {\emph {\bibinfo {title} {{Synchronization: A
  Universal Concept in Nonlinear Sciences}}}},\ \bibinfo {series} {Cambridge
  Nonlinear Science Series}, Vol.~\bibinfo {volume} {12}\ (\bibinfo
  {publisher} {Cambridge University Press},\ \bibinfo {address} {Cambridge,
  UK},\ \bibinfo {year} {2001})\BibitemShut {NoStop}%
\bibitem [{\citenamefont {Kuramoto}(1975)}]{Kuramoto1975}%
  \BibitemOpen
  \bibfield  {author} {\bibinfo {author} {\bibfnamefont {Y.}~\bibnamefont
  {Kuramoto}},\ }\bibfield  {title} {\bibinfo {title} {{Self-entrainment of a
  population of coupled non-linear oscillators}},\ }in\ \href
  {https://doi.org/10.1007/BFb0013365} {\emph {\bibinfo {booktitle}
  {International Symposium on Mathematical Problems in Theoretical Physics}}},\
  \bibinfo {series} {Lecture Notes in Physics}, Vol.~\bibinfo {volume} {39},\
  \bibinfo {editor} {edited by\ \bibinfo {editor} {\bibfnamefont
  {H.}~\bibnamefont {Araki}}}\ (\bibinfo  {publisher} {Springer, Berlin,
  Heidelberg},\ \bibinfo {year} {1975})\ pp.\ \bibinfo {pages}
  {420--422}\BibitemShut {NoStop}%
\bibitem [{\citenamefont {Acebr\'on}\ \emph {et~al.}(2005)\citenamefont
  {Acebr\'on}, \citenamefont {Bonilla}, \citenamefont {P\'erez~Vicente},
  \citenamefont {Ritort},\ and\ \citenamefont {Spigler}}]{RevModPhys.77.137}%
  \BibitemOpen
  \bibfield  {author} {\bibinfo {author} {\bibfnamefont {J.~A.}\ \bibnamefont
  {Acebr\'on}}, \bibinfo {author} {\bibfnamefont {L.~L.}\ \bibnamefont
  {Bonilla}}, \bibinfo {author} {\bibfnamefont {C.~J.}\ \bibnamefont
  {P\'erez~Vicente}}, \bibinfo {author} {\bibfnamefont {F.}~\bibnamefont
  {Ritort}},\ and\ \bibinfo {author} {\bibfnamefont {R.}~\bibnamefont
  {Spigler}},\ }\bibfield  {title} {\bibinfo {title} {{The Kuramoto model: A
  simple paradigm for synchronization phenomena}},\ }\href
  {https://doi.org/10.1103/RevModPhys.77.137} {\bibfield  {journal} {\bibinfo
  {journal} {Rev. Mod. Phys.}\ }\textbf {\bibinfo {volume} {77}},\ \bibinfo
  {pages} {137} (\bibinfo {year} {2005})}\BibitemShut {NoStop}%
\bibitem [{\citenamefont {Motter}\ \emph {et~al.}(2013)\citenamefont {Motter},
  \citenamefont {Myers}, \citenamefont {Anghel},\ and\ \citenamefont
  {Nishikawa}}]{Motter2013}%
  \BibitemOpen
  \bibfield  {author} {\bibinfo {author} {\bibfnamefont {A.~E.}\ \bibnamefont
  {Motter}}, \bibinfo {author} {\bibfnamefont {S.~A.}\ \bibnamefont {Myers}},
  \bibinfo {author} {\bibfnamefont {M.}~\bibnamefont {Anghel}},\ and\ \bibinfo
  {author} {\bibfnamefont {T.}~\bibnamefont {Nishikawa}},\ }\bibfield  {title}
  {\bibinfo {title} {{Spontaneous synchrony in power-grid networks}},\ }\href
  {https://doi.org/10.1038/nphys2535} {\bibfield  {journal} {\bibinfo
  {journal} {Nature Physics}\ }\textbf {\bibinfo {volume} {9}},\ \bibinfo
  {pages} {191} (\bibinfo {year} {2013})}\BibitemShut {NoStop}%
\bibitem [{\citenamefont {Nishikawa}\ and\ \citenamefont
  {Motter}(2015)}]{Nishikawa_2015}%
  \BibitemOpen
  \bibfield  {author} {\bibinfo {author} {\bibfnamefont {T.}~\bibnamefont
  {Nishikawa}}\ and\ \bibinfo {author} {\bibfnamefont {A.~E.}\ \bibnamefont
  {Motter}},\ }\bibfield  {title} {\bibinfo {title} {{Comparative analysis of
  existing models for power-grid synchronization}},\ }\href
  {https://doi.org/10.1088/1367-2630/17/1/015012} {\bibfield  {journal}
  {\bibinfo  {journal} {New Journal of Physics}\ }\textbf {\bibinfo {volume}
  {17}},\ \bibinfo {pages} {015012} (\bibinfo {year} {2015})}\BibitemShut
  {NoStop}%
\bibitem [{\citenamefont {van~der Pol}(1927)}]{vanderPol1927}%
  \BibitemOpen
  \bibfield  {author} {\bibinfo {author} {\bibfnamefont {B.}~\bibnamefont
  {van~der Pol}},\ }\bibfield  {title} {\bibinfo {title} {{Forced oscillations
  in a circuit with non-linear resistance. ({Reception} with reactive
  triode)}},\ }\href@noop {} {\bibfield  {journal} {\bibinfo  {journal} {The
  London, Edinburgh, and Dublin Philosophical Magazine and Journal of Science}\
  }\bibinfo {series} {7},\ \textbf {\bibinfo {volume} {3}},\ \bibinfo {pages}
  {65} (\bibinfo {year} {1927})}\BibitemShut {NoStop}%
\bibitem [{\citenamefont {Mirollo}\ and\ \citenamefont
  {Strogatz}(1990)}]{doi:10.1137/0150098}%
  \BibitemOpen
  \bibfield  {author} {\bibinfo {author} {\bibfnamefont {R.~E.}\ \bibnamefont
  {Mirollo}}\ and\ \bibinfo {author} {\bibfnamefont {S.~H.}\ \bibnamefont
  {Strogatz}},\ }\bibfield  {title} {\bibinfo {title} {{Synchronization of
  Pulse-Coupled Biological Oscillators}},\ }\href
  {https://doi.org/10.1137/0150098} {\bibfield  {journal} {\bibinfo  {journal}
  {SIAM Journal on Applied Mathematics}\ }\textbf {\bibinfo {volume} {50}},\
  \bibinfo {pages} {1645} (\bibinfo {year} {1990})},\ \Eprint
  {https://arxiv.org/abs/https://doi.org/10.1137/0150098}
  {https://doi.org/10.1137/0150098} \BibitemShut {NoStop}%
\bibitem [{\citenamefont {Haken}(2002)}]{haken2002brain}%
  \BibitemOpen
  \bibfield  {author} {\bibinfo {author} {\bibfnamefont {H.}~\bibnamefont
  {Haken}},\ }\href@noop {} {\emph {\bibinfo {title} {{Brain Dynamics:
  Synchronization and Activity Patterns in Pulse-Coupled Neural Nets with
  Delays and Noise}}}},\ Springer Series in Synergetics\ (\bibinfo  {publisher}
  {Springer},\ \bibinfo {address} {Berlin, Heidelberg},\ \bibinfo {year}
  {2002})\BibitemShut {NoStop}%
\bibitem [{\citenamefont {Ma}\ and\ \citenamefont {Tang}(2017)}]{Ma2017}%
  \BibitemOpen
  \bibfield  {author} {\bibinfo {author} {\bibfnamefont {J.}~\bibnamefont
  {Ma}}\ and\ \bibinfo {author} {\bibfnamefont {J.}~\bibnamefont {Tang}},\
  }\bibfield  {title} {\bibinfo {title} {{A review for dynamics in neuron and
  neuronal network}},\ }\href {https://doi.org/10.1007/s11071-017-3565-3}
  {\bibfield  {journal} {\bibinfo  {journal} {Nonlinear Dynamics}\ }\textbf
  {\bibinfo {volume} {89}},\ \bibinfo {pages} {1569} (\bibinfo {year}
  {2017})}\BibitemShut {NoStop}%
\bibitem [{\citenamefont {Laskar}\ \emph {et~al.}(2020)\citenamefont {Laskar},
  \citenamefont {Adhikary}, \citenamefont {Mondal}, \citenamefont {Katiyar},
  \citenamefont {Vinjanampathy},\ and\ \citenamefont
  {Ghosh}}]{laskar2020observation}%
  \BibitemOpen
  \bibfield  {author} {\bibinfo {author} {\bibfnamefont {A.~W.}\ \bibnamefont
  {Laskar}}, \bibinfo {author} {\bibfnamefont {P.}~\bibnamefont {Adhikary}},
  \bibinfo {author} {\bibfnamefont {S.}~\bibnamefont {Mondal}}, \bibinfo
  {author} {\bibfnamefont {P.}~\bibnamefont {Katiyar}}, \bibinfo {author}
  {\bibfnamefont {S.}~\bibnamefont {Vinjanampathy}},\ and\ \bibinfo {author}
  {\bibfnamefont {S.}~\bibnamefont {Ghosh}},\ }\bibfield  {title} {\bibinfo
  {title} {{Observation of Quantum Phase Synchronization in Spin-1 Atoms}},\
  }\href {https://doi.org/10.1103/PhysRevLett.125.013601} {\bibfield  {journal}
  {\bibinfo  {journal} {Phys. Rev. Lett.}\ }\textbf {\bibinfo {volume} {125}},\
  \bibinfo {pages} {013601} (\bibinfo {year} {2020})}\BibitemShut {NoStop}%
\bibitem [{\citenamefont {Parra-L\'opez}\ and\ \citenamefont
  {Bergli}(2020)}]{PhysRevA.101.062104}%
  \BibitemOpen
  \bibfield  {author} {\bibinfo {author} {\bibfnamefont {A.}~\bibnamefont
  {Parra-L\'opez}}\ and\ \bibinfo {author} {\bibfnamefont {J.}~\bibnamefont
  {Bergli}},\ }\bibfield  {title} {\bibinfo {title} {{Synchronization in
  two-level quantum systems}},\ }\href
  {https://doi.org/10.1103/PhysRevA.101.062104} {\bibfield  {journal} {\bibinfo
   {journal} {Phys. Rev. A}\ }\textbf {\bibinfo {volume} {101}},\ \bibinfo
  {pages} {062104} (\bibinfo {year} {2020})}\BibitemShut {NoStop}%
\bibitem [{\citenamefont {Roulet}\ and\ \citenamefont
  {Bruder}(2018{\natexlab{a}})}]{PhysRevLett.121.053601}%
  \BibitemOpen
  \bibfield  {author} {\bibinfo {author} {\bibfnamefont {A.}~\bibnamefont
  {Roulet}}\ and\ \bibinfo {author} {\bibfnamefont {C.}~\bibnamefont
  {Bruder}},\ }\bibfield  {title} {\bibinfo {title} {{Synchronizing the
  Smallest Possible System}},\ }\href
  {https://doi.org/10.1103/PhysRevLett.121.053601} {\bibfield  {journal}
  {\bibinfo  {journal} {Phys. Rev. Lett.}\ }\textbf {\bibinfo {volume} {121}},\
  \bibinfo {pages} {053601} (\bibinfo {year} {2018}{\natexlab{a}})}\BibitemShut
  {NoStop}%
\bibitem [{\citenamefont {Li}\ \emph {et~al.}(2025)\citenamefont {Li},
  \citenamefont {Xie}, \citenamefont {Yang}, \citenamefont {Li}, \citenamefont
  {Zhao}, \citenamefont {Cheng}, \citenamefont {Peng}, \citenamefont {Li},
  \citenamefont {Lutz}, \citenamefont {Lin},\ and\ \citenamefont
  {Du}}]{doi:10.1126/sciadv.ady5649}%
  \BibitemOpen
  \bibfield  {author} {\bibinfo {author} {\bibfnamefont {Y.}~\bibnamefont
  {Li}}, \bibinfo {author} {\bibfnamefont {Z.}~\bibnamefont {Xie}}, \bibinfo
  {author} {\bibfnamefont {X.}~\bibnamefont {Yang}}, \bibinfo {author}
  {\bibfnamefont {Y.}~\bibnamefont {Li}}, \bibinfo {author} {\bibfnamefont
  {X.}~\bibnamefont {Zhao}}, \bibinfo {author} {\bibfnamefont {X.}~\bibnamefont
  {Cheng}}, \bibinfo {author} {\bibfnamefont {X.}~\bibnamefont {Peng}},
  \bibinfo {author} {\bibfnamefont {J.}~\bibnamefont {Li}}, \bibinfo {author}
  {\bibfnamefont {E.}~\bibnamefont {Lutz}}, \bibinfo {author} {\bibfnamefont
  {Y.}~\bibnamefont {Lin}},\ and\ \bibinfo {author} {\bibfnamefont
  {J.}~\bibnamefont {Du}},\ }\bibfield  {title} {\bibinfo {title}
  {{Experimental realization and synchronization of a quantum van der Pol
  oscillator}},\ }\href {https://doi.org/10.1126/sciadv.ady5649} {\bibfield
  {journal} {\bibinfo  {journal} {Science Advances}\ }\textbf {\bibinfo
  {volume} {11}},\ \bibinfo {pages} {eady5649} (\bibinfo {year} {2025})},\
  \Eprint
  {https://arxiv.org/abs/https://www.science.org/doi/pdf/10.1126/sciadv.ady5649}
  {https://www.science.org/doi/pdf/10.1126/sciadv.ady5649} \BibitemShut
  {NoStop}%
\bibitem [{\citenamefont {Krithika}\ \emph {et~al.}(2022)\citenamefont
  {Krithika}, \citenamefont {Solanki}, \citenamefont {Vinjanampathy},\ and\
  \citenamefont {Mahesh}}]{PhysRevA.105.062206}%
  \BibitemOpen
  \bibfield  {author} {\bibinfo {author} {\bibfnamefont {V.~R.}\ \bibnamefont
  {Krithika}}, \bibinfo {author} {\bibfnamefont {P.}~\bibnamefont {Solanki}},
  \bibinfo {author} {\bibfnamefont {S.}~\bibnamefont {Vinjanampathy}},\ and\
  \bibinfo {author} {\bibfnamefont {T.~S.}\ \bibnamefont {Mahesh}},\ }\bibfield
   {title} {\bibinfo {title} {{Observation of quantum phase synchronization in
  a nuclear-spin system}},\ }\href
  {https://doi.org/10.1103/PhysRevA.105.062206} {\bibfield  {journal} {\bibinfo
   {journal} {Phys. Rev. A}\ }\textbf {\bibinfo {volume} {105}},\ \bibinfo
  {pages} {062206} (\bibinfo {year} {2022})}\BibitemShut {NoStop}%
\bibitem [{\citenamefont {Zhang}\ \emph {et~al.}(2023)\citenamefont {Zhang},
  \citenamefont {Wang}, \citenamefont {Wang}, \citenamefont {Zhang},
  \citenamefont {Wu}, \citenamefont {Jie},\ and\ \citenamefont
  {Lu}}]{zhang2023}%
  \BibitemOpen
  \bibfield  {author} {\bibinfo {author} {\bibfnamefont {L.}~\bibnamefont
  {Zhang}}, \bibinfo {author} {\bibfnamefont {Z.}~\bibnamefont {Wang}},
  \bibinfo {author} {\bibfnamefont {Y.}~\bibnamefont {Wang}}, \bibinfo {author}
  {\bibfnamefont {J.}~\bibnamefont {Zhang}}, \bibinfo {author} {\bibfnamefont
  {Z.}~\bibnamefont {Wu}}, \bibinfo {author} {\bibfnamefont {J.}~\bibnamefont
  {Jie}},\ and\ \bibinfo {author} {\bibfnamefont {Y.}~\bibnamefont {Lu}},\
  }\bibfield  {title} {\bibinfo {title} {{Quantum synchronization of a single
  trapped-ion qubit}},\ }\href
  {https://doi.org/10.1103/PhysRevResearch.5.033209} {\bibfield  {journal}
  {\bibinfo  {journal} {Phys. Rev. Res.}\ }\textbf {\bibinfo {volume} {5}},\
  \bibinfo {pages} {033209} (\bibinfo {year} {2023})}\BibitemShut {NoStop}%
\bibitem [{\citenamefont {Liu}\ \emph {et~al.}(2026)\citenamefont {Liu},
  \citenamefont {Wu}, \citenamefont {Moore}, \citenamefont {Haeffner},\ and\
  \citenamefont {W\"achtler}}]{waechtler2026}%
  \BibitemOpen
  \bibfield  {author} {\bibinfo {author} {\bibfnamefont {J.}~\bibnamefont
  {Liu}}, \bibinfo {author} {\bibfnamefont {Q.}~\bibnamefont {Wu}}, \bibinfo
  {author} {\bibfnamefont {J.~E.}\ \bibnamefont {Moore}}, \bibinfo {author}
  {\bibfnamefont {H.}~\bibnamefont {Haeffner}},\ and\ \bibinfo {author}
  {\bibfnamefont {C.~W.}\ \bibnamefont {W\"achtler}},\ }\bibfield  {title}
  {\bibinfo {title} {{Observation of Synchronization between Two Quantum van
  der Pol Oscillators in Trapped Ions}},\ }\href
  {https://doi.org/10.1103/w1bm-wjl4} {\bibfield  {journal} {\bibinfo
  {journal} {Phys. Rev. X}\ }\textbf {\bibinfo {volume} {16}},\ \bibinfo
  {pages} {021062} (\bibinfo {year} {2026})}\BibitemShut {NoStop}%
\bibitem [{\citenamefont {Schmolke}\ and\ \citenamefont
  {Lutz}(2026)}]{schmolke2026synchronizationquantumregime}%
  \BibitemOpen
  \bibfield  {author} {\bibinfo {author} {\bibfnamefont {F.}~\bibnamefont
  {Schmolke}}\ and\ \bibinfo {author} {\bibfnamefont {E.}~\bibnamefont
  {Lutz}},\ }\href {https://arxiv.org/abs/2606.21226} {\bibinfo {title}
  {{Synchronization in the quantum regime}}} (\bibinfo {year} {2026}),\ \Eprint
  {https://arxiv.org/abs/2606.21226} {arXiv:2606.21226 [quant-ph]} \BibitemShut
  {NoStop}%
\bibitem [{\citenamefont {Roulet}\ and\ \citenamefont
  {Bruder}(2018{\natexlab{b}})}]{PhysRevLett.121.063601}%
  \BibitemOpen
  \bibfield  {author} {\bibinfo {author} {\bibfnamefont {A.}~\bibnamefont
  {Roulet}}\ and\ \bibinfo {author} {\bibfnamefont {C.}~\bibnamefont
  {Bruder}},\ }\bibfield  {title} {\bibinfo {title} {{Quantum Synchronization
  and Entanglement Generation}},\ }\href
  {https://doi.org/10.1103/PhysRevLett.121.063601} {\bibfield  {journal}
  {\bibinfo  {journal} {Phys. Rev. Lett.}\ }\textbf {\bibinfo {volume} {121}},\
  \bibinfo {pages} {063601} (\bibinfo {year} {2018}{\natexlab{b}})}\BibitemShut
  {NoStop}%
\bibitem [{\citenamefont {Buča}\ \emph {et~al.}(2022)\citenamefont {Buča},
  \citenamefont {Booker},\ and\ \citenamefont {Jaksch}}]{SciPostPhys.12.3.097}%
  \BibitemOpen
  \bibfield  {author} {\bibinfo {author} {\bibfnamefont {B.}~\bibnamefont
  {Buča}}, \bibinfo {author} {\bibfnamefont {C.}~\bibnamefont {Booker}},\ and\
  \bibinfo {author} {\bibfnamefont {D.}~\bibnamefont {Jaksch}},\ }\bibfield
  {title} {\bibinfo {title} {{Algebraic theory of quantum synchronization and
  limit cycles under dissipation}},\ }\href
  {https://doi.org/10.21468/SciPostPhys.12.3.097} {\bibfield  {journal}
  {\bibinfo  {journal} {SciPost Phys.}\ }\textbf {\bibinfo {volume} {12}},\
  \bibinfo {pages} {097} (\bibinfo {year} {2022})}\BibitemShut {NoStop}%
\bibitem [{\citenamefont {Mari}\ \emph {et~al.}(2013)\citenamefont {Mari},
  \citenamefont {Farace}, \citenamefont {Didier}, \citenamefont {Giovannetti},\
  and\ \citenamefont {Fazio}}]{PhysRevLett.111.103605}%
  \BibitemOpen
  \bibfield  {author} {\bibinfo {author} {\bibfnamefont {A.}~\bibnamefont
  {Mari}}, \bibinfo {author} {\bibfnamefont {A.}~\bibnamefont {Farace}},
  \bibinfo {author} {\bibfnamefont {N.}~\bibnamefont {Didier}}, \bibinfo
  {author} {\bibfnamefont {V.}~\bibnamefont {Giovannetti}},\ and\ \bibinfo
  {author} {\bibfnamefont {R.}~\bibnamefont {Fazio}},\ }\bibfield  {title}
  {\bibinfo {title} {{Measures of Quantum Synchronization in Continuous
  Variable Systems}},\ }\href {https://doi.org/10.1103/PhysRevLett.111.103605}
  {\bibfield  {journal} {\bibinfo  {journal} {Phys. Rev. Lett.}\ }\textbf
  {\bibinfo {volume} {111}},\ \bibinfo {pages} {103605} (\bibinfo {year}
  {2013})}\BibitemShut {NoStop}%
\bibitem [{\citenamefont {Witthaut}\ \emph {et~al.}(2017)\citenamefont
  {Witthaut}, \citenamefont {Wimberger}, \citenamefont {Burioni},\ and\
  \citenamefont {Timme}}]{Witthaut2017}%
  \BibitemOpen
  \bibfield  {author} {\bibinfo {author} {\bibfnamefont {D.}~\bibnamefont
  {Witthaut}}, \bibinfo {author} {\bibfnamefont {S.}~\bibnamefont {Wimberger}},
  \bibinfo {author} {\bibfnamefont {R.}~\bibnamefont {Burioni}},\ and\ \bibinfo
  {author} {\bibfnamefont {M.}~\bibnamefont {Timme}},\ }\bibfield  {title}
  {\bibinfo {title} {{Classical synchronization indicates persistent
  entanglement in isolated quantum systems}},\ }\href
  {https://doi.org/10.1038/ncomms14829} {\bibfield  {journal} {\bibinfo
  {journal} {Nature Communications}\ }\textbf {\bibinfo {volume} {8}},\
  \bibinfo {pages} {14829} (\bibinfo {year} {2017})}\BibitemShut {NoStop}%
\bibitem [{\citenamefont {Ameri}\ \emph {et~al.}(2015)\citenamefont {Ameri},
  \citenamefont {Eghbali-Arani}, \citenamefont {Mari}, \citenamefont {Farace},
  \citenamefont {Kheirandish}, \citenamefont {Giovannetti},\ and\ \citenamefont
  {Fazio}}]{PhysRevA.91.012301}%
  \BibitemOpen
  \bibfield  {author} {\bibinfo {author} {\bibfnamefont {V.}~\bibnamefont
  {Ameri}}, \bibinfo {author} {\bibfnamefont {M.}~\bibnamefont
  {Eghbali-Arani}}, \bibinfo {author} {\bibfnamefont {A.}~\bibnamefont {Mari}},
  \bibinfo {author} {\bibfnamefont {A.}~\bibnamefont {Farace}}, \bibinfo
  {author} {\bibfnamefont {F.}~\bibnamefont {Kheirandish}}, \bibinfo {author}
  {\bibfnamefont {V.}~\bibnamefont {Giovannetti}},\ and\ \bibinfo {author}
  {\bibfnamefont {R.}~\bibnamefont {Fazio}},\ }\bibfield  {title} {\bibinfo
  {title} {{Mutual information as an order parameter for quantum
  synchronization}},\ }\href {https://doi.org/10.1103/PhysRevA.91.012301}
  {\bibfield  {journal} {\bibinfo  {journal} {Phys. Rev. A}\ }\textbf {\bibinfo
  {volume} {91}},\ \bibinfo {pages} {012301} (\bibinfo {year}
  {2015})}\BibitemShut {NoStop}%
\bibitem [{\citenamefont {Koppenh\"ofer}\ and\ \citenamefont
  {Roulet}(2019)}]{PhysRevA.99.043804}%
  \BibitemOpen
  \bibfield  {author} {\bibinfo {author} {\bibfnamefont {M.}~\bibnamefont
  {Koppenh\"ofer}}\ and\ \bibinfo {author} {\bibfnamefont {A.}~\bibnamefont
  {Roulet}},\ }\bibfield  {title} {\bibinfo {title} {{Optimal synchronization
  deep in the quantum regime: Resource and fundamental limit}},\ }\href
  {https://doi.org/10.1103/PhysRevA.99.043804} {\bibfield  {journal} {\bibinfo
  {journal} {Phys. Rev. A}\ }\textbf {\bibinfo {volume} {99}},\ \bibinfo
  {pages} {043804} (\bibinfo {year} {2019})}\BibitemShut {NoStop}%
\bibitem [{\citenamefont {Tan}\ \emph {et~al.}(2022)\citenamefont {Tan},
  \citenamefont {Bruder},\ and\ \citenamefont
  {Koppenh{\"{o}}fer}}]{Tan2022halfintegervs}%
  \BibitemOpen
  \bibfield  {author} {\bibinfo {author} {\bibfnamefont {R.}~\bibnamefont
  {Tan}}, \bibinfo {author} {\bibfnamefont {C.}~\bibnamefont {Bruder}},\ and\
  \bibinfo {author} {\bibfnamefont {M.}~\bibnamefont {Koppenh{\"{o}}fer}},\
  }\bibfield  {title} {\bibinfo {title} {{Half-integer vs. integer effects in
  quantum synchronization of spin systems}},\ }\href
  {https://doi.org/10.22331/q-2022-12-29-885} {\bibfield  {journal} {\bibinfo
  {journal} {{Quantum}}\ }\textbf {\bibinfo {volume} {6}},\ \bibinfo {pages}
  {885} (\bibinfo {year} {2022})}\BibitemShut {NoStop}%
\bibitem [{\citenamefont {Koppenh\"ofer}\ \emph {et~al.}(2020)\citenamefont
  {Koppenh\"ofer}, \citenamefont {Bruder},\ and\ \citenamefont
  {Roulet}}]{PhysRevResearch.2.023026}%
  \BibitemOpen
  \bibfield  {author} {\bibinfo {author} {\bibfnamefont {M.}~\bibnamefont
  {Koppenh\"ofer}}, \bibinfo {author} {\bibfnamefont {C.}~\bibnamefont
  {Bruder}},\ and\ \bibinfo {author} {\bibfnamefont {A.}~\bibnamefont
  {Roulet}},\ }\bibfield  {title} {\bibinfo {title} {{Quantum synchronization
  on the IBM Q system}},\ }\href
  {https://doi.org/10.1103/PhysRevResearch.2.023026} {\bibfield  {journal}
  {\bibinfo  {journal} {Phys. Rev. Res.}\ }\textbf {\bibinfo {volume} {2}},\
  \bibinfo {pages} {023026} (\bibinfo {year} {2020})}\BibitemShut {NoStop}%
\bibitem [{\citenamefont {Tao}\ \emph {et~al.}(2025)\citenamefont {Tao},
  \citenamefont {Schmolke}, \citenamefont {Hu}, \citenamefont {Huang},
  \citenamefont {Zhou}, \citenamefont {Zhang}, \citenamefont {Chu},
  \citenamefont {Zhang}, \citenamefont {Sun}, \citenamefont {Guo},
  \citenamefont {Niu}, \citenamefont {Weng}, \citenamefont {Liu}, \citenamefont
  {Zhong}, \citenamefont {Tan}, \citenamefont {Yu},\ and\ \citenamefont
  {Lutz}}]{Tao2025}%
  \BibitemOpen
  \bibfield  {author} {\bibinfo {author} {\bibfnamefont {Z.}~\bibnamefont
  {Tao}}, \bibinfo {author} {\bibfnamefont {F.}~\bibnamefont {Schmolke}},
  \bibinfo {author} {\bibfnamefont {C.-K.}\ \bibnamefont {Hu}}, \bibinfo
  {author} {\bibfnamefont {W.}~\bibnamefont {Huang}}, \bibinfo {author}
  {\bibfnamefont {Y.}~\bibnamefont {Zhou}}, \bibinfo {author} {\bibfnamefont
  {J.}~\bibnamefont {Zhang}}, \bibinfo {author} {\bibfnamefont
  {J.}~\bibnamefont {Chu}}, \bibinfo {author} {\bibfnamefont {L.}~\bibnamefont
  {Zhang}}, \bibinfo {author} {\bibfnamefont {X.}~\bibnamefont {Sun}}, \bibinfo
  {author} {\bibfnamefont {Z.}~\bibnamefont {Guo}}, \bibinfo {author}
  {\bibfnamefont {J.}~\bibnamefont {Niu}}, \bibinfo {author} {\bibfnamefont
  {W.}~\bibnamefont {Weng}}, \bibinfo {author} {\bibfnamefont {S.}~\bibnamefont
  {Liu}}, \bibinfo {author} {\bibfnamefont {Y.}~\bibnamefont {Zhong}}, \bibinfo
  {author} {\bibfnamefont {D.}~\bibnamefont {Tan}}, \bibinfo {author}
  {\bibfnamefont {D.}~\bibnamefont {Yu}},\ and\ \bibinfo {author}
  {\bibfnamefont {E.}~\bibnamefont {Lutz}},\ }\bibfield  {title} {\bibinfo
  {title} {{Noise-induced quantum synchronization with entangled
  oscillations}},\ }\href {https://doi.org/10.1038/s41467-025-63196-6}
  {\bibfield  {journal} {\bibinfo  {journal} {Nature Communications}\ }\textbf
  {\bibinfo {volume} {16}},\ \bibinfo {pages} {8457} (\bibinfo {year}
  {2025})}\BibitemShut {NoStop}%
\bibitem [{\citenamefont {Reiter}\ and\ \citenamefont
  {S\o{}rensen}(2012)}]{PhysRevA.85.032111}%
  \BibitemOpen
  \bibfield  {author} {\bibinfo {author} {\bibfnamefont {F.}~\bibnamefont
  {Reiter}}\ and\ \bibinfo {author} {\bibfnamefont {A.~S.}\ \bibnamefont
  {S\o{}rensen}},\ }\bibfield  {title} {\bibinfo {title} {{Effective operator
  formalism for open quantum systems}},\ }\href
  {https://doi.org/10.1103/PhysRevA.85.032111} {\bibfield  {journal} {\bibinfo
  {journal} {Phys. Rev. A}\ }\textbf {\bibinfo {volume} {85}},\ \bibinfo
  {pages} {032111} (\bibinfo {year} {2012})}\BibitemShut {NoStop}%
\bibitem [{\citenamefont {Molenda}\ \emph {et~al.}(2026)\citenamefont
  {Molenda}, \citenamefont {Zhong}, \citenamefont {Viswanathan}, \citenamefont
  {Li}, \citenamefont {Yan}, \citenamefont {Marino},\ and\ \citenamefont
  {Blume}}]{molenda}%
  \BibitemOpen
  \bibfield  {author} {\bibinfo {author} {\bibfnamefont {X.}~\bibnamefont
  {Molenda}}, \bibinfo {author} {\bibfnamefont {S.}~\bibnamefont {Zhong}},
  \bibinfo {author} {\bibfnamefont {B.}~\bibnamefont {Viswanathan}}, \bibinfo
  {author} {\bibfnamefont {X.}~\bibnamefont {Li}}, \bibinfo {author}
  {\bibfnamefont {Y.}~\bibnamefont {Yan}}, \bibinfo {author} {\bibfnamefont
  {A.~M.}\ \bibnamefont {Marino}},\ and\ \bibinfo {author} {\bibfnamefont
  {D.}~\bibnamefont {Blume}},\ }\bibfield  {title} {\bibinfo {title}
  {{Auxiliary-state-facilitated phase synchronization phenomena in isolated
  spin systems}},\ }\href {https://doi.org/10.1103/kxtk-r6hn} {\bibfield
  {journal} {\bibinfo  {journal} {Phys. Rev. A}\ }\textbf {\bibinfo {volume}
  {113}},\ \bibinfo {pages} {023711} (\bibinfo {year} {2026})}\BibitemShut
  {NoStop}%
\bibitem [{\citenamefont {Verstraete}\ \emph {et~al.}(2009)\citenamefont
  {Verstraete}, \citenamefont {Wolf},\ and\ \citenamefont
  {Ignacio~Cirac}}]{Verstraete2009}%
  \BibitemOpen
  \bibfield  {author} {\bibinfo {author} {\bibfnamefont {F.}~\bibnamefont
  {Verstraete}}, \bibinfo {author} {\bibfnamefont {M.~M.}\ \bibnamefont
  {Wolf}},\ and\ \bibinfo {author} {\bibfnamefont {J.}~\bibnamefont
  {Ignacio~Cirac}},\ }\bibfield  {title} {\bibinfo {title} {{Quantum
  computation and quantum-state engineering driven by dissipation}},\ }\href
  {https://doi.org/10.1038/nphys1342} {\bibfield  {journal} {\bibinfo
  {journal} {Nature Physics}\ }\textbf {\bibinfo {volume} {5}},\ \bibinfo
  {pages} {633} (\bibinfo {year} {2009})}\BibitemShut {NoStop}%
\bibitem [{\citenamefont {Harrington}\ \emph {et~al.}(2022)\citenamefont
  {Harrington}, \citenamefont {Mueller},\ and\ \citenamefont
  {Murch}}]{Harrington2022}%
  \BibitemOpen
  \bibfield  {author} {\bibinfo {author} {\bibfnamefont {P.~M.}\ \bibnamefont
  {Harrington}}, \bibinfo {author} {\bibfnamefont {E.~J.}\ \bibnamefont
  {Mueller}},\ and\ \bibinfo {author} {\bibfnamefont {K.~W.}\ \bibnamefont
  {Murch}},\ }\bibfield  {title} {\bibinfo {title} {{Engineered dissipation for
  quantum information science}},\ }\href
  {https://doi.org/10.1038/s42254-022-00494-8} {\bibfield  {journal} {\bibinfo
  {journal} {Nature Reviews Physics}\ }\textbf {\bibinfo {volume} {4}},\
  \bibinfo {pages} {660} (\bibinfo {year} {2022})}\BibitemShut {NoStop}%
\bibitem [{\citenamefont {B\'egoc}\ \emph {et~al.}(2025)\citenamefont
  {B\'egoc}, \citenamefont {Cichelli}, \citenamefont {Singh}, \citenamefont
  {Bensch}, \citenamefont {Amico}, \citenamefont {Perciavalle}, \citenamefont
  {Rossini}, \citenamefont {Amico},\ and\ \citenamefont {Morsch}}]{w3x9-ll79}%
  \BibitemOpen
  \bibfield  {author} {\bibinfo {author} {\bibfnamefont {B.}~\bibnamefont
  {B\'egoc}}, \bibinfo {author} {\bibfnamefont {G.}~\bibnamefont {Cichelli}},
  \bibinfo {author} {\bibfnamefont {S.~P.}\ \bibnamefont {Singh}}, \bibinfo
  {author} {\bibfnamefont {F.}~\bibnamefont {Bensch}}, \bibinfo {author}
  {\bibfnamefont {V.}~\bibnamefont {Amico}}, \bibinfo {author} {\bibfnamefont
  {F.}~\bibnamefont {Perciavalle}}, \bibinfo {author} {\bibfnamefont
  {D.}~\bibnamefont {Rossini}}, \bibinfo {author} {\bibfnamefont
  {L.}~\bibnamefont {Amico}},\ and\ \bibinfo {author} {\bibfnamefont
  {O.}~\bibnamefont {Morsch}},\ }\bibfield  {title} {\bibinfo {title}
  {{Controlled dissipation for Rydberg atom experiments}},\ }\href
  {https://doi.org/10.1103/w3x9-ll79} {\bibfield  {journal} {\bibinfo
  {journal} {Phys. Rev. A}\ }\textbf {\bibinfo {volume} {112}},\ \bibinfo
  {pages} {023312} (\bibinfo {year} {2025})}\BibitemShut {NoStop}%
\bibitem [{\citenamefont {Rossini}\ and\ \citenamefont
  {Vicari}(2021)}]{ROSSINI20211}%
  \BibitemOpen
  \bibfield  {author} {\bibinfo {author} {\bibfnamefont {D.}~\bibnamefont
  {Rossini}}\ and\ \bibinfo {author} {\bibfnamefont {E.}~\bibnamefont
  {Vicari}},\ }\bibfield  {title} {\bibinfo {title} {{Coherent and dissipative
  dynamics at quantum phase transitions}},\ }\href
  {https://doi.org/https://doi.org/10.1016/j.physrep.2021.08.003} {\bibfield
  {journal} {\bibinfo  {journal} {Physics Reports}\ }\textbf {\bibinfo {volume}
  {936}},\ \bibinfo {pages} {1} (\bibinfo {year} {2021})},\ \bibinfo {note}
  {coherent and dissipative dynamics at quantum phase transitions}\BibitemShut
  {NoStop}%
\bibitem [{\citenamefont {Banchi}\ \emph {et~al.}(2014)\citenamefont {Banchi},
  \citenamefont {Giorda},\ and\ \citenamefont {Zanardi}}]{PhysRevE.89.022102}%
  \BibitemOpen
  \bibfield  {author} {\bibinfo {author} {\bibfnamefont {L.}~\bibnamefont
  {Banchi}}, \bibinfo {author} {\bibfnamefont {P.}~\bibnamefont {Giorda}},\
  and\ \bibinfo {author} {\bibfnamefont {P.}~\bibnamefont {Zanardi}},\
  }\bibfield  {title} {\bibinfo {title} {{Quantum information-geometry of
  dissipative quantum phase transitions}},\ }\href
  {https://doi.org/10.1103/PhysRevE.89.022102} {\bibfield  {journal} {\bibinfo
  {journal} {Phys. Rev. E}\ }\textbf {\bibinfo {volume} {89}},\ \bibinfo
  {pages} {022102} (\bibinfo {year} {2014})}\BibitemShut {NoStop}%
\bibitem [{\citenamefont {Carmichael}(2015)}]{PhysRevX.5.031028}%
  \BibitemOpen
  \bibfield  {author} {\bibinfo {author} {\bibfnamefont {H.~J.}\ \bibnamefont
  {Carmichael}},\ }\bibfield  {title} {\bibinfo {title} {{Breakdown of Photon
  Blockade: A Dissipative Quantum Phase Transition in Zero Dimensions}},\
  }\href {https://doi.org/10.1103/PhysRevX.5.031028} {\bibfield  {journal}
  {\bibinfo  {journal} {Phys. Rev. X}\ }\textbf {\bibinfo {volume} {5}},\
  \bibinfo {pages} {031028} (\bibinfo {year} {2015})}\BibitemShut {NoStop}%
\bibitem [{\citenamefont {Degen}\ \emph {et~al.}(2017)\citenamefont {Degen},
  \citenamefont {Reinhard},\ and\ \citenamefont
  {Cappellaro}}]{RevModPhys.89.035002}%
  \BibitemOpen
  \bibfield  {author} {\bibinfo {author} {\bibfnamefont {C.~L.}\ \bibnamefont
  {Degen}}, \bibinfo {author} {\bibfnamefont {F.}~\bibnamefont {Reinhard}},\
  and\ \bibinfo {author} {\bibfnamefont {P.}~\bibnamefont {Cappellaro}},\
  }\bibfield  {title} {\bibinfo {title} {{Quantum sensing}},\ }\href
  {https://doi.org/10.1103/RevModPhys.89.035002} {\bibfield  {journal}
  {\bibinfo  {journal} {Rev. Mod. Phys.}\ }\textbf {\bibinfo {volume} {89}},\
  \bibinfo {pages} {035002} (\bibinfo {year} {2017})}\BibitemShut {NoStop}%
\bibitem [{\citenamefont {Vaidya}\ \emph {et~al.}(2025)\citenamefont {Vaidya},
  \citenamefont {J\"ager},\ and\ \citenamefont
  {Shankar}}]{PhysRevA.111.012410}%
  \BibitemOpen
  \bibfield  {author} {\bibinfo {author} {\bibfnamefont {G.~M.}\ \bibnamefont
  {Vaidya}}, \bibinfo {author} {\bibfnamefont {S.~B.}\ \bibnamefont
  {J\"ager}},\ and\ \bibinfo {author} {\bibfnamefont {A.}~\bibnamefont
  {Shankar}},\ }\bibfield  {title} {\bibinfo {title} {{Quantum synchronization
  and dissipative quantum sensing}},\ }\href
  {https://doi.org/10.1103/PhysRevA.111.012410} {\bibfield  {journal} {\bibinfo
   {journal} {Phys. Rev. A}\ }\textbf {\bibinfo {volume} {111}},\ \bibinfo
  {pages} {012410} (\bibinfo {year} {2025})}\BibitemShut {NoStop}%
\bibitem [{\citenamefont {Lai}\ \emph {et~al.}(2025)\citenamefont {Lai},
  \citenamefont {Miranowicz},\ and\ \citenamefont {Nori}}]{Lai2025}%
  \BibitemOpen
  \bibfield  {author} {\bibinfo {author} {\bibfnamefont {D.-G.}\ \bibnamefont
  {Lai}}, \bibinfo {author} {\bibfnamefont {A.}~\bibnamefont {Miranowicz}},\
  and\ \bibinfo {author} {\bibfnamefont {F.}~\bibnamefont {Nori}},\ }\bibfield
  {title} {\bibinfo {title} {{Nonreciprocal quantum synchronization}},\ }\href
  {https://doi.org/10.1038/s41467-025-63408-z} {\bibfield  {journal} {\bibinfo
  {journal} {Nature Communications}\ }\textbf {\bibinfo {volume} {16}},\
  \bibinfo {pages} {8491} (\bibinfo {year} {2025})}\BibitemShut {NoStop}%
\bibitem [{\citenamefont {Dutta}\ and\ \citenamefont
  {Cooper}(2019)}]{PhysRevLett.123.250401}%
  \BibitemOpen
  \bibfield  {author} {\bibinfo {author} {\bibfnamefont {S.}~\bibnamefont
  {Dutta}}\ and\ \bibinfo {author} {\bibfnamefont {N.~R.}\ \bibnamefont
  {Cooper}},\ }\bibfield  {title} {\bibinfo {title} {{Critical Response of a
  Quantum van der Pol Oscillator}},\ }\href
  {https://doi.org/10.1103/PhysRevLett.123.250401} {\bibfield  {journal}
  {\bibinfo  {journal} {Phys. Rev. Lett.}\ }\textbf {\bibinfo {volume} {123}},\
  \bibinfo {pages} {250401} (\bibinfo {year} {2019})}\BibitemShut {NoStop}%
\bibitem [{\citenamefont {Zhong}\ \emph {et~al.}(2026)\citenamefont {Zhong},
  \citenamefont {Sudler}, \citenamefont {Blume},\ and\ \citenamefont
  {Marino}}]{zhong2026light}%
  \BibitemOpen
  \bibfield  {author} {\bibinfo {author} {\bibfnamefont {S.}~\bibnamefont
  {Zhong}}, \bibinfo {author} {\bibfnamefont {A.}~\bibnamefont {Sudler}},
  \bibinfo {author} {\bibfnamefont {D.}~\bibnamefont {Blume}},\ and\ \bibinfo
  {author} {\bibfnamefont {A.~M.}\ \bibnamefont {Marino}},\ }\bibfield  {title}
  {\bibinfo {title} {{Light storage, retrieval, and controllable interference
  in an atomic tripod system}},\ }\href
  {https://link.aps.org/doi/10.1103/2175-ztdj} {\bibfield  {journal} {\bibinfo
  {journal} {Physical Review A}\ }\textbf {\bibinfo {volume} {113}},\ \bibinfo
  {pages} {013103} (\bibinfo {year} {2026})}\BibitemShut {NoStop}%
\bibitem [{\citenamefont {Galve}\ \emph {et~al.}(2017)\citenamefont {Galve},
  \citenamefont {Giorgi},\ and\ \citenamefont {Zambrini}}]{galve2017quantum}%
  \BibitemOpen
  \bibfield  {author} {\bibinfo {author} {\bibfnamefont {F.}~\bibnamefont
  {Galve}}, \bibinfo {author} {\bibfnamefont {G.~L.}\ \bibnamefont {Giorgi}},\
  and\ \bibinfo {author} {\bibfnamefont {R.}~\bibnamefont {Zambrini}},\
  }\bibfield  {title} {\bibinfo {title} {{Quantum Correlations and
  Synchronization Measures}},\ }in\ \href
  {https://doi.org/10.1007/978-3-319-53412-1_18} {\emph {\bibinfo {booktitle}
  {Lectures on General Quantum Correlations and their Applications}}}\
  (\bibinfo  {publisher} {Springer},\ \bibinfo {year} {2017})\ pp.\ \bibinfo
  {pages} {393--420}\BibitemShut {NoStop}%
\bibitem [{\citenamefont {Sudler}\ \emph {et~al.}(2024)\citenamefont {Sudler},
  \citenamefont {Talukdar},\ and\ \citenamefont {Blume}}]{PhysRevE.109.054207}%
  \BibitemOpen
  \bibfield  {author} {\bibinfo {author} {\bibfnamefont {A.~J.}\ \bibnamefont
  {Sudler}}, \bibinfo {author} {\bibfnamefont {J.}~\bibnamefont {Talukdar}},\
  and\ \bibinfo {author} {\bibfnamefont {D.}~\bibnamefont {Blume}},\ }\bibfield
   {title} {\bibinfo {title} {{Driven generalized quantum Rayleigh--van der Pol
  oscillators: Phase localization and spectral response}},\ }\href
  {https://doi.org/10.1103/PhysRevE.109.054207} {\bibfield  {journal} {\bibinfo
   {journal} {Phys. Rev. E}\ }\textbf {\bibinfo {volume} {109}},\ \bibinfo
  {pages} {054207} (\bibinfo {year} {2024})}\BibitemShut {NoStop}%
\bibitem [{foo()}]{footnote_notation}%
  \BibitemOpen
  \href@noop {} {\bibinfo {title} {{Note that the labeling of the states and
  the phase conventions in the present work differ from
  Ref.~\protect\cite{molenda}}}}\BibitemShut {NoStop}%
\bibitem [{SM(2026)}]{SM}%
  \BibitemOpen
  \href@noop {} {\bibinfo {title} {Supplemental material}} (\bibinfo {year}
  {2026})\BibitemShut {NoStop}%
\bibitem [{\citenamefont {Mok}\ \emph {et~al.}(2020)\citenamefont {Mok},
  \citenamefont {Kwek},\ and\ \citenamefont
  {Heimonen}}]{PhysRevResearch.2.033422}%
  \BibitemOpen
  \bibfield  {author} {\bibinfo {author} {\bibfnamefont {W.-K.}\ \bibnamefont
  {Mok}}, \bibinfo {author} {\bibfnamefont {L.-C.}\ \bibnamefont {Kwek}},\ and\
  \bibinfo {author} {\bibfnamefont {H.}~\bibnamefont {Heimonen}},\ }\bibfield
  {title} {\bibinfo {title} {{Synchronization boost with single-photon
  dissipation in the deep quantum regime}},\ }\href
  {https://doi.org/10.1103/PhysRevResearch.2.033422} {\bibfield  {journal}
  {\bibinfo  {journal} {Phys. Rev. Res.}\ }\textbf {\bibinfo {volume} {2}},\
  \bibinfo {pages} {033422} (\bibinfo {year} {2020})}\BibitemShut {NoStop}%
\bibitem [{\citenamefont {Pegg}\ and\ \citenamefont {Barnett}(1988)}]{pegg}%
  \BibitemOpen
  \bibfield  {author} {\bibinfo {author} {\bibfnamefont {D.~T.}\ \bibnamefont
  {Pegg}}\ and\ \bibinfo {author} {\bibfnamefont {S.~M.}\ \bibnamefont
  {Barnett}},\ }\bibfield  {title} {\bibinfo {title} {{Unitary Phase Operator
  in Quantum Mechanics}},\ }\href {https://doi.org/10.1209/0295-5075/6/6/002}
  {\bibfield  {journal} {\bibinfo  {journal} {Europhysics Letters}\ }\textbf
  {\bibinfo {volume} {6}},\ \bibinfo {pages} {483} (\bibinfo {year}
  {1988})}\BibitemShut {NoStop}%
\bibitem [{\citenamefont {Shapiro}\ and\ \citenamefont
  {Shepard}(1991)}]{PhysRevA.43.3795}%
  \BibitemOpen
  \bibfield  {author} {\bibinfo {author} {\bibfnamefont {J.~H.}\ \bibnamefont
  {Shapiro}}\ and\ \bibinfo {author} {\bibfnamefont {S.~R.}\ \bibnamefont
  {Shepard}},\ }\bibfield  {title} {\bibinfo {title} {{Quantum phase
  measurement: A system-theory perspective}},\ }\href
  {https://doi.org/10.1103/PhysRevA.43.3795} {\bibfield  {journal} {\bibinfo
  {journal} {Phys. Rev. A}\ }\textbf {\bibinfo {volume} {43}},\ \bibinfo
  {pages} {3795} (\bibinfo {year} {1991})}\BibitemShut {NoStop}%
\bibitem [{\citenamefont {Joshi}\ and\ \citenamefont
  {Xiao}(2005)}]{PhysRevA.71.041801}%
  \BibitemOpen
  \bibfield  {author} {\bibinfo {author} {\bibfnamefont {A.}~\bibnamefont
  {Joshi}}\ and\ \bibinfo {author} {\bibfnamefont {M.}~\bibnamefont {Xiao}},\
  }\bibfield  {title} {\bibinfo {title} {{Generalized dark-state polaritons for
  photon memory in multilevel atomic media}},\ }\href
  {https://doi.org/10.1103/PhysRevA.71.041801} {\bibfield  {journal} {\bibinfo
  {journal} {Phys. Rev. A}\ }\textbf {\bibinfo {volume} {71}},\ \bibinfo
  {pages} {041801(R)} (\bibinfo {year} {2005})}\BibitemShut {NoStop}%
\bibitem [{\citenamefont {Fleischhauer}\ and\ \citenamefont
  {Lukin}(2002)}]{PhysRevA.65.022314}%
  \BibitemOpen
  \bibfield  {author} {\bibinfo {author} {\bibfnamefont {M.}~\bibnamefont
  {Fleischhauer}}\ and\ \bibinfo {author} {\bibfnamefont {M.~D.}\ \bibnamefont
  {Lukin}},\ }\bibfield  {title} {\bibinfo {title} {{Quantum memory for
  photons: Dark-state polaritons}},\ }\href
  {https://doi.org/10.1103/PhysRevA.65.022314} {\bibfield  {journal} {\bibinfo
  {journal} {Phys. Rev. A}\ }\textbf {\bibinfo {volume} {65}},\ \bibinfo
  {pages} {022314} (\bibinfo {year} {2002})}\BibitemShut {NoStop}%
\bibitem [{\citenamefont {Beck}\ and\ \citenamefont
  {Mazets}(2017)}]{PhysRevA.95.013818}%
  \BibitemOpen
  \bibfield  {author} {\bibinfo {author} {\bibfnamefont {S.}~\bibnamefont
  {Beck}}\ and\ \bibinfo {author} {\bibfnamefont {I.~E.}\ \bibnamefont
  {Mazets}},\ }\bibfield  {title} {\bibinfo {title} {{Propagation of coupled
  dark-state polaritons and storage of light in a tripod medium}},\ }\href
  {https://doi.org/10.1103/PhysRevA.95.013818} {\bibfield  {journal} {\bibinfo
  {journal} {Phys. Rev. A}\ }\textbf {\bibinfo {volume} {95}},\ \bibinfo
  {pages} {013818} (\bibinfo {year} {2017})}\BibitemShut {NoStop}%
\bibitem [{\citenamefont {L\"orch}\ \emph {et~al.}(2017)\citenamefont
  {L\"orch}, \citenamefont {Nigg}, \citenamefont {Nunnenkamp}, \citenamefont
  {Tiwari},\ and\ \citenamefont {Bruder}}]{PhysRevLett.118.243602}%
  \BibitemOpen
  \bibfield  {author} {\bibinfo {author} {\bibfnamefont {N.}~\bibnamefont
  {L\"orch}}, \bibinfo {author} {\bibfnamefont {S.~E.}\ \bibnamefont {Nigg}},
  \bibinfo {author} {\bibfnamefont {A.}~\bibnamefont {Nunnenkamp}}, \bibinfo
  {author} {\bibfnamefont {R.~P.}\ \bibnamefont {Tiwari}},\ and\ \bibinfo
  {author} {\bibfnamefont {C.}~\bibnamefont {Bruder}},\ }\bibfield  {title}
  {\bibinfo {title} {{Quantum Synchronization Blockade: Energy Quantization
  Hinders Synchronization of Identical Oscillators}},\ }\href
  {https://doi.org/10.1103/PhysRevLett.118.243602} {\bibfield  {journal}
  {\bibinfo  {journal} {Phys. Rev. Lett.}\ }\textbf {\bibinfo {volume} {118}},\
  \bibinfo {pages} {243602} (\bibinfo {year} {2017})}\BibitemShut {NoStop}%
\bibitem [{\citenamefont {Solanki}\ \emph {et~al.}(2023)\citenamefont
  {Solanki}, \citenamefont {Mehdi}, \citenamefont {Hajdu\ifmmode~\check{s}\else
  \v{s}\fi{}ek},\ and\ \citenamefont {Vinjanampathy}}]{PhysRevA.108.022216}%
  \BibitemOpen
  \bibfield  {author} {\bibinfo {author} {\bibfnamefont {P.}~\bibnamefont
  {Solanki}}, \bibinfo {author} {\bibfnamefont {F.~M.}\ \bibnamefont {Mehdi}},
  \bibinfo {author} {\bibfnamefont {M.}~\bibnamefont
  {Hajdu\ifmmode~\check{s}\else \v{s}\fi{}ek}},\ and\ \bibinfo {author}
  {\bibfnamefont {S.}~\bibnamefont {Vinjanampathy}},\ }\bibfield  {title}
  {\bibinfo {title} {{Symmetries and synchronization blockade}},\ }\href
  {https://doi.org/10.1103/PhysRevA.108.022216} {\bibfield  {journal} {\bibinfo
   {journal} {Phys. Rev. A}\ }\textbf {\bibinfo {volume} {108}},\ \bibinfo
  {pages} {022216} (\bibinfo {year} {2023})}\BibitemShut {NoStop}%
\bibitem [{\citenamefont {Gilmore}\ \emph {et~al.}(1975)\citenamefont
  {Gilmore}, \citenamefont {Bowden},\ and\ \citenamefont
  {Narducci}}]{PhysRevA.12.1019}%
  \BibitemOpen
  \bibfield  {author} {\bibinfo {author} {\bibfnamefont {R.}~\bibnamefont
  {Gilmore}}, \bibinfo {author} {\bibfnamefont {C.~M.}\ \bibnamefont
  {Bowden}},\ and\ \bibinfo {author} {\bibfnamefont {L.~M.}\ \bibnamefont
  {Narducci}},\ }\bibfield  {title} {\bibinfo {title} {{Classical-quantum
  correspondence for multilevel systems}},\ }\href
  {https://doi.org/10.1103/PhysRevA.12.1019} {\bibfield  {journal} {\bibinfo
  {journal} {Phys. Rev. A}\ }\textbf {\bibinfo {volume} {12}},\ \bibinfo
  {pages} {1019} (\bibinfo {year} {1975})}\BibitemShut {NoStop}%
\bibitem [{\citenamefont {Hush}\ \emph {et~al.}(2015)\citenamefont {Hush},
  \citenamefont {Li}, \citenamefont {Genway}, \citenamefont {Lesanovsky},\ and\
  \citenamefont {Armour}}]{PhysRevA.91.061401}%
  \BibitemOpen
  \bibfield  {author} {\bibinfo {author} {\bibfnamefont {M.~R.}\ \bibnamefont
  {Hush}}, \bibinfo {author} {\bibfnamefont {W.}~\bibnamefont {Li}}, \bibinfo
  {author} {\bibfnamefont {S.}~\bibnamefont {Genway}}, \bibinfo {author}
  {\bibfnamefont {I.}~\bibnamefont {Lesanovsky}},\ and\ \bibinfo {author}
  {\bibfnamefont {A.~D.}\ \bibnamefont {Armour}},\ }\bibfield  {title}
  {\bibinfo {title} {{Spin correlations as a probe of quantum synchronization
  in trapped-ion phonon lasers}},\ }\href
  {https://doi.org/10.1103/PhysRevA.91.061401} {\bibfield  {journal} {\bibinfo
  {journal} {Phys. Rev. A}\ }\textbf {\bibinfo {volume} {91}},\ \bibinfo
  {pages} {061401(R)} (\bibinfo {year} {2015})}\BibitemShut {NoStop}%
\end{thebibliography}

\end{document}